\documentclass[11pt]{article}
\usepackage{epsf,epsfig,float,amssymb,latexsym,amsmath,amsthm,fancyhdr,pst-all}
\usepackage{graphics,psfrag}
\usepackage{multirow}
\usepackage{bbold}
\usepackage{cite}
\usepackage[normalem]{ulem} 
\usepackage[numbers,compress]{natbib}
\usepackage{cite}

\usepackage{doi}
\def\XXint#1#2#3{{\setbox0=\hbox{$#1{#2#3}{\int}$}
     \vcenter{\hbox{$#2#3$}}\kern-.5\wd0}}

\newcommand{\kp}{\kappa}
\newcommand{\lm}{\lambda}
\newcommand{\ep}{\epsilon}
\newcommand{\vep}{\varepsilon}

\newcommand{\vp}{{\mathbf{p}}}
\newcommand{\hvp}{{\hat{\mathbf{p}}}}
\newcommand{\vq}{{\mathbf{q}}}

\newcommand{\vz}{{\mathbf{z}}}
\newcommand{\hvz}{{\hat{\mathbf{z}}}}
\renewcommand{\S}{{S}}
\newcommand{\n}{{N}}

\newcommand{\mfl}{\mathfrak{L}}
\newcommand{\mfs}{\mathfrak{S}}

\newcommand{\la}{\langle}
\newcommand{\ra}{\rangle}
\newcommand{\las}{{_S\langle}}
\newcommand{\ras}{{\rangle_S}}
\newcommand{\nn}{\nonumber}

\hypersetup{backref, colorlinks=false} 

\newcommand{\email}[1]{\footnote{{\em } \texttt{#1}}}

\begin{document}

\thispagestyle{empty}

\title{Unitarization of infinite-range forces: graviton-graviton 
scattering}
\author{D.~Blas$^a$\email{diego.blas@kcl.ac.uk},\, J.~Mart\'{\i}n 
Camalich$^{b,c}$\email{jcamalich@iac.es}, \,
J.~A. Oller$^d$\email{oller@um.es}   \\[0.5em]
{\it \small $^a$Theoretical Particle Physics and Cosmology Group, Department of 
Physics,}\\
{\it \small King's College London, Strand, London WC2R 2LS, UK.}\\
{\it \small $^b$Instituto de Astrof\'{\i}sica de Canarias, C/ V\'{\i}a L\'actea,
s/n E38205 - La Laguna, Tenerife, Spain.}\\
{\it \small $^c$Universidad de La Laguna, Departamento de Astrof\'{\i}sica, La 
Laguna, Tenerife, Spain.}\\
{\it \small $^d$Departamento de F\'{\i}sica. Universidad de Murcia. E-30071,
Murcia. Spain.}
}  
\maketitle
\begin{abstract}
\vspace{-10cm}
\begin{flushright}
 KCL-2020-57
\end{flushright}
\vspace{9.6cm}

A method to unitarize the scattering amplitude produced by infinite-range forces 
is developed and applied to Born terms. In order to apply $S$-matrix techniques, 
based on unitarity and analyticity, we first derive  an 
$S$-matrix free of infrared divergences. This is 
achieved by removing a divergent phase factor due to the interactions mediated 
by the massless particles in the crossed channels, a procedure that is related 
to previous formalisms to treat infrared divergences. 
We apply this method in detail by unitarizing the Born terms for 
graviton-graviton scattering in pure gravity and we find a scalar 
graviton-graviton resonance with vacuum quantum numbers ($J^{PC}=0^{++}$) that 
we call the \textit{graviball}.
Remarkably, this resonance is located below the Planck mass but deep in the 
complex $s$-plane (with $s$ the usual Mandelstam variable), so that its effects 
along the physical real $s$ axis peak for values significantly lower than this 
scale.
This implies that the corrections to the leading-order amplitude in the 
gravitational effective field theory are larger than expected from  naive 
dimensional analysis for $s$ around and above the peak position. We   argue 
that the position and width of the graviball are reduced when including extra 
light fields in the theory. This  could  lead to phenomenological consequences 
in  scenarios of quantum gravity with a large number of such fields or, in 
general, with a low-energy ultraviolet completion. We also apply  this formalism 
to two non-relativistic potentials with exact known solutions for the scattering 
amplitudes: Coulomb scattering and an energy-dependent potential obtained from 
the Coulomb one with a zero at threshold. This latter case shares the same $J=0$ 
partial-wave projected Born term as the graviton-graviton case, except for a 
global factor. We find that the relevant resonance structure of these examples 
is reproduced by our methods, which represents a strong indication  of their 
robustness.

\end{abstract}


\newpage
\tableofcontents

\newpage

\section{Introduction}
\label{sec:intro}
\def\theequation{\arabic{section}.\arabic{equation}}
\setcounter{equation}{0}

Several techniques from quantum field theory have been successfully applied to 
scattering processes in gravitational theories, see e.g. 
\cite{Bern:2002kj,Giddings:2011xs,Porto:2016pyg,Cheung:2017pzi,
Strominger:2017zoo,Donoghue:2017pgk,Ciafaloni:2018uwe} for recent reviews, and 
references therein. 
The study of an $S$-matrix formulation has been a particularly fruitful direction of research where results from black-hole physics and string theory 
have found a rich niche, e.g. 
\cite{Gross:1968in,Amati:1987wq,tHooft:1996rdg,Amati:2007ak,Giddings:2009gj,
Bezrukov:2015ufa,Dvali:2014ila}. Furthermore, unitarity arguments have been used 
in the past as a guide to build ultraviolet completions of general relativity 
{\it above the Planck mass}, see e.g.~\cite{Alonso:2019ptb} for a recent 
effort. 
The purpose of this work and the related \cite{Blas:2020och} is to initiate the 
exploration of a  complementary direction in this program by considering the 
unitarization of Born amplitudes, the lowest-order graviton-graviton scattering, 
at energies {\it below the Planck mass}. In this program, we will also derive 
generic results that can be used to unitarize other theories with long-range 
interaction. 

One of the  motivations for our study is to clarify whether the scattering in 
this kinematic region is resonant.  Recall that in the analogous case  of pure 
Yang-Mills, there is theoretical evidence that
 the self-interactions of the gluons 
 generate  massive states (\textit{glueballs}) filling its spectrum 
~\cite{Fritzsch:1975tx,Bali:1993fb,Morningstar:1999rf,Lucini:2004my}.  These 
states lead to a new form of matter in full quantum chromodynamics (QCD) and 
have been identified with resonances measured in 
experiment~\cite{albaladejo.200207,Chanowitz:2005du,Brodsky:2018snc,
Mathieu:2008me}. Therefore, it is interesting and pertinent to ask whether 
exotic resonance states of two gravitons (\textit{graviballs}) could arise 
because of their gravitational interactions in a quantum theory. These may even 
lead to phenomenological consequences at relatively low-energy scales. 

In order to address this question, we deal with the quantum formulation of  
general  relativity within the framework of effective field theory (EFT) where 
gravitational interactions are organized in a derivative (or momentum) expansion 
\cite{weinphys,Donoghue:1993eb,donoghue.200205.1,BjerrumBohr:2002kt,burgess,
Donoghue:2017pgk}. This is valid as long as the energies considered are well 
below the Planck mass $M_P=G^{-1/2}\sim 1 \times 10^{19}~\text{GeV}$,
which is the natural cutoff of the gravitational 
EFT~\cite{han.200204.1,Veneziano:2001ah,Donoghue:1999qh,burgess}. Interestingly, 
another energy expansion serves as the  basis for the EFT of QCD at low energy. 
This is called chiral perturbation theory 
(ChPT)~\cite{Weinberg:1966fm,Gasser:1983yg,Pich:1995bw} and it describes the 
self-interactions of pions and other hadrons with a cutoff $\Lambda\sim1$~GeV. 
Since the low-energy limits of QCD and quantum gravity can be treated with 
similar EFTs, analogies between the two could be expected. 
With this in mind, one may recall that in QCD there is a state with a mass and 
width of the order of $\Lambda/2$, the so-called $\sigma$-meson or $f_0(500)$, 
which is the lightest resonance known in this theory~\cite{pdg}. The presence of 
this state can be rigorously determined by combining amplitudes calculated in 
the EFT with $S$-matrix methods 
\cite{Oller:1997ti,Caprini:2005zr,Pelaez:2015qba}, or only from the latter ones 
\cite{GarciaMartin:2011jx}.  
In fact, clear resonance peaks associated with the $\sigma$ have been observed  
experimentally 
in the two-pion invariant mass distributions of $D^+\to\pi^-\pi^+\pi^+$
\cite{Aitala:2000xu,Link:2003gb,Bonvicini:2007tc} or $J/\Psi\to 
\omega\pi^+\pi^-$ \cite{Augustin:1988ja,Ablikim:2004qna}. Furthermore, this 
indicates that perturbative calculations for some processes, like the 
scalar-isoscalar $\pi\pi$ phase shifts, $\gamma\gamma\to \pi\pi$, $\eta\to 3\pi$ 
decays, etc, are affected by strong unitarity corrections  
~(see~\cite{orev,Oller:2020guq,Pelaez:2015qba} for reviews). 

In this paper we apply a combination of EFT and $S$-matrix methods to the 
scattering of gravitons at low energies to shed light on the possible presence 
of graviballs, as discussed by us recently in Ref.~\cite{graviballshort}.
Our study is complementary to other  analyses that considered the scattering of 
particles with different flavors induced by graviton exchange in the 
$s$-channel~\cite{han.200204.1,donoghue.200204.1,calmet.200204.1}.
These works employ the one-loop calculation of the graviton self-energy in the 
EFT in the presence of $N$ light degrees of freedom~\cite{donoghue.200205.1}, 
and emphasize the importance of nonperturbative effects in quantum gravity in 
order to restore unitarity at $s\sim N\,G^{-1}$. Ref.~\cite{calmet.200204.1} 
also studies the $J=2$ resonances that result from the resummed formula of the 
graviton propagator.
However, none of these references address the scattering of particles with the 
same flavor.  
In this case, there are contributions from the exchange of gravitons in the {\it 
crossed channels} leading to {\it infrared (IR)  divergences} in the 
partial-wave amplitudes (PWAs). 
These difficulties already appear at the level of the partial-wave projected 
Born amplitude,
and stem from the angular projection in the angular region of forward or 
backward scattering, depending on whether the graviton exchange is in the $t$ or 
$u$ channel, respectively (see also \cite{Giddings:2009gj}).

IR divergences are a well understood feature of theories with infinite-range 
interactions~\cite{Bloch:1937pw,Yennie:1961ad,Weinberg:1965nx,Kulish.200121}. 
Their contributions to scattering matrix elements were derived for gravitons in 
the
classic reference~\cite{Weinberg:1965nx}, showing that the techniques used to 
remove them from physical observables are common to quantum electrodynamics 
(QED) and a quantum theory of gravitons. It is also possible to tackle the IR 
divergences so as to end with finite $S$-matrix amplitudes. 
The general procedure for quantum electrodynamics (QED) was elaborated by Kulish 
and Faddeev in Ref.~\cite{Kulish.200121}, and it allows one to define a new 
unitary $S$-matrix operator with well defined matrix elements free of IR 
divergences. Ref.~\cite{Ware:2013zja} provided an explicit  extension of the 
formalism to gravity (see also 
\cite{Hirai:2020kzx,Hannesdottir:2019rqq,Himwich:2020rro} for related recent 
works).

We will use these methods to  remove an  IR divergent global phase from the 
$S$-matrix, which is common to all partial waves. This phase, conjectured long 
before by Dalitz for the case of elastic non-relativistic Coulomb scattering 
\cite{Dalitz:1951}, stems from the resummation of the diagrams that conform the 
Born series by iterating the one-graviton exchange and it  first appeared  in 
Ref.~\cite{Weinberg:1965nx}.
The new $S$-matrix allows us to implement nonperturbative unitarity methods for 
partial-wave amplitudes (PWAs) based on $S$-matrix theory, so that our formalism 
can be regarded as an extension to infinite-range interactions of the $S$-matrix 
techniques so popular in QCD. The study of PWAs is typically the most adequate 
method to impose restrictions of unitarity to amplitudes at low energies. 
As an illustration of the method developed here, and
to clarify the new subtleties generated by the IR divergences, we will introduce 
a toy model in non-relativistic Quantum Mechanics, that we call Adler-Coulomb 
(AC) scattering, which has
the same partial-wave projected Born amplitudes as graviton-graviton scattering 
for {the} same final and initial helicities and that can be solved exactly. We 
also study with the same techniques the unitarization of pure Coulomb 
scattering, which presents interesting new aspects, like, for instance, that the 
unitarization becomes valid for large (instead of low) three-momenta.

 By applying the previous $S$-matrix 
techniques to pure gravity, the resulting $S$-wave PWA has a resonant pole 
corresponding to a graviball with vacuum quantum numbers 
($J^{PC}=0^{++}$).~\footnote{Bound states of gravitational  waves, known as {\it 
geons}, have been studied in classical general relativity in 
\cite{Brill:1964zz,Anderson_1997}. Also the recent work \cite{Guiot:2020pku} 
discussed the possibility of bound states of systems of gravitons from the 
effective potential at large distances. These states have no relation with our resonance corresponding to  different physical situations.} One important 
peculiarity of this resonance is that its pole position $s_P$ (in the Mandelstam 
variable $s$) is \textit{almost} a purely imaginary number whose absolute value 
lies below $G^{-1}$ if one considers natural assumptions. As a result, the 
dynamics of graviton-graviton scattering would be driven by this resonance even 
at energies significantly lower than the ultraviolet cutoff of the theory.
This may open up several phenomenological implications to test quantum gravity 
predictions at energies lower than the cutoff. Namely, resonant two-graviton 
exchange in $S$-wave could induce rescattering effects in different processes 
as, for instance, the production of multiple gravitons by some energetic or 
massive source. As we will discuss, this may be particularly interesting for 
theories with large number of light degrees of freedom for which this resonance 
is expected to become narrower and lighter. Our present study derives in  detail 
the results presented in Ref.~\cite{graviballshort} to deal with the IR 
divergences for infinite-range interactions and the unitarization of the 
corresponding Born terms. It also largely extends this analysis in several 
directions. In particular, we also present the calculations for varying 
space-time dimensions $d$, make a thorough study of the robustness of the 
methods and exploit many of its applications to gravitational and Coulomb-like 
scattering.

Finally, we would like to remark that other results to 
unitarize the high-energy scattering amplitudes through black-hole production 
have been derived
in the self-completeness scenario of gravity~\cite{Dvali:2010bf,Dvali:2014ila}.
The latter are interpreted as bound states of gravitons with masses $ \gg M_P$, 
and may start to leave traces at energies below the cut-off  
\cite{Dvali:2011aa,Dvali:2008fd}.
Furthermore, microscopic  black holes which become lighter with the number $N$ 
of particle species as $M_P/\sqrt{N}$ are predicted in 
Refs.~\cite{Dvali:2008fd,Antoniadis:1998ig}.
Though the precise connection between these results and the {\it graviball} is 
not clear, it is certainly intriguing that both non-perturbative methods agree 
qualitatively on the onset of non trivial states.

The contents of the manuscript are organized as follows. 
The formalism needed to calculate the partial-wave projected Born amplitudes for 
graviton-graviton scattering
is derived in Sec.~\ref{sec:dis}. Some extra technical details in this respect 
are given in the App. \ref{app_states}.
Our treatment of infinite-range interactions leading to partial-wave amplitudes 
(PWAs) free of IR divergences is discussed in Sec.~\ref{sec.200122.1}.
The partial-wave projected Born terms are then  implemented in the unitarization 
method developed in Sec.~\ref{sec.200531.1}. 
The Sec.~\ref{sec.200605.1} applies this unitarization method to $\pi\pi$ with 
$J=0$ and the appearance of the $\sigma$ resonance is treated.   
 Sec.~\ref{sec.200131.1} is then devoted to the discussion of a 
graviton-graviton scalar  resonance or graviball 
which emerges as a pole in the $J=0$ PWA and which manifests in the  rescattering of two 
gravitons with these quantum numbers. Subsec.~\ref{sec.200813.1} is dedicated to the robustness of the graviball, where we discuss the importance of the interplay 
between the fundamental and unitarity cutoffs.
In section \ref{app.170723.2} 
we apply our method to the exactly soluble AC model and non-relativistic Coulomb 
scattering.
We develop a comparison with the known exact solutions, and the  higher-order 
contributions to the unitarized Born terms are worked out. 
We dedicate 
Sec.~\ref{sec.210115.1} to study the PWAs for $d>4$ because 
in that case there are no IR divergences
and this fact can be used to  elaborate a method 
to estimate quantitatively 
the pole positions. This method is tested by applying it to
recover successfully the exact solution of the AC model and is then applied to the graviball. 
Concluding remarks, prospects for future work and new directions  are 
gathered in Sec.~\ref{sec:conclu}. Some technical material is given in 
App.~\ref{app.200805.1} on PWAs in $d$ dimensions,  while $S$-wave Coulomb 
scattering for varying $d$ is treated in App.~\ref{sec.200812.1}.

\section{Partial-wave projection of the graviton-graviton scattering amplitudes}
\label{sec:dis}
\def\theequation{\arabic{section}.\arabic{equation}}
\setcounter{equation}{0}

In this work, we focus on the  the graviton-graviton scattering process
\begin{align}
\label{200119.2}
|\vp_1,\lm_1\ra |\vp_2,\lm_2\ra\to |\vp_3,\lm_3\ra|\vp_4,\lm_4\ra~.
\end{align}
The energy of each graviton is $p_i^0=|\vp_i|$ and the corresponding 
four-momentum is indicated by
$p_i$. Here we denote by $|\vp, \lm\rangle$ the one-graviton state of 
three-momentum $\vp$ and helicity $\lm$. These states are normalized as
\begin{align}
\label{200116.1}
\langle \vp', \lm'|\vp ,\lm\rangle = 2p_0 
(2\pi)^3\delta(\vp'-\vp)\delta_{\lm'\lm}~,
\end{align}
where $p_0=|\vp|$ is the energy of the massless graviton.
Our definitions for the one- and two-graviton states are given in detail in 
App.~\ref{app_states}, where the relation between the basis states with 
well-defined three-momentum and total angular momentum is also worked out. 

The $S$- and $T$-matrix operators are related by 
\begin{align}
\label{200119.1}
S=I+i (2\pi)^4 \delta^{(4)}(P_f-P_i) T~.
\end{align}
states, respectively.
Their matrix elements for two-graviton scattering are expressed in terms of the 
usual Mandelstam variables $s$, $t$ and $u$, defined as 
\begin{align}
\label{200119.3}
s=&(p_1+p_2)^2=(p_3+p_4)^2~,\\
t=&(p_1-p_3)^2=(p_2-p_4)^2~,\nn\\
u=&(p_1-p_4)^2=(p_2-p_3)^2~.\nn
\end{align}
Because of the null mass of a graviton they fulfill that  
\begin{align}
\label{200121.1}
s+t+u&=0~.
\end{align}

The basic building blocks for our study of the graviton-graviton scattering are  
tree-level or Born amplitudes.
Within an EFT calculation of two-graviton scattering these are the lowest-order 
amplitudes
in powers of $G$,  where $G$ is the Newton constant.
We adapt them from the calculation in Ref.~\cite{grisaru.170513.1}.
The normalization in this paper for the Born amplitudes only differs from ours 
by an extra factor $i$. 
The expressions for the Born amplitudes $ F_{\lm_3\lm_4,\lm_1\lm_2}\equiv T^{\rm 
Born}_{\lm_3\lm_4,\lm_1\lm_2}$ are given according to the number of gravitons 
with  helicity $-2$:

\begin{enumerate}
\item $F_{\lm_3\lm_4,\lm_2\lm_1}=0$ if only one $\lm_i=-2(2)$ and the others are 
$2(-2)$.
\item When  all helicities are equal then   
\begin{align}
\label{200119.6}
F_{22,22}(s,t,u)&=F_{-2-2,-2-2}(s,t,u)=\frac{\kp^2}{4}\frac{s^4}{stu}~,
\end{align}
with $\kp^2=32\pi G$.

\item Finally, if there are two positive and two negative helicities the 
expressions 
are all of them related by parity and crossing transformations (among themselves 
and with case 2 above) and can be written as
\begin{align}
\label{200119.7}
F_{-22,-22}(s,t,u)&=F_{2-2,2-2}(s,t,u)=\frac{\kp^2}{4}\frac{u^4}{stu}~,\\
F_{2-2,-22}(s,t,u)&=F_{2-2,2-2}(s,u,t)=F_{-22,2-2}(s,t,u)=\frac{\kp^2}{4}\frac{
t^4}{stu}~,
\end{align}
with any other combination having a zero Born scattering amplitude. 
\end{enumerate}

\subsection{Partial-wave amplitudes and unitarity}
\label{sec.200121.1}

We now express the previous amplitudes in the
basis with well defined angular momentum $|pJ,\lm_1\lm_2\ra_\S$, where $\S$ 
denotes the Bose symmetry of the state (see App.~\ref{app_states}).
The partial-wave amplitude (PWA) 
is given by 
\begin{align}
\label{200119.4}
\bar{T}^{(J)}_{\lm'_1\lm'_2,\lm_1\lm_2}(s)\equiv {_\S\la} 
pJ,\lm_1'\lm_2'|T|pJ,\lm_1\lm_2\ra_\S&=
\frac{1}{8\pi^2} \int_{-1}^{+1} d\!\cos\theta' \, d^J_{\lm\lm'}(\theta')\,
_\S\la \vp'_{xz},\lm'_1\lm'_2|T|p\vz,\lm_1\lm_2\ra_\S~, 
\end{align}
where we have defined  coordinates such that the final three-momentum $\vp'$
lies in the $xz$ plane with $\phi=0$, namely 
$\vp'=\vp'_{xz}=(\sin\theta,0,\cos\theta)$ and $d^{(J)}_{\lm\lm'}(\theta)$ is 
the  Wigner (small) d-matrix.

 The PWA series for graviton-graviton scattering does not converge\footnote{Note 
that several studies on gravitational scattering have been devoted to processes 
with large impact parameter, where the eikonal analysis would require an 
infinite number of PWAs, e.g. \cite{Ciafaloni:2018uwe,Giddings:2011xs}. Our 
approach is complementary to these studies.} due to the collapse of the Lehmann 
ellipse  and the coalescence of the right- and left-handed cuts at 
$s=0$~\cite{Lehmann:1958ita,Giddings:2009gj,martin.200705.1}.
However, since the Hamiltonian $H$ and the angular momentum $\mathbf{J}$ 
commute,  one is naturally driven to study the PWAs as the scattering amplitudes 
in the basis of states that diagonalize simultaneously both operators. PWAs are 
important,  for instance,  to study the physical effects due to the {\it 
rescattering} of two gravitons with the quantum numbers of a partial-wave. In 
this regard, one could consider external probes that select these quantum 
numbers for a two-graviton system, so that they rescatter with a given total 
angular momentum $J$, and hence, according to the corresponding subset of PWAs. 
The analysis from few PWAs may also be relevant when  considering the 
interference effects between the gravitational and other interactions. This  
formalism is used  in nuclear physics  to disentangle the interference effects 
between the Coulomb and the strong interactions e.g. in proton-proton or 
proton-$\pi^\pm$ interactions \cite{landau.170517.1}.\footnote{In the low-energy 
region only a few strong PWAs enter into play, and they select for the 
interference effects the corresponding Coulomb PWAs. } 

The unitarity of the $S$-matrix, $SS^\dagger=S^\dagger S=I$  for positive values 
of $s$, 
in the approximation of keeping only two-graviton intermediate 
states\footnote{The single graviton channel vanishes from helicity conservation 
and because $s=0$ on-shell.}, 
implies that 
\begin{align}
\label{200120.1}
_\S\la \vp',\lm'_1\lm'_2|T|p\vz,\lm_1\lm_2\ra_\S&-
{_\S\la \vp',\lm'_1\lm'_2|T^\dagger|p\vz,\lm_1\lm_2\ra_\S}\\
&=\frac{i}{64\pi^2}\sum_{\mu_1,\mu_2}\int d\hat{q}\,
 {_\S\la 
\vp',\lm'_1\lm'_2|T|\vq,\mu_1\mu_2\ra_\S}\,_\S\la\vq,\mu_1\mu_2|T^\dagger|p\vz,
\lm_1\lm_2\ra_\S~,\nn
\end{align}
where a symmetry factor 1/2 has been included in the right-hand side because of 
the symmetric states used. 
Since the gravitons have zero mass there are also other contributions in the 
right-hand side of the unitarity relation 
involving a larger number of gravitons in the intermediate states (three and 
more). However, these terms  
are at least two-loop contributions that, in the low-energy EFT of gravity, are 
${\cal O}((Gs)^3)$, and we will ignore them in the following. 

In terms of the graviton-graviton PWAs the two-body unitarity relation of 
Eq.~\eqref{200120.1} becomes
\begin{align}
\label{200120.2}
\Im _\S\la p J,\lm_1'\lm_2'|T|p J,\lm_1\lm_2\ra_\S
=\frac{\pi}{8}\sum_{\mu_1,\mu_2} \,
_\S\la p J,\lm'_1\lm'_2|T|p J,\mu_1\mu_2\ra_\S\,_\S\la p 
J,\mu_1\mu_2|T^\dagger|p J,\lm_1\lm_2\ra_\S \theta(s)~,
\end{align}
where the symbol $\Im $ denotes the imaginary part and  $\theta(x)$ is the 
Heaviside or step function. 
Here we have also taken into account that graviton-graviton PWAs are symmetric 
under the exchange between the final and initial states because of time-reversal 
invariance.

The $S$-matrix in partial waves $\bar{S}^{(J)}(s)$ is given by
\begin{align}
\label{200121.11}  
\bar{S}^{(J)}(s)&=I+i\frac{\pi 2^{|\lm|/4}}{4}\bar{T}^{(J)}(s)~,
\end{align}
being a matrix in the space of initial and final helicities for given $s$, and 
where $|\lm|=|\lm'_2-\lm'_1|=|\lm_2-\lm_1|=0$ or 4.  The factor $2^{|\lm|/4}$ 
appears because of the different normalization between the states having equal 
or different helicities, as discussed at the end of App.~\ref{app_states}. From 
the
unitarity relation in PWAs, Eq.~\eqref{200120.2}, it can be readily obtained 
that
\begin{align}
\label{200121.12}
\bar{S}^{(J)}\bar{S}^{(J)\dagger}=I~.
\end{align}
Since this matrix is of finite order then this last equation also implies that 
 $\bar{S}^{(J)\dagger}\bar{S}^{(J)}=I$.

Regarding the set of PWAs under consideration, we notice that 
$|pJM,\lambda\lambda\ras$ only involves
even $J$ because under the exchange of the two helicities, cf. 
Eq.~\eqref{170517.2},
\begin{align}
\label{200122.1}
|pJM,\lambda\lambda\ras&=(-1)^J|pJM,\lambda\lambda\ras~,  
\end{align}
so that it is zero if $J$ is odd.
Furthermore, since
\begin{align}
\label{200528.4}
|pJM,\lm_2\lm_1\ras&=(-1)^J|pJM,\lm_1\lm_2\ras~,
\end{align}
we do not consider the scattering amplitudes 
${_\S\la}pJM,\lm_2\lm_1|T|pJM,\lm_1\lm_2{\ra_\S}$
and ${_\S\la}pJM,\lm_2\lm_1|T|pJM,\lm_2\lm_1{\ra_\S}$, because they are 
$(-1)^J{_\S\la}pJM,\lm_1\lm_2|T|pJM,\lm_1\lm_2{\ra_\S}$ and 
${_\S\la}pJM,\lm_1\lm_2|T|pJM,\lm_1\lm_2{\ra_\S}$,
respectively. 
As a result, we dedicate our study to the partial waves
\begin{align}
\label{200125.1}
&{_\S\la}pJ,22|T|pJ,22{\ra_\S}~(\text{for even }J)~\text{and}~~
\las pJ,2-2|T|pJ,2-2\ras~. 
\end{align}
Let us also indicate that due to the parity symmetry, $\las 
pJ,-2-2|T|pJ,-2-2\ras=\las pJ,22|T|pJ,22\ras$ 
and this PWA is not treated separately.

\section{Treatment of the infinite-range nature of the interactions}
\label{sec.200122.1}
\def\theequation{\arabic{section}.\arabic{equation}}
\setcounter{equation}{0}

The direct application of the previous PWA formalism to the amplitudes 
\eqref{200119.6} and \eqref{200119.7} is hindered by divergences reflecting the 
infinite-range character of the gravitational interaction  
\cite{Giddings:2009gj}.
In this section we provide a  treatment and interpretation of this difficulty, 
based on well known results about the IR properties of the  amplitudes of 
infinite-range interactions \cite{Weinberg:1965nx,Kulish.200121,Ware:2013zja}. 

Let us first illustrate the issue by considering the PWA of the Born amplitude 
$F_{22,22}(s,t,u)$ with  $J=0$, for
which one has the angular projection
\begin{align}
\label{200121.2}
F^{(0)}_{22,22}(s)&=\frac{\kappa^2 
s^3}{32\pi^2}\int_{-1}^{+1}d\!\cos\theta\frac{1}{tu}
=-\frac{\kappa^2 s^2}{16\pi^2}\int_{-1}^{+1} \frac{d\!\cos\theta}{t}~.
\end{align}
Here we have used Eq.~\eqref{200121.1} 
and the explicit expression for $t$ and $u$ in terms of the scattering angle 
$\theta$
\begin{align}
\label{200121.3}
t=-(\vp-\vp')^2&=-2p^2(1-\cos\theta)~,\\
u=-(\vp+\vp')^2&=-2p^2(1+\cos\theta)~.\nn
\end{align}
We follow the notation that $\vp=\vp_1$ is the initial  center-of-mass (CM) 
three-momentum, $\vp'=\vp_3$ is the final one and
$p$ is the common modulus of all CM three-momenta.

The angular integration in Eq.~\eqref{200121.2} has a logarithmic divergence in 
the upper limit
of integration for $\cos\theta\to 1$.
This is actually an IR divergence  because the four-momentum squared
of the $t$-channel exchanged graviton, which is equal to $t$, vanishes for 
$\cos\theta\to 1$.
As we show in the following, this IR divergence is associated to a virtual soft graviton that connects
two external on-shell graviton lines, following the classification of IR 
divergences of
Ref.~\cite{Weinberg:1965nx}. Weinberg in this reference studies in detail this type of 
divergences and shows that 
they can be resummed correcting the $S$-matrix by a factor
\begin{align}\label{eq:WeinbergGen}
\exp{\left[\int^{\mathfrak{L}} \frac{d^4q}{(2\pi)^4} B(q,\mu)\right]},
\end{align}
where $B(q,\mu)$ represents the correction to the processes given by single soft 
(virtual) graviton exchange and where the IR divergence is regularized by a 
graviton mass $\mu$ \eqref{eq:WeinbergGen}. The variable $\mfl$ is a cutoff in 
$|\mathbf{q}|$ that separates the regions of ``hard'' and ``soft'' graviton 
momenta and is chosen to be low enough to justify the iteration of diagrams 
leading to Eq.~\eqref{eq:WeinbergGen}.  

 The integral in this equation gives a diverging imaginary contribution for every pair of 
either initial or final particles in the scattering process that  is independent 
of their spins\footnote{The divergence of the real part is not of concern here since it does not appear in the Born series. Its cancellation in  inclussive processes is due to the radiation of soft photons \cite{Weinberg:1965nx}.}.
This contribution can be entirely expressed in terms of the Lorentz invariant
relative velocity $\beta_{ab}$ of the two particles involved, $a$ and $b$, in 
the rest frame of either, 
\begin{align}
\label{200121.4}
\beta_{ab}&=\frac{[(p_ap_b)^2-(m_am_b)^2]^{1/2}}{p_ap_b}~.
\end{align}
where $p_a$ and $p_b$ are the four-vectors of the particles with masses $m_a$ 
and $m_b$, respectively. 
The phase factor of interest that stems from Eq.~\eqref{eq:WeinbergGen} for a given pair of initial or final 
particles is \cite{Weinberg:1965nx}
\begin{align}
\label{200121.5}
\exp\left[-i\frac{Gm_am_b(1+\beta_{ab}^2)}{\beta_{ab}[1-\beta_{ab}^2]^{1/2}}
\log\frac{\mu}{\mathfrak{L}}\right]~.
\end{align}
The total exponential factor $S_c(s)$ in our case is given by  
\begin{align}
\label{200121.6}
S_c(s)&=\lim_{m\to 0}
\exp
\left[-i2\frac{Gm^2(1+\beta^2)}{\beta[1-\beta^2]^{1/2}}\log\frac{\mu}{\mathfrak{
L}}\right]
=\exp\left[-i2 G s \log\frac{\mu}{\mathfrak{L}}\right]~,  
\end{align}
where we have taken the massless limit replacing $\beta_{ab}$ by 
$\beta=1-2m^4/s^2+{\cal O}(m^8)$, and have taken into account that $(p_a+p_b)^2=s$ for both pairs of either incoming or outgoing gravitons. As we 
will see below in Sec.~\ref{sec.200805.2} the IR divergence can also be derived 
using dimensional regularization (see also \cite{Naculich:2011ry,Ware:2013zja}).

From the  linear term in the expansion of the phase factor $S_c(s)$ in powers of 
$G$, 
one can read its (divergent) contribution to the Born scattering amplitude,
\begin{align}
\label{200121.8}
S_c(s)&=1-i2 G s \log\frac{\mu}{\mfl}
+{\cal O}(G^2),
\end{align}
that gives the following contribution to the PWA
\begin{align}
\label{200121.8b}
\delta F^{(J)}_{\lm_1'\lm_2',\lambda_1\lambda_2}=-\frac{1}{2^{\lm/4}}\frac{8 
Gs}{\pi}
\log\frac{\mu}{\mfl}~,    
\end{align}
where the factor $1/2^{\lm/4}$ reflects the different normalization of the 
symmetrized states with $\lambda=0$ or $\lambda=4$, cf. Eq.~\eqref{200121.11}. 
As expected, this is precisely the divergence that stems in the $J=0$ 
partial-wave projection of Eq.~\eqref{200121.2}, 
after giving a mass $\mu\to 0^+$ to the exchanged graviton,
\begin{align}
\label{200121.9}
F^{(0)}_{22,22}(s)&=-\frac{\kappa^2 s^2}{16\pi^2}\int_{-1}^{+1} 
\frac{d\!\cos\theta}{t-\mu^2}
=\frac{\kp^2s}{8\pi^2}\log\left(1+\frac{4p^2}{\mu^2}\right)
\to \frac{\kp^2s}{4\pi^2}\log\frac{2p}{\mu}
=\frac{8Gs}{\pi}\log\frac{2p}{\mu}~.
\end{align}

It is important to emphasize that this contribution (as well as the whole phase 
space factor) is independent of $J$, being common to all PWAs with the given set 
of helicities $\lm_i$ and $\lm_i'$. 
For  instance, for the case $\lambda=0$ 
we have
\begin{align}
\label{200121.10}
F^{(J)}_{22,22}(s)&=-\frac{2Gs^2}{\pi}\int_{-1}^{+1}d\!\cos\theta\frac{
P_J(\cos\theta)}{t-\mu^2}~,
\end{align}
where we have taken into account that $d_{00}^{(J)}(\theta)=P_J(\cos\theta)$ 
\cite{rose.170517.1},
with $P_J(\cos\theta)$ the Legendre polynomials. In this equation we can isolate 
the divergent piece
by adding and subtracting $P_J(1)=1$ to the numerator of the integrand. Then,
\begin{align}
\label{200125.2}
F^{(J)}_{22,22}(s)&
=-\frac{2Gs^2}{\pi}\int_{-1}^{+1}d\!\cos\theta\frac{P_J(\cos\theta)-1}{t}
+\frac{8Gs}{\pi}\log\frac{2p}{\mu}~.
\end{align}
where the last term is
the diverging contribution of Eq.~\eqref{200121.8b}.

For $\lambda=4$ and arbitrary $J$ the partial-wave projection is
\begin{align}
\label{200125.3}
F^{(J)}_{2-2,2-2}(s)&=\frac{\kp^2}{32\pi^2s}\int_{-1}^{+1}d\!\cos\theta 
\frac{d_{44}^{(J)}(\theta)u^3}{t}~.
\end{align}
In order to isolate the diverging piece we sum and subtract 
$-s^3d_{44}^{(J)}(0)$ to the numerator of
the integrand and, since $d_{44}^{(J)}(0)=1$ \cite{rose.170517.1}, we have
\begin{align}
\label{200125.4}
F^{(J)}_{2-2,2-2}(s)&=\frac{G}{\pi 
s}\int_{-1}^{+1}\frac{d\!\cos\theta}{t}\left[d_{44}^{(J)}(\theta)u^3+s^3\right]
+\frac{4Gs}{\pi}\log\frac{2p}{\mu}~.
\end{align}
Again, it is clear the appearance of the diverging factor according to 
Eq.~\eqref{200121.8b}. One can also explicitly calculate the result for the case 
$\lambda=-4$ which receives a divergent contribution from $S_c(s)$ that is 
related to the one in $\lambda=4$ by $\delta F^{(J)}_{2-2,-22}=(-1)
^J\delta F^{(J)}_{2-2,2-2}$. Calculating the corresponding PWA and using the 
property $d_{4-4}^{(J)}(\theta)=(-1)^Jd_{44}^{(J)}(\pi-\theta)$ one, indeed, 
obtains the same divergence as in Eq.~\eqref{200125.4} but multiplied by 
$(-1)^J$. 

 Given the previous results, we redefine the $S$ matrix in PWAs by extracting 
the divergent phase $S_c(s)$ in Eq.~\eqref{200121.6} from $\bar{S}^{(J)}$,
\begin{align}
\label{200121.13}
\bar S^{(J)}&= S_cS^{(J)}=\exp
\left[-i2 G s \log\frac{\mu}{\mfl} \right]S^{(J)}~,
\end{align}
and where the new $S^{(J)}$ includes virtual contributions from gravitons with 
momenta only greater than $\mfl$. This global phase is not ``physical'' (not 
observable) because it does not contribute
to the cross section for the scattering process of Eq.~\eqref{200119.2}.   
Related to this, $S_c(s)$ is trivial from the analytical point of view  and it 
does not lead to
interesting features in the full $S$ matrix such as the presence of poles 
corresponding to bound, virtual   
or resonance states.  

Since we removed a global phase,   $S^{(J)}$  fulfills also the unitarity 
relation in PWAs
\begin{align}
\label{200121.14}
S^{(J)}S^{(J)\dagger}=S^{(J)\dagger}S^{(J)}=I~.
\end{align}
The $S$ matrix $S^{(J)}(s)$ generates a new $T$ matrix in
PWAs by taking into account 
Eq.~\eqref{200121.11},
\begin{align}
\label{200121.15}
T^{(J)}(s)&\equiv -i\frac{4(S^{(J)}-I)}{\pi 2^{|\lm|/4}}~.
\end{align}
Given that $S^{(J)}(s)$ is a unitary matrix, 
Eq.~\eqref{200121.14},
$T^{(J)}$ also satisfies a two-body unitarity relation analogous to that in 
Eq.~\eqref{200120.2}.
Namely, 
\begin{align}
\label{200121.16}
\Im T^{(J)}=\frac{\pi 2^{|\lm|/4}}{8}T^{(J)}T^{(J)\dagger}\theta(s)~.
\end{align}
This equation can also be recast as 
\begin{align}
\label{200121.17}
\Im \frac{1}{T^{(J)}}=-\frac{\pi 2^{|\lm|/4}}{8}\theta(s)~.
\end{align}

We will denote by $V^{(J)}$ the partial-wave projected Born amplitudes 
corresponding to $S^{(J)}$. They  
are free of the IR divergences  and carry an explicit dependence on the IR scale 
$\mfl$.~\footnote{This connection between the soft infrared divergences of the one-loop scattering amplitudes and the kinematic divergences in the tree-level PWAs has been independently demonstrated for gauge theories and gravity in Ref.~\cite{Baratella:2020dvw}}
They can be obtained from \eqref{200121.13} by expanding  up to ${\cal O}(G)$, 
\begin{align}
\label{200121.18}
S^{(J)}(s)&=I+i2^{\lm/4}\frac{\pi }{4}V^{(J)}(s)+{\cal O}(G^2)
=\left[I+i 2^{\lm/4}\frac{\pi}{4}F^{(J)}(s)\right]\left[1+i 
2Gs\log\frac{\mu}{\mfl}\right]+{\cal O}(G^2)~,
\end{align}
from where,
\begin{align}
\label{200121.18b}
V^{(J)}(s)&=F^{(J)}(s)+\frac{1}{2^{\lm/4}}\frac{8Gs}{\pi}\log\frac{\mu}{\mfl}~.
\end{align} 
For instance,  from Eq.~\eqref{200121.9} one obtains the IR-safe result
\begin{align}
\label{200121.19}
V^{(0)}_{22,22}(s)&=\frac{8Gs}{\pi}\log\frac{2p}{\mfl}~.
\end{align}

Recall that the parameter $\mfl$ is used to define the soft gravitons that are 
resummed to generate $S_c(s)$.  A change of this cutoff produces just a 
rescaling of the $S^{(J)}$-matrix,
\begin{align}
\label{eq:phasedepLambda}
S^{(J)}_{\mfl^\prime}=\left(\frac{\mfl^\prime}{\mfl}\right)^{2i G s}S^{(J)}.
\end{align}
Thus, although the dependence on $\mfl$ is physically spurious, it will show up 
in a truncated perturbative expansion of $S^{(J)}$. On dimensional grounds it is 
clear that $\mfl\propto p$ because at ${\cal O}(G)$ this is the only magnitude 
with dimension of momentum that is available to enter inside the propagators in 
the Feynman diagrams giving rise to the Born term~\cite{Weinberg:1965nx}.
Therefore, we define
\begin{align}
\label{200121.20}
\mfl=\sqrt{s}/a~,
\end{align}
where one should expect that $a>1$ so that $\mfl$ is significantly smaller than 
$\sqrt{s}$. Thus $\log a$, which is the parameter finally entering in the 
partial-wave projected Born term, is ${\cal O}(1)$ because, once the IR 
divergences are removed, the scattering amplitudes do not involve any other 
pathological scale.
Indeed, this ambiguity can  only be resolved conclusively by the exact solution, 
or by  higher order calculations (which may also open the possibility of other 
methods, e.g. \cite{Stevenson:1981vj,Brodsky:1982gc,Su:2012iy}. 
Along these lines, besides the naturalness criteria, we develop a method for 
estimating $\log a$ in  Sec.~\ref{sec.210115.2}, which is also tested  in models 
where the exact solution is known.\footnote{In Ref.~\cite{Kulish.200121}, Kulish and Faddeev dealt with the IR divergences 
affecting QED by defining an $S$ matrix which is free of them  and with the same 
physical cross sections. A similar method was explicitly implemented for gravity 
recently in \cite{Ware:2013zja} (see also \cite{Strominger:2017zoo} for a recent 
review).}

In summary, our final expressions for $V^{(J)}_{2\pm2,2\pm 2}(s)$ are
\begin{align}
\label{200121.21}
V^{(J)}_{22,22}(s)&=-\frac{2Gs^2}{\pi}\int_{-1}^{+1}d\!\cos\theta\frac{
P_J(\cos\theta)-1}{t}
+\frac{8Gs}{\pi}\log a~,\\
\label{200121.22}
V^{(J)}_{2-2,2-2}(s)&=\frac{G}{\pi 
s}\int_{-1}^{+1}\frac{d\!\cos\theta}{t}\left[d_{44}^{(J)}(\theta)u^3+s^3\right]
+\frac{4Gs}{\pi}\log a~.
\end{align}
For the important particular case $J=0$, analyzed in detail below,   
\begin{align}
\label{200121.23}
V^{(0)}_{22,22}(s)&=\frac{8Gs}{\pi}\log a~.
\end{align}
The first higher partial wave with $\lm_i=2$ has $J=2$  and
\begin{align}
\label{200204.1}
V^{(2)}_{22,22}(s)&=\frac{8Gs}{\pi}\left(\log a-\frac{3}{2}\right)~.
\end{align}
For $\log a\gtrsim 1$ we see that the interaction is  weaker than for $J=0$, and 
may even change its attractive nature.

It is notorious that the resulting (modified) partial-wave projected Born 
scattering amplitudes
$V^{(J)}_{22,22}(s)$ and $V^{(J)}_{2-2,2-2}(s)$, Eqs.~\eqref{200121.21} and 
\eqref{200121.22}, respectively,
are free of any cut in the complex-$s$ plane.
We will show below that this is also the case for other models with 
infinite-range interactions that we will  explore (cf. Sec.~\ref{app.170723.2}, 
and Sec.~\ref{app.200127.1}). In a sense, the presence of IR divergences 
eventually simplifies the analytical structure of PWAs. By the same token it 
also paves the way for the applicability of the unitarization methods, cf. 
Sec.~\ref{sec.200531.1}.

\section{Unitarization of the graviton-graviton PWAs}
\def\theequation{\arabic{section}.\arabic{equation}}
\setcounter{equation}{0}
\label{sec.200531.1}

\begin{figure}
\begin{center}
\includegraphics[width=110mm]{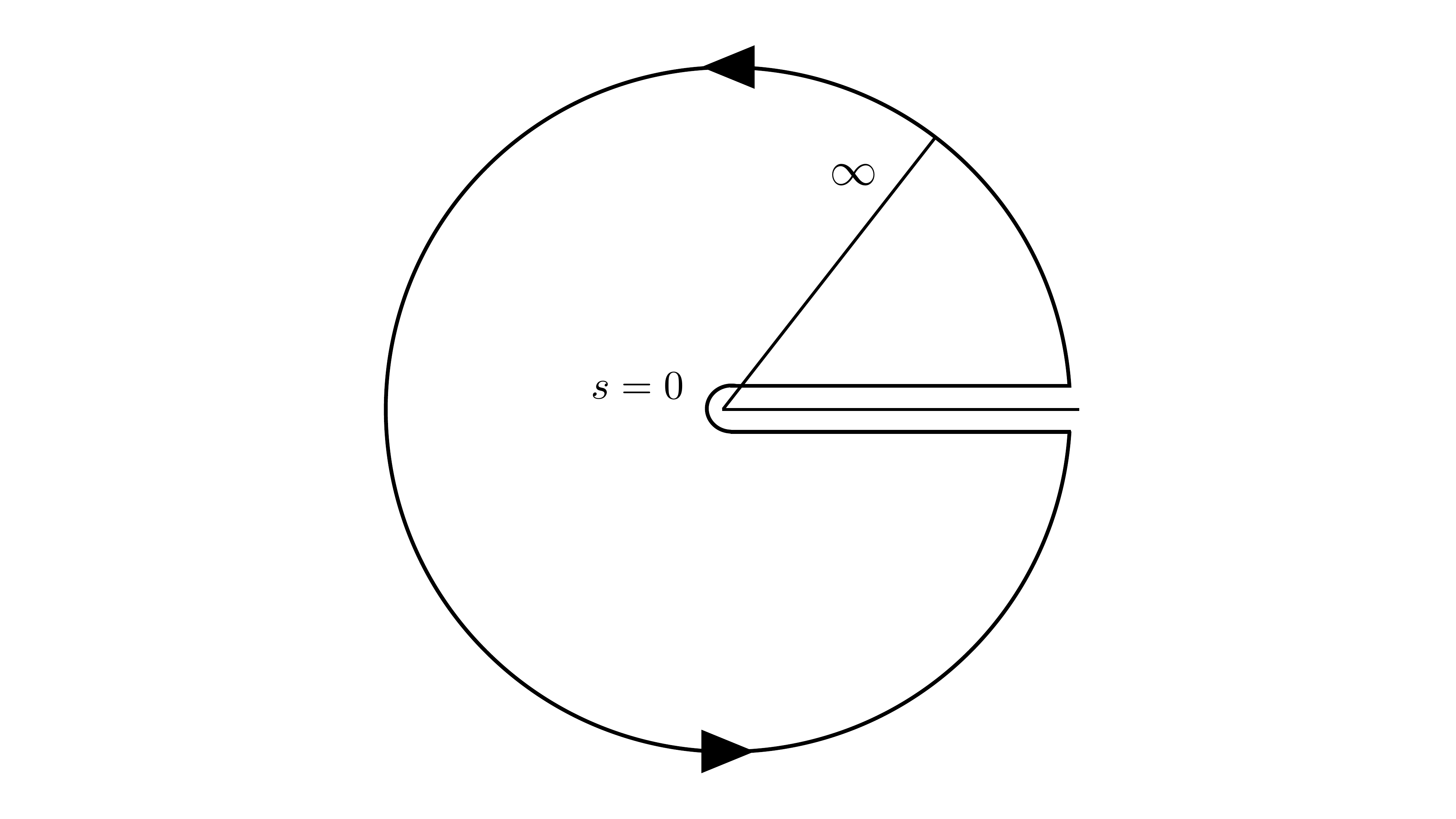}
\caption{{\small The integration contour ${\cal C}$ used for the DR of 
$T^{(J)}(s)^{-1}$ in Eq.~\eqref{200530.1}}
\label{fig.200530.1}}
\end{center}
\end{figure}

One important consequence of the treatment of IR singularities discussed in the 
previous section is the disappearance of the left-hand cut (LHC) from the 
partial-wave-projected Born amplitudes. Indeed, the expressions
in Eqs.~\eqref{200121.21} and \eqref{200121.22} are free from any cut. As a 
consequence,  the corresponding $T$ matrix in PWAs, obtained from 
the iteration of the Born terms, only has the right-hand cut (RHC) or unitarity 
cut, cf. Eq.~\eqref{200121.16}. { In a sense, the presence of IR divergences 
eventually simplifies the analytical structure of PWAs.} 

Because of the Schwarz reflection principle, the discontinuity of the inverse of 
a PWA across the RHC is given by the right-hand side  of 
Eq.~\eqref{200121.17} multiplied by $2i$. 
This enables a simple parametrization of the PWAs which fulfills exact two-body 
unitarity. To simplify the notation let us focus first on the scattering with 
$\lm=0$. The generalization of the result to $\lm=\pm4$ is straightforward and 
will be given afterwards. It is enough to introduce an analytical function 
$g(s)$ in the complex-$s$ plane with only the RHC and whose discontinuity along 
it is the same as that of $1/T^{J}(s)$. 
To build it, let us consider a one-subtracted dispersion relation (DR) by taking 
the integration contour ${\cal C}$ drawn in Fig.~\ref{fig.200530.1} and apply 
the Cauchy's integration theorem to $g(z)/[(z-s)(z+s_0)]$, where $z$ is the 
integration variable and $s_0$ is the subtraction point. 
 Namely,
\begin{align}
\label{200716.1}
\oint_{{\cal C}}\frac{g(z)}{(z-s)(z+s_0)}dz=2\pi i 
\left(\frac{g(s)}{s+s_0}-\frac{g(-s_0)}{s+s_0}\right)~.
\end{align}
By adding one subtraction, the integration along the circle at infinity vanishes 
and  we are left with the following expression for $g(s)$, 
\begin{align}
\label{200716.2}
g(s)&=a(s_0)-\frac{s+s_0}{8}\int_0^\infty \frac{ds'}{(s'-s)(s'+s_0)}~,
\end{align}
where $a(s_0)\equiv g(-s_0)$ is called the subtraction constant.
The integration can be done explicitly and we rewrite $g(s)$ as\footnote{The 
$\log z$ is defined with ${\rm arg}z\in[-\pi,\pi]$. If the latter is chosen 
between
$[0,2\pi)$ then $g(s)=a(s_0)+\frac{1}{8}\log \frac{s}{s_0}-i\frac{\pi}{8}$.} 
\begin{align}
\label{200531.1}
g(s)=a(s_0)+\frac{1}{8}\log\frac{-s}{s_0}~,
\end{align}
In summary, by construction $g(s)$ only has the RHC with an imaginary part 
\begin{align}
\label{200531.2} 
\Im g(s)&=-\frac{\pi}{8}\theta(s)~,~s\in \mathbb{R}~,
\end{align}
as required by unitarity, cf. Eq.~\eqref{200121.17}.

We then take the following general expression for a PWA 
$T_{\lm'_1\lm'_2,\lm_1\lm_2}^{(J)}(s)$~\cite{nd}, which automatically fulfills 
two-body unitarity, Eq.~\eqref{200121.16},  
\begin{align}
\label{200530.1}
T_{\lm'_1\lm'_2,\lm_1\lm_2}^{(J)}(s)&=
\left[\frac{1}{R_{\lm'_1\lm'_2,\lm_1\lm_2}^{(J)}(s)}+2^{|\lm|/4}g(s)\right]^{-1}
~.
\end{align} 
Unless strictly necessary, in the following we suppress the helicity subscripts 
of $T^{(J)}$ and $R^{(J)}$ for simplicty in the writing.
The function $R^{(J)}(s)$ has no two-body unitarity cut because this is fully 
accounted for by $g(s)$, as it is clear by comparing with 
Eq.~\eqref{200121.17}. 
Following  Refs.~\cite{oww,Oller:2000fj} we obtain $R^{(J)}(s)$ by matching  
$T^{(J)}(s)$ in Eq.~\eqref{200530.1} to 
its perturbative calculation order by order within the low-energy EFT for 
gravity. 
Noting that the contribution from $g(s)$ enters at $\mathcal O((G s)^2)$ in the 
expansion of $T^{(J)}(s)$ in powers of $s$,  Eq.~\eqref{200530.1}, one obtains 
that  
\begin{equation}
\label{200530.2}
R^{(J)}=V^{(J)}+{\cal O}\left((G s)^2\right).
\end{equation}
In general $R^{(J)}(s)$ comprises LHCs due to multi-graviton exchanges, as well 
as extra RHCs involving intermediate states with more than 2 gravitons. However, 
all these contributions are ${\cal O}\left((G s)^2\right)$. For the present case 
we have just one pole in $T^{(J)}(s)^{-1}$ because $V^{(J)}(s)$ has only one 
simple zero at $s\to0$, as found in general relativity. Scenarios including 
higher-energy degrees of freedom can be implemented in the dispersion relation 
by including more poles in $R^{(J)}$, called Castillejo-Dalitz-Dyson (CDD) poles 
after Ref.~\cite{Castillejo:1956}.

The issue of determining {\it a priori} a value for the subtraction constant 
$a(s_0)$ at the scale $s_0$ has been already discussed in QCD. 
In the absence of fine tuning, the typical value of $a(s_0)$ should be of the 
same order of magnitude as its variation under changes of order 1 in the 
subtraction point $s_0$.\footnote{This is equivalent to the estimation of 
higher-order effects of perturbative calculations in QCD or ChPT obtained by 
variations of the renormalization scale.} Notice that the combination 
$a(s_0)+\frac{1}{8}\log\frac{-s}{s_0}$ is independent of the value of $s_0$.
This fixes the running of $a(s_0)$  with $s_0$ as
\begin{align}
\label{200531.3} 
a(s_0)-a(s_0')&=\frac{1}{8}\log\frac{s_0}{s'_0}~,
\end{align}
where $s'_0$ is another subtraction point. Therefore, according to Eq.~\eqref{200531.3}, 
we would have that
\begin{align}
\label{200531.4}
|a(s_0)|\sim |a(s_0)-a(s'_0)|=\frac{1}{8}\left|\log \frac{s_0}{s'_0}\right|\ll 
1~.
\end{align}
Hence, an assessment on the smallness of the previous estimate can be inferred by 
comparing it with $V^{(J)}(s)^{-1}$. Namely, from Eq.~\eqref{200121.23} we have 
that
\begin{align}
\label{200531.5}
\frac{1}{V^{(0)}_{22,22}}&=\frac{\pi/8}{Gs\log a}~, 
\end{align}
which is much larger than $1/8$ for $s\ll \pi G^{-1}$. 

Another method to choose $a(s_0)$ is adopting the so-called {\it natural values} 
for both $a(s_0)$ and $s_0$ \cite{Oller:2000fj}. This is based on comparing 
$g(s)$ in Eq.~\eqref{200531.1} with the same integral regularized by using a 
cutoff $\Lambda^2$, that we call $g_c(s)$~\cite{iam.cc},
\begin{align}
\label{200531.6} 
g_c(s)&=-\frac{1}{8}\int_0^{\Lambda^2}\frac{ds'}{s'-s}=\frac{1}{8}\log\frac{-s}{
\Lambda^2}+\mathcal O(s/\Lambda^2)~.  
\end{align}
Identifying $s_0$ with the cutoff
in Eq.~\eqref{200531.1} and matching the expressions for $g(s)$ and $g_c(s)$, 
one concludes that
\begin{align}
\label{200127.6}
a(\Lambda^2)&=0~,
\end{align}
which is the same value as the one obtained above with the other method in 
Eq.~\eqref{200531.4}. Therefore, in the following we fix $a(\Lambda^2)=0$ (with 
$s_0=\Lambda^2$) and one could vary $\Lambda$ to tentatively determine the 
uncertainty in our results.

It is also important to give an estimate for the cutoff $\Lambda$ that, 
generally speaking, corresponds to the scale suppressing the higher-dimension 
operators in the EFT~\cite{Georgi:1984,Donoghue:1999qh}. Combining 
Eqs.~\eqref{200530.1} and \eqref{200530.2} the unitarized amplitude can be 
expanded to next-to-leading order as,
\begin{align}
\label{eq:TJNLO}
T^{(J)}=V^{(J)}\left(1-\frac{V^{(J)}}{8}\log\frac{-s}{\Lambda^2}\right)+\mathcal 
O\left((Gs)^3\right).    
\end{align}
A value of the cutoff $\Lambda^2$ can be thus estimated by assuming that the 
correction has the expected size in the EFT expansion,  $s/\Lambda^2$. 
For definiteness, let us take $V^{(0)}(s)$, Eq.~\eqref{200121.23}, and then  
for $\log(-s/s_0)\simeq1$, we obtain that 
\begin{align}
\label{200602.1}
s_0={\Lambda}^2=\pi(G\log a)^{-1}\sim G^{-1}~.
\end{align}
Let us stress again that the unitarization procedure only resums the higher 
order corrections needed to reproduce exactly the two-graviton cut. In 
particular, the next-to-leading order (NLO) contribution in Eq.~\eqref{eq:TJNLO} 
displays a dependency in $\Lambda$ that 
is cancelled by other terms in $V^{(J)}(s)$ when matching with the perturbative amplitude at NLO, which is UV-finite in pure gravity~\cite{Donoghue:1999qh,Dunbar:1994bn,Dunbar:1995ed}.

Another method to determine the scale of new physics based on unitarity of PWA 
is considering the violation of perturbative 
unitarity~\cite{quigg,han.200204.1,donoghue.200204.1}. This stems from the 
fundamental relations Eqs.~\eqref{200121.14} and \eqref{200121.15},  which allow 
one to express the PWA in $J=0$ for $s>0$ in terms of a phase shift $\delta$ and 
the inelasticity parameter $\eta$ ($0\leq \eta\leq 1$) as
\begin{align}
\label{eq:phaseshiftJ0}
T^{(0)}_{22,22}=-\frac{4i}{\pi}(\eta e^{2i\delta}-1)~.    
\end{align}
This leads to the constraint,
\begin{align}
\label{eq:constraintPU}
|{\rm Re}\,T^{(0)}_{22,22}|\leq \frac{4}{\pi},    
\end{align}
which is violated by the Born amplitude at,
\begin{align}
s_{\rm pu}=\frac{1}{2}(G\log a)^{-1}\sim G^{-1}.
\end{align}
This is of the same order although somewhat smaller than the unitarity 
correction in eq.~\eqref{200602.1}. Note, however, that $s_{\rm pu}$ represents the energy at which unitarity corrections are needed to fulfill the 
constraint~\eqref{eq:constraintPU}, but they can still be small compared to the 
leading order. For instance, ${\rm Re}\,T^{(0)}_{22,22}$ evaluated with the 
unitarized amplitude at $s_{\rm pu}$ is only $\sim 5\%$ smaller than the 
leading-order contribution from the Born term, which is consistent with a 
correction of order $s_{\rm pu}/\Lambda^2=1/(2\pi)$. 

It is important to point out that cutoffs derived from unitarity considerations 
within the EFT do not correspond, necessarily, to the \textit{fundamental} scale 
of its ultraviolet completion (see e.g. Ref.~\cite{donoghue.200204.1}). For 
gravity, this scale could be as low as the TeV range in scenarios with extra 
dimensions~\cite{ArkaniHamed:1998rs,Antoniadis:1998ig,Randall:1999ee,
Randall:1999vf,Giudice:2016yja} or, for example, gravitation could be properly 
described by a quantum field theory with an ultraviolet fixed 
point~\cite{weinberg:safe,Niedermaier:2006wt,Codello:2008vh,Falls:2014tra} or 
with new non-trivial dynamics at energies below $G^{-1/2}$ 
\cite{Blas:2009qj,Steinwachs:2020jkj}. 
In the following we assume that the fundamental scale of gravity is of the same 
order as the one derived from unitarity considerations. We will discuss later 
the implications of  other scenarios where this is not the case.

Finally, the PWA in Eq.~\eqref{200530.1} are defined in the \textit{physical 
sheet}, where the physical region is the positive real axis of $s$ approached 
from above. The structure of the amplitude in the unphysical Riemann sheet (RS) 
can be obtained by analytical continuation of Eq.~\eqref{200530.1}. This is 
particularly relevant for resonances, which are identified with poles in the 
second RS reached from the physical region by burrowing down through the 
RHC~\cite{Eden:1966dnq,martin.200705.1,Oller:2017alp}.

\section{Connection with hadronic resonances}
\label{sec.200605.1}

After having derived the dispersion relations for graviton-graviton scattering, 
it seems worth pausing and recalling how the same formalism predicts some 
non-trivial features of hadronic resonances. This will give 
more solid 
basis to some of the methods above, illustrating in a nontrivial way our 
approach for unitarizing amplitudes and estimating the {\it natural} values of 
$a(s_0)$ and $s_0\simeq {\Lambda}^2$. It also serves as motivation towards the 
possible presence of similar phenomena in gravity, the focus of 
Sec.~\ref{sec.200131.1}.

\subsection{PW unitarization of $\pi\pi$ scattering in the chiral limit}

The lightest resonance in QCD, the $f_0(500)$ or  $\sigma$ meson, has vacuum 
quantum numbers, $J^{PC}=0^{++}$.
This resonance was studied in pion-pion scattering with a unitarization method 
similar to the one employed here in the
pioneering works of Refs.~\cite{Oller:1997ti,oo98}. In the chiral limit the 
pions become massless and the
unitarity loop function is the same as in graviton-graviton scattering, 
Eq.~\eqref{200531.1}, 
except for a factor $1/(2\pi^2)$ which appears in the standard 
normalization.\footnote{This stems from the normalization used for the 
two-graviton states, cf. Eq.~\eqref{200116.8} in App.~\ref{app_states}.} We call 
it
$g_\pi(s)$ and its expression is 
\begin{align}
\label{200603.2}
g_\pi(s)&=\frac{1}{2\pi^2}g(s)=\frac{1}{(4\pi)^2}\left(a_\pi(s_0)+\log\frac{-s}{
s_0}\right)~,
\end{align}
with $a(s_0)\to a_\pi(s_0)/8$.
The $S$-wave isoscalar $\pi\pi$ scattering amplitude in the chiral limit at 
leading
order  is $s/f_\pi^2$, with $f_\pi$ the pion decay constant.
This tree-level amplitude is analogous to $V^{(0)}(s)$, the $J=0$ 
graviton-graviton amplitude in
Eq.~\eqref{200121.23} after curing the IR divergence. The zero at $s=0$ in the 
case of $\pi\pi$ scattering
is referred to as an Adler zero and follows from the Goldstone theorem. 
The unitarized expression for the isoscalar scalar $\pi\pi$ PWA in the chiral 
limit is \cite{okenky}
\begin{align}
\label{200603.3}
T^{(0)}_\pi(s)&=\left[\frac{f_\pi^2}{s}+g_\pi(s)\right]^{-1}~.
\end{align}

A subtraction constant in $g_\pi(s)$ of natural size can be determined by using 
a 
cutoff 
$\Lambda$ as in Eq.~\eqref{200531.6}. Thus,
\begin{align}
\label{200603.4}
a_\pi(s_0)=\log\frac{s_0}{\Lambda^2}~.
\end{align}
In the following 
$s_0=\Lambda^2$, then $a_\pi(\Lambda^2)=0$
and Eq.~\eqref{200603.3} simplifies to 
\begin{align}
\label{200603.5}
T^{(0)}_\pi(s)&=\left[\frac{f_\pi^2}{s}+\frac{1}{(4\pi)^2}\log\frac{-s}{
\Lambda^2} \right]^{-1}~.
\end{align}

We are now interested in the emergence of resonances corresponding to poles in 
the second RS of the unitarized amplitude Eq.~\eqref{200603.5}. As discussed 
above, the second RS is reached by analytical continuation of $g(s)$, crossing 
the unitarity cut from the physical region above the real $s$ axis ($\Im s>0$) 
towards the lower half plane with $\Im 
s<0$~\cite{Eden:1966dnq,martin.200705.1,oller.book}. 
Thus,\footnote{In order to keep the desirable and standard property in 
$S$-matrix theory for massive particles that $g_{II}(s^*)=g_{II}(s)^*$, 
one should use instead
\begin{align}
\label{200209.1}
g_{\pi,II}(s)=g_\pi(s)+i\frac{1}{8\pi}\sqrt{\frac{s-4\mu^2}{s}}
\end{align}
where the $\sqrt{z}$ is evaluated for ${\rm arg}z\in[0,2\pi)$, and then the 
limit $\mu\to 0$ has to be taken at the end of the calculation. With this 
definition $g_{\pi,II}(s)$ is real for $s\in[0,4\mu^2]$ and, therefore, it 
satisfies the Schwarz reflection principle. 
Nonetheless, we do not insist on this point and conform ourselves with the 
simpler Eq.~\eqref{170725.4} 
because for $\Im s<0$ they give the same result, and the resonance poles lie in 
complex conjugate positions.    
As it actually happens if the proper Eq.~\eqref{200209.1} were used for 
$g_{\pi,II}(s)$.}
\begin{align}
\label{170725.4}
g_{\pi,II}(s)&=g_\pi(s)-i\frac{1}{8\pi}\,.
\end{align}

The cutoff in ChPT is usually taken as  $\Lambda\simeq 4\pi f_\pi$ which 
agrees with the one obtained by studying the size of the leading unitarity 
contribution, as explained above in connection to Eq.~\eqref{200602.1}.
Indeed, this correction is $(s/f_\pi^2)^2/(16\pi^2)$,
that equated to $s/\Lambda^2$ times the leading order, $s/f_\pi^2$, leads also 
to $\Lambda=4\pi f_\pi$~\cite{Georgi:1984}.~\footnote{Note that this coincidence 
between the fundamental and unitarity scales in ChPT is rather exceptional and  
breaks down for other values of the number of colors $N_c$ in the  large-$N_c$ 
QCD  (see Ref.~\cite{donoghue.200204.1} and discussion below).}

\subsection{Postdiction of the $\sigma$ meson}
\label{sec.201008.1}

We concluded in the previous section that the graviton-graviton and $\pi\pi$ 
scatterings share a similar structure for their DRs. 
As a motivation of the possible consequences for graviton-graviton scattering, 
let us first recall how Eq.~\eqref{200603.5}  postdicts the existence of the 
$\sigma$ meson. 

The PWA $T^{(0)}_\pi(s)$ continued to the second RS has a pole in $s_\sigma$ 
satisfying the secular equation 
\begin{align}
\label{eq:secular.sigma0}
\frac{(4\pi f_\pi)^2}{s_\sigma}+\log\frac{-s_\sigma}{\Lambda^2}-2i\pi=0.
\end{align}
We can rewrite this expression as 
\begin{align}
\label{eq:secular.sigma1}
\frac{1}{x}&+\log(-x)-2i\pi=0,~~ \mathrm{where}~~
x=\frac{s_\sigma}{\Lambda^2}~,~\Lambda=4\pi f_\pi~.
\end{align}
The solution to this equation can be found easily by an iterative method. We 
first neglect $\log(-x)$ and obtain that $x_1=-i/(2\pi)$. Then, in a second 
iteration, instead of neglecting the $\log$ term, we keep only its imaginary 
part, $\log(-x_1)=i\pi/2+\ldots$, so that
\begin{align}
\label{200206.2}
x&\simeq-i \frac{2}{3\pi}=-i\,0.20.
\end{align}
A direct numerical evaluation of Eq.~\eqref{eq:secular.sigma1} gives
\begin{align}
\label{200602.4}
x&=0.07-i\,0.20
\end{align}
This pole is located deep in the complex-$s$ plane and has a small real part. 
Therefore, the dynamics of $S$-wave $\pi\pi$ scattering are strongly influenced 
by this resonance even at energies significantly lower than the cutoff of the 
EFT. In Table~\ref{tab.sigma} we show the poles in the variable 
$\sqrt{s_\sigma}$ (which is the one quoted in phenomenology) and units of GeV 
that are obtained numerically by solving Eq.~\eqref{eq:secular.sigma0} using the 
nominal value of the cutoff $\Lambda=4\pi f_\pi$ and the value of $f_\pi$ in the 
chiral limit $f_\pi=f\simeq 86.1$~MeV~\cite{Aoki:2019cca}. We also investigate 
variations of the results by changing the cutoff by a factor 2 around the 
nominal value, which is a range exaggerated compared to the actual one 
constrained by ChPT phenomenology. Nonetheless, the pole position has a very 
mild dependence on it, changing only by $\sim 20\%$.

\begin{table}
\begin{center}
\begin{tabular}{c|ccc}
\hline
&  $\Lambda$/2 &  ${\Lambda}$ & $2\,\Lambda$ \\
\hline
Chiral limit & $0.36 - i 0.35$ & $0.41 - i 0.29$ & $0.43 -i 0.22$ \\
{  Physical point} & $0.41 - i 0.33$ & $ {0.46 - i 0.26}$ & $0.47 - i 0.18$ \\
\hline
\end{tabular}
\caption{Lightest pole positions in the variable $\sqrt{s_\sigma}$ and in units 
of GeV of the $\sigma$-resonance in QCD and using the unitarized $S$-wave 
$\pi\pi$ PWA at leading order (Born term) in ChPT. We show results for the 
chiral limit and the physical point, and for the nominal cutoff of the EFT, 
$\Lambda=4\pi f_\pi$, and variations of factor $2$.
\label{tab.sigma}}
\end{center}
\end{table}

The previous prediction was based on the chiral limit of massless pions. For 
physical (massive) pions the isoscalar scalar $\pi\pi$ PWA at leading order in 
chiral perturbation theory is $(s-m_\pi^2/2)/f_\pi^2$, and the unitarity loop 
function becomes
\begin{align}
\label{200603.7}
g_\pi(s)&=\frac{1}{(4\pi)^2}\left(\log\frac{m_\pi^2}{\Lambda^2}
+\sigma(s)\log\frac{\sigma(s)+1}{\sigma(s)-1}\right)~,
\end{align}
with $\sigma(s)=\sqrt{1-4m_\pi^2/s}$. The unitarized amplitude is now
\begin{align}
\label{200603.8}
T^{(0)}_\pi(s)&=\left[\frac{f_\pi^2}{s-m_\pi^2/2}+g_\pi(s)\right]^{-1}~,
\end{align}
and its analytical continuation to the second RS unveils the poles shown for the 
``physical point'' in Table~\ref{tab.sigma}, where we have used the physical 
values for the pion mass, $m_\pi=139$~MeV, and the decay constant $f_\pi=92.4$~MeV. As in the chiral limit, the pole position shows little sensitivity to the 
value of the cutoff.  
It is certainly remarkable that these results, completely fixed by chiral 
dynamics, unitarity and analyticity, agree with the experimental range quoted by 
the Particle Data Group (PDG), $\sqrt{s_\sigma}=(400 - 550) - i(200 - 350)$ 
MeV~\cite{pdg} and, within a 10\%, with more sophisticated theoretical 
determinations based on dispersion relations~\cite{martin.200207}, 
$\sqrt{s_\sigma}=457\pm 14-i\,279\pm 11$~MeV.
In particular, the pole position evaluated at the physical point and using the 
nominal cutoff of ChPT, $\Lambda=4\pi f_\pi$, is completely consistent with this 
determination.

To further appreciate this result as a way to reproduce resonances that are 
dynamically generated by the degrees of freedom considered, let us briefly 
discuss what happens when extending this procedure to other PWAs. In particular, 
in the $\pi\pi$ $P$-wave scattering amplitude one identifies experimentally the 
$\rho(770)$ resonance. However, unitarizing as before the leading $\pi\pi$ 
scattering amplitude in this channel one only obtains a pole consistent with the 
$\rho(770)$ using unnatural values for the cutoff, $\Lambda\sim1$ 
TeV~\cite{oller.book}. Thus, this resonance cannot be explained dynamically 
using pions as explicit degrees of freedom, clearly indicating that the $\rho$ 
and the $\sigma$ are resonances of very different nature. This difference can be 
examined by varying the number of colors $N_c$ in QCD while letting its coupling 
constant scale as $1/\sqrt{N_c}$ to ensure a smooth large $N_c$ 
limit~\cite{tHooft:1973alw,Witten:1979kh}. The meson decay constant scales as 
$f_\pi\sim\mathcal O(\sqrt{N_c})$ and, therefore, the position of the pole of 
the $\sigma$ resonance is very sensitive to the value of $N_c$, becoming heavier 
and broader as $N_c$ increases~\cite{nd,Pelaez:2003dy}. On the other hand, 
fundamental quark-antiquark mesons such as the $\rho$-resonance become stable 
particles with a definite $\mathcal O(1\text{ GeV})$ mass in the large-$N_c$ 
limit~\cite{Witten:1979kh,Pelaez:2003dy}. In summary, while some resonances are 
naturally produced by unitarizing the dynamics of the low-energy degrees of 
freedom in the EFT, other states are ``elementary'' and they encode information 
about the UV completion. As discussed above in Sec.~\ref{sec.200531.1}, these 
states can be added as CDD poles to the inverse of the amplitude, which requires 
using additional experimental (or theoretical) information for their positions 
and residues. This distinction between nature of resonances is also very relevant for gravity. In addition to our results we refer to the related 
  self-completeness scenario 
advocated in Ref.~\cite{Dvali:2010bf} for gravity, in which a tower of black 
holes is required by unitarity while other ``elementary'' states would have mass 
near the cutoff of the theory. See also the related 
Refs.~\cite{Dvali:2008fd,Antoniadis:1998ig}.

The unitarized ChPT has found many other successful applications in the strong 
regime of QCD. A remarkable example concerning the hadronic spectrum is the 
$\Lambda(1405)$ resonance in $S$-wave meson-baryon scattering with strangeness 
$-1$ \cite{kw,or,Oller:2000fj}. 
This resonance was already predicted in Ref.~\cite{Dalitz:1959} but, only after 
the unitarization of the chiral EFT PWAs, it was   understood that it 
corresponds to two poles \cite{Oller:2000fj}. This important topic deserves a 
separate review in the PDG \cite{pdg}.


\section{Resonances of the graviton-graviton scattering: the graviball}
\label{sec.200131.1}
\def\theequation{\arabic{section}.\arabic{equation}}
\setcounter{equation}{0}

We proceed now with the application of Eqs.~\eqref{200530.1} and 
\eqref{200530.2} to study graviton-graviton scattering in PWAs. The latter were 
obtained by unitarizing  the Born scattering amplitude, properly modified to 
take care of the IR divergences,
as discussed in Sec.~\ref{sec.200122.1}. Specifically, we consider the emergence 
of resonances in these PWAs, which correspond to poles in the second RS. 
Needless to say, we do not find bound-state poles in the physical RS.
In terms of the analytical continuation of $g(s)$ for graviton-graviton 
scattering into the second RS 
\begin{align}
\label{200827.1}
g_{II}(s)=g(s)-i\frac{\pi}{4},
\end{align} 
the PWAs in this sheet, 
$T^{(J)}_{II;\lm_3\lm_4,\lm_1\lm_2}(s)$, read
\begin{align}
\label{200209.1b}
T^{(J)}_{II;22,22}(s)&=\left[V^{(J)}_{22,22}(s)^{-1}+g_{II}(s)\right]^{-1}~,\\
T^{(J)}_{II;2-2,2-2}(s)&=\left[V^{(J)}_{2-2,2-2}(s)^{-1}+2g_{II}(s)\right]^{-1}
~.\nn
\end{align}

Let us discuss the pole positions in $J=0$, for which $V^{(0)}(s)$ is given in 
Eq.~\eqref{200121.19}. 
We suppress from now on the helicity indices because there is a contribution to 
this PWA only when all the graviton helicities are equal. Indeed, in the other 
possible configurations with helicities $\pm 2$ in the initial and final states 
angular momentum conservation requires that $J\geq 4$.
The PWA $T^{(0)}_{II}(s)$ has a pole in the second RS at $s_P$, whose position 
satisfies the secular equation
\begin{align}
\label{200206.1a}
\frac{\pi}{Gs_P\log a}+\log\frac{-s_P}{\Lambda^2}-i2\pi&=0~.
\end{align}
We rewrite the previous equation in terms of dimensionless variables as
\begin{align}
\label{200206.1}
\frac{1}{\omega x}&+\log (-x)-i2\pi=0,
\end{align}
where
\begin{align}
    x&=\frac{s_P}{\Lambda^2},~~ ~~~~
\omega=\Lambda^2 \frac{G\log a}{\pi}.\nn
\end{align}
When the cutoff derived from unitarity in Eq.~\eqref{200603.4} is chosen this 
equation becomes the same secular equation as for the scattering of massless 
pions, Eq.~\eqref{eq:secular.sigma1}. Hence, we find a resonance with the 
quantum numbers of the vacuum in the gravitational EFT which is  analogous to 
the $\sigma$ resonance of QCD,
\begin{align}
\label{eq:grav.pole.app}
s_P=(0.07-i0.20)\,\Lambda^2\simeq-i\frac{2}{3\pi}\Lambda^2,
\end{align} 
when $\Lambda^2=\pi(G\log a)^{-1}$. Following the analogy with glueballs, we 
will call this resonance the {\it graviball}. 

The pole position $s_P$ is almost purely imaginary, with a real part that is 
smaller, by approximately a factor 3, than its imaginary part. 
Let us write $s_P$ as $s_P= (\varkappa-i2/(3\pi))\Lambda^2$, where $\varkappa 
\Lambda^2$
is the real part of $s_P$.
A Laurent-series expansion around the pole at $s_P$ is dominated by the 
pole-term contribution, 
\begin{align}
\label{200602.2}
\frac{1}{s-s_P}&=\frac{1}{s-\varkappa \Lambda^2+i\frac{2\Lambda^2}{3\pi} },
\end{align}
whose modulus squared is
\begin{align}
\label{200602.3}
\frac{1}{|s-s_P|^2}&=\frac{1}{(s-\varkappa \Lambda^2)^2+\frac{4 
\Lambda^4}{9\pi^2} }~.
\end{align}
Remarkably its peak is located at $\varkappa \Lambda^2\ll \Lambda^2$, because 
$|\varkappa|\ll 1$, and its strength
spreads over a wide region. For instance, the point in $s$ at which its value is 
half the value at
the peak is $s\sim2 \Lambda^2/(3\pi)$. It is also asymmetric in  the physical 
axis for $s>0$  because
its width is much larger than the position of the peak.

\begin{figure}
\begin{center}
\begin{tabular}{ll}
\includegraphics[width=0.45\textwidth]{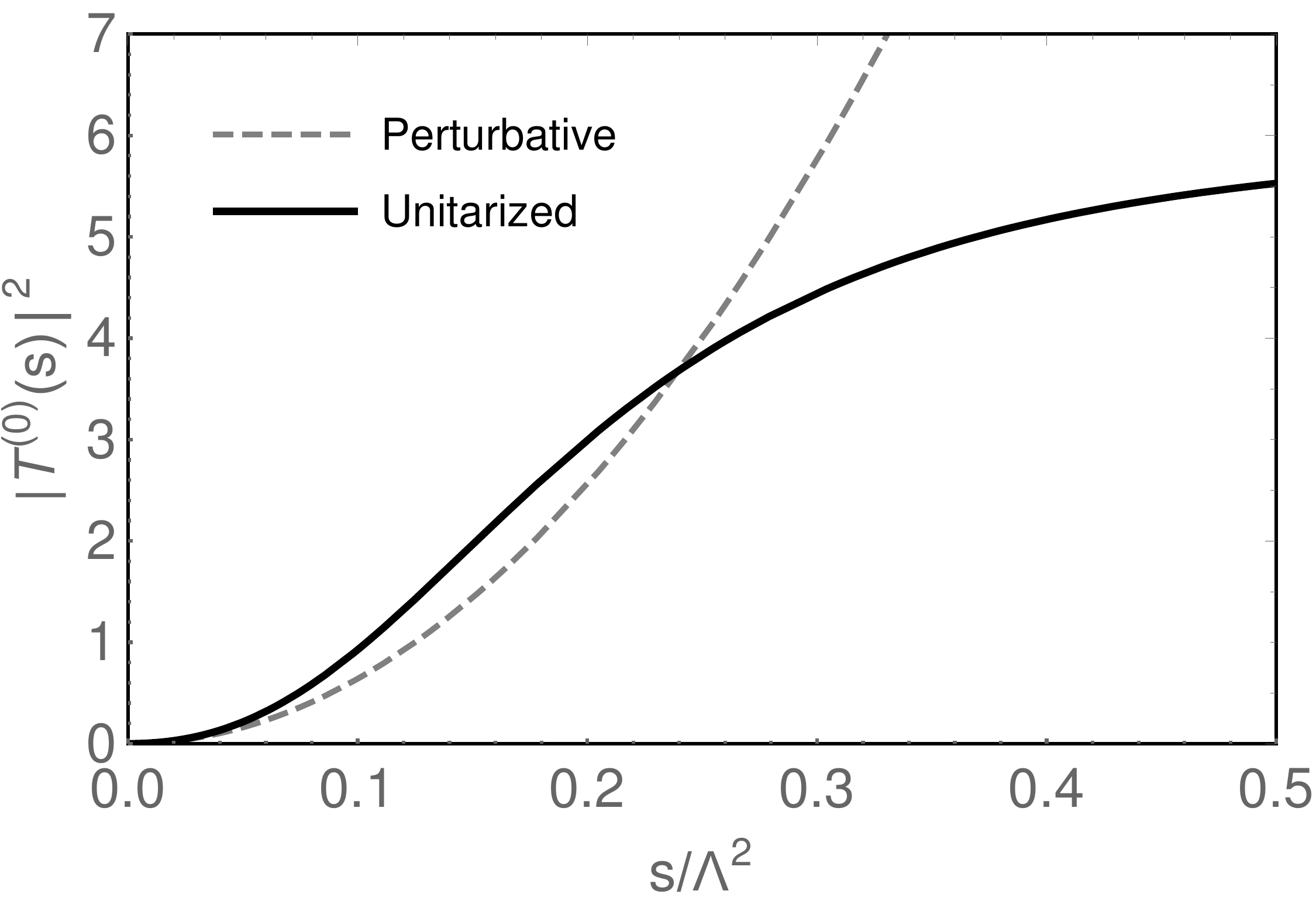} & 
\includegraphics[width=0.45\textwidth]{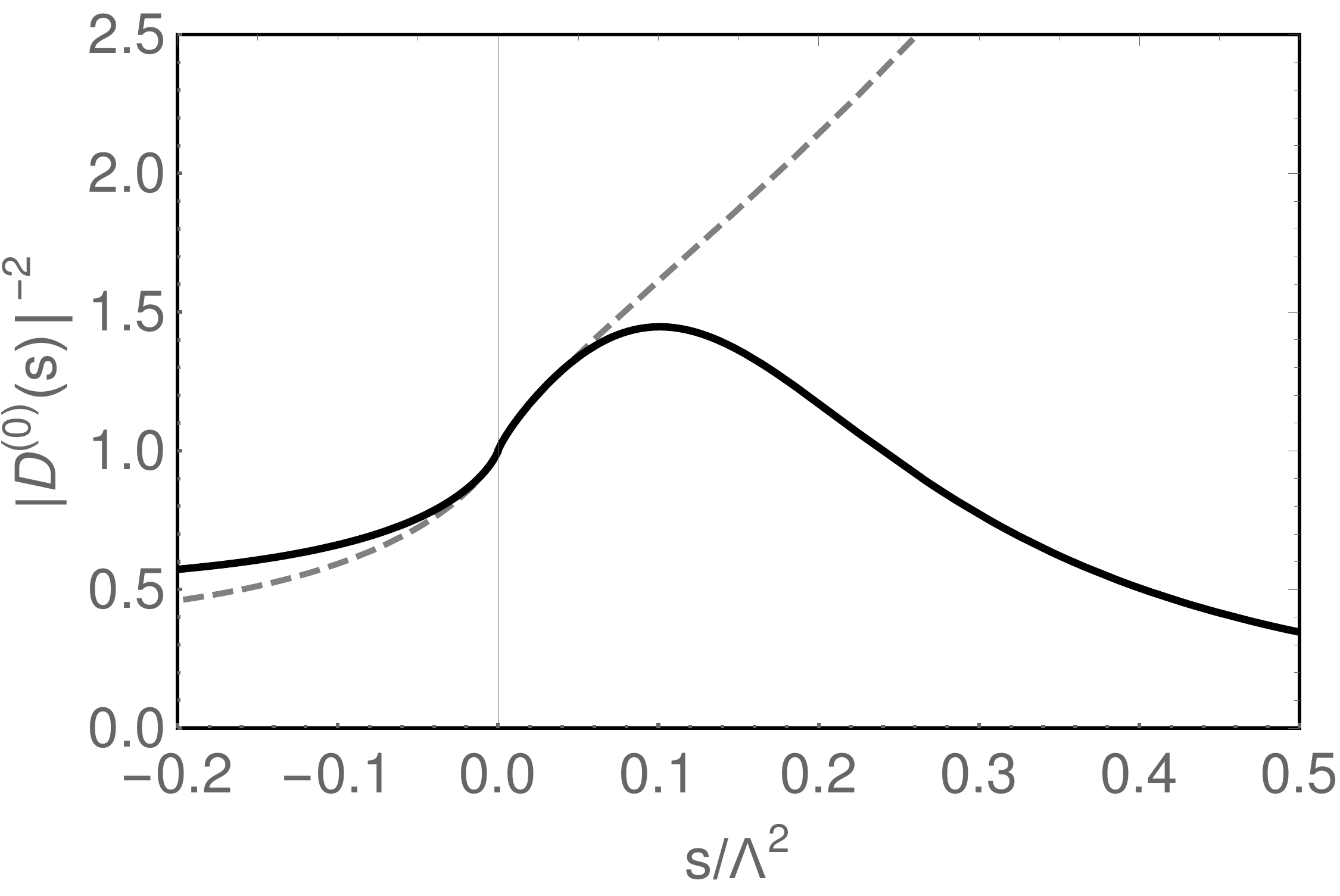} 
\end{tabular}
\caption{{\small The moduli squared of  $T^{(0)}(s)$ (left) and $1/D^{(0)}(s)$ 
(right panel) 
are plotted for $J=0$  as a function of $s$ in units of $\Lambda^2$.   
The presence of the resonance pole is clearly appreciated in $|1/D^{(0)}(s)|^2$ 
for values of $s$ much
lower than $|s_P|$.
The resonance shape is strongly distorted for $|T^{(0)}(s)|^2$ because of the 
zero at $s=0$.The solid(dashed) lines correspond to the full(perturbative)  
results.
\label{fig.170725.1} }} 
\end{center}
\end{figure}

The pole-term contribution in Eq.~\eqref{200602.2} is distorted in the shape of  
$|T^{(0)}(s)|^2$
because of the zero at $s=0$, which implies an energy-dependent driving factor 
$s^2$.
As a result $|T^{(0)}(s)|^2$ has no resonance shape as shown in the left panel 
of Fig.~\ref{fig.170725.1}.
This situation is, again, analogous to the $\sigma/f_0(500)$ resonance in 
pion-pion scattering in QCD \cite{Oller:2004xm} because of the Adler zero, cf. 
Eq.~\eqref{200603.3}.

The resonance effects manifest themselves more clearly in the modulus squared of 
the function that drives  the final- and initial-state interactions
in the corresponding graviton-graviton PWA.
For instance, one could think of a situation where multiple gravitons are 
produced by an energetic source so that a pair of
gravitons with quantum numbers $(J, \lm_{1-4})$ rescatter with certain  energy.
The function   that controls the exchange of
graviton-graviton states with such quantum numbers, 
or from external sources that couple to this system,  
$1/D^{(J)}_{\lm_3\lm_4,\lm_1\lm_2}(s)$, is given by~\cite{oller.book} (also 
called an Omn\`es function)
\begin{align}
\label{170725.7}
\frac{1}{D^{(J)}_{2\pm,2\pm 2}(s)}&=\left[1+2^{\lm/4}V^{(J)}_{2\pm2,2\pm2}(s) 
g(s)\right]^{-1}~,
\end{align}
where $\lambda=\lambda_1-\lambda_2$.
Notice also that $1/D^{(J)}=T^{(J)}/V^{(J)}$. 

We show in the right panel of Fig.~\ref{fig.170725.1} the modulus squared of 
$|1/D^{(0)}(s)|^2$ as a function of $s$ in units of $\Lambda^2$.
We clearly observe a resonance shape with an $s$ dependence driven by the  pole 
$s_P$ in the complex-$s$ plane,  Eq.~\eqref{200206.2}.
The  peak in $|1/D^{(0)}(s)|^2$ is located at $s\simeq\, 0.1 \Lambda^2$, which 
is much smaller than the modulus of $s_P$, but 
displaced towards larger values than $\varkappa \Lambda^2$.
Of course, in the Laurent expansion of $1/D^{(J)}(s)$ there are other 
contributions in addition to the just the pole-term one (this is discussed explicitly for the $\sigma$ resonance in Refs.~\cite{Oller:2004xm,Oller:2003ca}). 
The value around 1 of $|1/D^{(0)}(s)|^2$  at the peak, as shown in 
Fig.~\ref{fig.170725.1}, can be understood  by identifying  the residue of 
$1/D^{(0)}(s)$ at $s_P$. 
Then,  the isolated pole-term contribution to $1/|D^{(0)}(s)|^2$ is 
approximately  $(4/(9\pi^2))/((s/\Lambda^2)^2+4/(9\pi^2))$, whose maximum value
is~1.

Therefore, we conclude that 
the physical effects of the pole could be felt at values quite lower  than 
$\Lambda^2$ because its resonant peak occurs at  $s \ll \Lambda^2$. 
Needless to say that a pure perturbative expansion within the EFT of gravity 
could be severely affected by this fact.\footnote{For instance, this has been the case for low-energy $\pi\pi$ scattering where the isoscalar scalar $\pi\pi$ phase shifts receives large higher-order corrections which amount around a 40\% of the leading result, while they are negligible for the isotensor scalar scattering length \cite{Leutwyler:2008ma}. The nucleon $\sigma$ term is also strongly affected by the non-perturbative physics due to the $\sigma$ resonance because of the pion-scalar form factor through the $\Delta_\sigma$ term \cite{Gasser:1990ap}.}
To illustrate this point let us consider the first two terms in the expansion of 
$1/D^{(0)}$ in powers of $s$, which we call
$1/D_2^{(0)}(s)$,
\begin{align}
\label{200603.1}
\frac{1}{D_2^{(0)}(s)}&=1-\frac{s}{\Lambda^2}\log\frac{-s}{\Lambda^2}~.
\end{align}
We compare in Fig.~\ref{fig.170725.1}, right panel, the modulus squared of this
function with $1/|D^{(0)}(s)|^2$ in the lower-energy region up
to $s/\Lambda^2=0.5$. The former is indicated by the dashed line and the latter 
by the solid one.
For negative $s$ the series expansion is much better behaved. 
However, the difference increases rapidly for positive $s$, where the peak lies,
so that it rises from near a 1\% at $s/\Lambda^2=0.06$    o more than 100\% at 
$s/\Lambda^2\gtrsim 0.2$.
In addition, $|1/D_2^{(0)}(s)|^2$ completely fails to provide the right resonant 
shape of $1/|D^{(0)}(s)|^2$ as
a function of $s/\Lambda^2$ above threshold. We will comment more on possible 
phenomenological consequences in Sec.~\ref{sec:conclu}, though we leave a more 
detailed analysis for future work. 
For clarification, with the dashed line in the left panel of 
Fig.~\ref{fig.170725.1} we actually plot $|V^{(0)}(s)/D_2^{(0)}(s)|^2$.

\vskip 10pt

We do not give the pole positions for $J=2$ because the partial-wave projected 
Born amplitude given in Eq.~\eqref{200204.1} 
is much weaker than for $J=0$, so that the poles lie too deep inside the 
complex-$s$ plane to consider them
reliable.  
Given the fact that, as $J$ increases, the corresponding pole positions move 
further away from the physical $s$ axis  
than the discussed pole for $J=0$, we stress the role of the latter in our 
exploratory purposes of this work. This trend will be  also manifest in the AC 
toy-model of scattering, developed in Sec.~\ref{app.170723.2}, with higher $J$ 
poles
moving deeper in the complex plane as $J=\ell$ increases.

\subsection{On the robustness of the graviball}
\label{sec.200813.1}
\def\theequation{\arabic{section}.\arabic{equation}}
\setcounter{equation}{0}



In the analysis of the graviball done above we have assumed that the cutoff of 
gravity is close to the unitarity one. Namely we have taken $\omega=1$ in 
Eq.~\eqref{200206.1} leading to a resonance in gravity resembling the $\sigma$ 
of QCD. The difference between  the two theories, however, is twofold: 
First, the IR divergences generated the parameter $\log a$ in our expressions; 
second, for gravity we do not know the fundamental cutoff of the EFT and the 
parameter $\omega$ can be significantly different from 1.

Changing $\omega$ only slightly, we find that  $s_P$ is rescaled by the same 
amount, as can be easily deduced from the derivation of 
Eq.~\eqref{eq:grav.pole.app}.
Increasing $\omega$ from 1 moves the pole closer to the real-$s$ axis and the 
resonance becomes narrower. One can also check that for $\omega\gtrsim10$ the 
real and imaginary parts of the pole are of the same order of magnitude and its 
absolute value monotonically decreases with $\omega$ reaching the limit $x\to0$ 
when $\omega\to \infty$.

An attempt to move towards large values of $\omega$  may be to add new `light' 
degrees of freedom to the problem. 
In fact, the strength of the interaction effectively increases when adding new 
fields because their corresponding two-particle cuts will contribute as 
intermediate states to $g(s)$ \cite{Veneziano:2001ah,Dvali:2007wp}.
In Eq.~\eqref{200530.1} we already noticed a factor 2 multiplying $g(s)$ for 
$\lambda=4$ because the two channels combining helicities $\pm2$ are coupled 
with each other. At the level of the denominator of the PWA, where 
$1/V^{(J)}(s)$ enters, this is equivalent to multiplying the coupling $G$ by the 
same factor and leaving $g(s)$ untouched. Thus, we can parameterize the strength 
of the gravitational interaction by replacing $\log a$ by $N  \log a$ in 
Eqs.~\eqref{200206.1a} and \eqref{200206.1}, where $N$ is the number of channels 
coupled to the two gravitons in $J=0$, assuming that all the corresponding 
fields have masses much lower then $G^{-1/2}$, so that the same $g(s)$ results 
for these new channels. Therefore, a (naive) realization of the case $\omega\gg 
1$ would  correspond to  increasing $N\gg 1$ while keeping the fundamental 
cutoff of the theory fixed. 

However, this might not be the case. As shown in  Ref.~\cite{han.200204.1}  the 
scale at which perturbative unitarity is violated in the EFT of quantum gravity 
coupled to $N_S$ scalar, $N_V$ vector and
$N_f$ fermion light fields, is  $20(G N)^{-1}$, with 
$N=\frac{2}{3}N_S+N_f+4N_V$. This may indicate a similar
reduction of the
fundamental scale of gravity. Note also that a similar reduction has been 
suggested from   other non-perturbative arguments based on black-hole physics, 
see e.g.   
Ref.~\cite{Dvali:2001gx,Dvali:2008fd,gomez,Dvali:2007hz,Dvali:2007wp,
Dvali:2014ila}. In this scenario, changes in $\Lambda^2$ due to the coupling of 
gravity to lighter fields, as those in the Standard Model, would not modify the 
value of the ratio $x=s_P/\Lambda^2$ within our approach. In absolute terms, 
this means that the resonance pole becomes narrower and lighter as $N$ 
increases, though the relation to the cutoff of the theory remains basically 
untouched.  On the other hand, scenarios with $\omega\gg 1$, corresponding to  
theories with   fundamental  cutoff   larger  than  the  unitarity  cutoff, may 
be achieved within the so-called self-healing mechanism of 
Ref.~\cite{donoghue.200204.1}. In this case,  the broken perturbative unitarity 
is ``healed'' by a resummation yielding a unitary non-perturbative result. 
There are also theories in which  gravitation could be properly described by a 
quantum field theory with an ultraviolet fixed 
point~\cite{weinberg:safe,Niedermaier:2006wt,Codello:2008vh,Falls:2014tra}. In 
this case, one would take $\omega=1$ because the cutoff of the theory would be 
fixed to the unitarity one,  since we are neglecting multi-particles 
intermediate states (by our unitarization procedure only two-body unitarity is 
restored), whose contributions are suppressed by powers of $(Gs)^3$ at least 
within the low-energy EFT of gravity 
\cite{Donoghue:1993eb,donoghue.200205.1,BjerrumBohr:2002kt,burgess,
Donoghue:2017pgk}.

For $\omega<1$ the pole moves further out in the complex-$s$ plane and for 
$\omega\lesssim 0.2$  its position is $|x_P|\gtrsim1$, where our approach rooted 
in the EFT becomes more model dependent. Notice that assuming $\log a\sim O(1)$ 
then $\omega< 1$ is realized in any scenario of (pure) gravity with a 
fundamental scale $\Lambda^2$ lower than $ \pi G^{-1}$. Nonetheless, as 
discussed above, 
even in this last case one could  restore that the pole position $s_P$ is below 
the fundamental cutoff $\Lambda^2$ by increasing the number of light degrees of 
freedom  in hypothetical scenarios in which $\Lambda$ stayed fixed (as in the 
presence of self-healing)  or, at least, decreased  with $N$ less strongly than 
$1/N$.


\section{IR divergences and unitarization of the exactly soluble AC model}
\label{app.170723.2}
\def\theequation{\arabic{section}.\arabic{equation}}
\setcounter{equation}{0}

Let us illustrate the method for the treatment of the IR singularities in 
Sec.~\ref{sec.200122.1}
for the crossed-channel exchange of massless force-carriers by applying them to 
an \emph{exactly}
solvable model in Quantum Mechanics that shares some similarities with 
gravitational scattering. 
For this, we  introduce   a modified Coulomb potential, with a zero in momentum 
space for $p\to 0$, vanishing as $p^2$. 
This mimics the behavior of graviton-graviton scattering.
This feature is also shared by $S$- and $P$-wave scattering of pions in low 
energy QCD and in the chiral limit.
For the actual pion masses the zero of the $S$-wave shifts by an amount of 
${\cal O}(m_\pi^2)$, with $m_\pi$ the pion mass, and it is called an Adler zero  
\cite{Adler:1964um,Adler:1965ga}, cf. \eqref{200603.8}.
This is the inspiration for the name {\it Adler-Coulomb} (AC) scattering for the 
toy model that we want to discuss now.

The energy-dependent AC potential is defined as\footnote{For 
definiteness, one can consider the Hermitian operator 
\begin{equation}
    V^{AC}(r)=-\frac{\alpha}{r}\frac{\partial_t^2}{M^2}~.\nn
\end{equation}
}
\begin{align}
\label{170723.1}
V^{AC}(r)=\frac{\alpha}{r}\frac{E^2}{M^2}~,
\end{align}
where $r$ is the interparticle distance,
$E=p^2/2m$ is the non-relativistic kinetic energy,
$m$ is the reduced mass,
$\alpha$ is a dimensionless constant 
  and $M$ is some (ultraviolet) scale introduced for dimensional reasons.

The scattering by an AC potential can be solved exactly because
it differs from the pure Coulomb one only by the energy dependent factor 
$E^2/M^2$. 
The energy enters parametrically in the differential Schr\"odinger equation,  
so that we can use the exact solutions in Quantum Mechanics for the Coulomb 
scattering and then make the replacement 
\begin{align}
\label{170723.2}
\alpha\to \alpha \frac{E^2}{M^2}~.
\end{align}
The kinematics for non-relativistic scattering is different from that discussed 
in Sec.~\ref{sec.200121.1}, cf. App.~\ref{app.200805.1}. 
In particular, the relation between the 
PWA $T_\ell(p^2)$  and the $S$ matrix $\mfs_\ell(p^2)$ is 
\begin{align}
\label{200716.3}
\mfs_\ell&=1+i\frac{ m p}{\pi}T_\ell~.
\end{align}
As a result, the unitarity relation for a PWA $T_\ell(p^2)$ reads
\begin{align}
\label{200128.3b}
\Im T_\ell(p^2)&=\frac{mp}{2\pi}|T_\ell(p^2)|^2~.
\end{align}
A detailed account of scattering in partial waves in non-relativistic scattering 
can be found in Ref.~\cite{Oller:2018zts}, whose conventions we used in the 
previous expression. 

Let us first recall that the Coulomb wave function with orbital angular momentum 
$\ell$ tends asymptotically, for $r\to \infty$, to \cite{landau.170517.1} 
\begin{align}
\label{170517.4b}
u^{C}_\ell(r)\sim A\sin\large(pr-\ell\frac{\pi}{2}+\sigma^C_\ell(p)+\gamma \log 
2pr\large)~,
\end{align}
where $\gamma$ is the Sommerfeld parameter 
\begin{align}
\label{170517.6b}
\gamma&=\frac{\alpha m}{p}~,
\end{align}
and $\sigma^C_\ell(p)$ is  the Coulomb phase shifts. On the other hand, for a 
finite-range potential the asymptotic wave function is
\begin{align}
\label{200130.6}
u_\ell(r)\sim A\sin(pr-\ell\frac{\pi}{2}+\delta_\ell(p))~.
\end{align}
The difference between the asymptotic forms in Eqs.~\eqref{170517.4b} and  
\eqref{200130.6} is clear.
In the case of an infinite-range potential 
the phase shifts are ill-defined because of the $r$ dependence stemming from the 
contribution
$\gamma\log 2pr$ in the phase of $u^{C}_\ell(r)$.
Precisely, the Coulomb phase is defined once the diverging
phase $\log 2pr$ for $r\to \infty$ is removed. The removal of this 
phase corresponds to the multiplication of the Dyson $S$ matrix for finite-range 
interactions by $S_c^{-1}$, so as to remove the IR divergences associated to the 
infinite-range interactions as discussed in Sec.~\ref{sec.200122.1}. 

The Coulomb phases are connected to the Coulomb $S$ matrix, $\mfs^C_\ell$, by  
\begin{align}
\label{200128.1}
\mfs^C_\ell(p^2)&=e^{2i\sigma^C_\ell(p)}=\frac{\Gamma(1+\ell-i\gamma)}{
\Gamma(1+\ell+i\gamma)}~,
\end{align}
so that
\begin{align}
\label{170517.7b}
\sigma^C_\ell(p)=\rm{arg}\,\Gamma(1+\ell-i\gamma)~.
\end{align}

Taking into account the rule of Eq.~\eqref{170723.2}, it follows that the 
asymptotic wave function
in the AC model is 
\begin{align}
\label{200130.3}
u^{AC}_\ell(r)\sim 
A\sin(pr-\ell\frac{\pi}{2}+\sigma^{AC}_\ell(p)+\frac{p^4}{(2mM)^2}\gamma \log 
2pr)~,
\end{align}
with $\sigma^{AC}_\ell(p)$ given now by
\begin{align}
\label{200130.4}
\sigma^{AC}_\ell(p)={\rm 
arg}\,\Gamma\left(1+\ell-i\frac{p^4}{(2mM)^2}\gamma\right)~,
\end{align}
corresponding to the AC-model $S$ matrix, $\mfs^{AC}_\ell$, given by  
\begin{align}
\label{200130.5}
\mfs^{AC}_\ell(p)&=e^{2i\sigma^{AC}_\ell(p)}=
\frac{\Gamma(1+\ell-i\frac{p^4}{(2mM)^2}\gamma)}{\Gamma(1+\ell+i\frac{p^4}{
(2mM)^2}\gamma)}~.
\end{align}

Let us now try to unveil some of the non-perturbative properties of 
\eqref{200130.5} by using unitarity methods. For this, let us consider
a  version of the AC potential screened for distances $r>R$ (such that $pR\gg 1$ 
and taking at the end the limit $R\to \infty$),
\begin{align}
\label{200130.7}
V^{AC}(r)=\frac{E^2}{M^2}\frac{\alpha}{r}\theta(R-r)~.
\end{align}
The previous potential is analogous to introducing a Yukawa potential with a 
non-vanishing photon mass~\cite{taylor.scattering}.  The introduction of a sharp 
IR cutoff is indeed a technique used to solve numerically the Coulomb potential 
in partial waves~\cite{landau.170517.1},
because with this finite-range potential one can  solve for the wave function 
$u_\ell(r)$ 
(and hence for $\delta_\ell(p)$), and match with the asymptotic behavior of the 
Coulomb wave function
$u_\ell^C(r)$ at a distance $R$.
As a result, the pure Coulomb phase shifts $\sigma^C_\ell(p)$  stems from the 
relation 
\begin{align}
\label{200130.8}
\sigma^C_\ell(p)=\delta_\ell(p)-\gamma\log 2 pR~.
\end{align}
Another useful feature is that one can proceed purely in momentum space and 
solve for the Lippmann-Schwinger equation in the
presence of another short-range potential, e.g. due to strong interactions, get 
$\delta_\ell(p)$, remove
from it the  divergent phase $\gamma\log  2kR+\sigma^C_\ell(p)$ and finally take 
the limit $R\to \infty$.
This is the standard procedure in nuclear physics to calculate the strong phase 
shifts
in the presence of a Coulomb potential in momentum space \cite{landau.170517.1}.

The Fourier transform of the AC screened potential in Eq.~\eqref{200130.7} is  
\begin{align}
\label{200130.9}
V^{AC}(q^2)&=\frac{p^4}{(mM)^2}\frac{\pi\alpha}{q^2}\left(1-\cos q R \right)~.
\end{align}
where $\vq=\vp'-\vp$ is the momentum transferred in the scattering events and 
$q^2=2p^2(1-\cos\theta)$.
From the form of the asymptotic wave function $u^{AC}_\ell(r)$ one concludes 
that
the diverging phase  $S_c^{AC}(p^2)$ is  
\begin{align}
\label{200130.10}
S_c^{AC}(p^2)&=e^{2i\frac{p^4}{(2mM)^2}\gamma \log2p R}~.
\end{align}
Then, at ${\cal O}(\alpha)$ it implies the following
diverging contribution to any partial-wave projected Born amplitude
\begin{align}
\label{200130.11}
\delta F^{AC}_\ell(p^2)&=\frac{p^2\pi\alpha}{2(mM)^2}\log 2pR
~.
\end{align}
This contribution is independent of the PWA, as it should, and 
has to be removed from the partial-wave projected Born amplitudes. 
Let us work out explicitly the PWA with $\ell=0$.
The projection in $S$-wave of $V^{AC}(q^2)$ is
\begin{align}
\label{200130.12}
F^{AC}_0(p^2)&=\frac{1}{2}\int_{-1}^{+1}d\!\cos\theta 
\frac{p^4\pi\alpha}{(mM)^2q^2}\left(1-\cos q R \right)
=\frac{p^2\pi\alpha}{2(mM)^2}\left[\gamma_E-{\rm ci}(2pR)+\log(2pR)\right]\\
&=\frac{p^2\pi\alpha}{2(mM)^2}\left[\gamma_E+\log 2pR\right]+{\cal 
O}(R^{-2})~,\nn
\end{align}
where $\gamma_E$ is the Euler constant and, as expected, it diverges as $R\to  \infty$.
After removing the diverging term of Eq.~\eqref{200130.11}, which is analogous 
to 
Eq.~\eqref{200121.18b} for graviton-graviton scattering, we are left with the 
final expression for
the $S$-wave projected Born amplitude, $V^{AC}_0(p^2)$, 
\begin{align}
\label{200130.13}
V^{AC}_0(p^2)&=\frac{p^2\gamma_E \pi\alpha }{2(mM)^2}~.
\end{align}
It is remarkable that  the right $S$-wave Born
amplitude is obtained in this way, as it can be
concluded by comparing with the expansion in powers of $\alpha$ of 
$\mfs^{AC}_\ell(p^2)$, Eq.~\eqref{200130.5},
\begin{align}
\label{200130.14}
\mfs^{AC}_0&=1+2i\frac{m \alpha}{p}\frac{\gamma_E p^4}{(2mM)^2}+{\cal 
O}(\alpha^2)=
1+i\frac{mp}{\pi}V^{AC}_0(p^2)+{\cal O}(p^2)~.
\end{align}

Had we proceeded as explained in Sec.~\ref{sec.200122.1}, we would have taken 
from  Ref.~\cite{Weinberg:1965nx} the phase factor $S_c$, which for Coulomb 
scattering reads \cite{Weinberg:1965nx}
\begin{align}
\label{200811.2}
S_c=\exp\left[i2\alpha\gamma \log\frac{\mfl}{\mu}\right]~.
\end{align}
Analogously to Eq.~\eqref{200121.8b}, we then have now that for the AC model  
$\delta F^{AC}_\ell=\frac{p^2\pi\alpha}{2(mM)^2} \log\frac{\mu}{\mfl}$, and 
$\log 2pR$ in Eq.~\eqref{200130.11} is replaced by $\log\mfl/\mu$. It is clear 
that the radius $R$ of the screening is proportional to $1/\mu$, and then we are 
left again with $\mfl\propto 2p/a$, $a\sim {\cal O}(1)$, as explained in 
Eq.~\eqref{200121.20}. This term has to be added to the partial-wave projection 
of the AC potential in momentum space with a finite photon mass $\mu$. Its 
$S$-wave projection reads
\begin{align}
\label{200810.1}    
F_0^{AC}&=\frac{p^4\pi\alpha}{(mM)^2}\int_{-1}^1\frac{d\cos\theta}{q^2+\mu^2}
=\frac{p^2\pi\alpha}{(mM)^2}\log\frac{2p}{\mu}~.
\end{align}
After summing it with $\delta F^{AC}_0(p^2)$ we are then left with
\begin{align}
\label{200810.2}    
V_0^{AC}&=\frac{p^2\pi\alpha}{(mM)^2}\log a~.
\end{align}
By comparing this expression with the exact result for $V^{AC}(p^2)$ in 
Eq.~\eqref{200130.13} we then identify $\log a\to \gamma_E/2$ at LO in the $p
^2/(2m M)^2$ expansion.  This value is somewhat smaller than the natural values 
we suggested before, though
it is derived by comparing perturbative results only. We will soon see that a 
larger value of $\log a$ is indeed found when matching non-perturbative 
predictions for the spectroscopy of the model.  
\subsection{Unitarization of the AC scattering and comparison with its exact 
solution}
\label{sec.200530.1}

For the unitarization of $V^{AC}_\ell(p^2)$ in the AC model we use 
an expression analogous to Eq.~\eqref{200716.2} but with a different integrand 
in
the integral of the RHC because the phase-space factor is different for massive 
particles
in non-relativistic scattering. Now we have,
\begin{align}
\label{200130.15a}
g(p^2)&=a(C)-\frac{m(p^2-C)}{2\pi^2}\int_0^\infty 
dk^2\frac{k}{(k^2-p^2)(k^2-C)}\\
&=a(C)-\frac{m\sqrt{-C}}{2\pi}-\frac{im\sqrt{p^2}}{2\pi}~.\nn
\end{align}
The subtraction point $C$  is negative ($C<0$), so that 
the associated subtraction constant $a(C)$ is real.
The corresponding expression for the unitarized PWAs is then
\begin{align}
\label{200130.15b}
T^{AC}_\ell(p^2)&=\left[\frac{1}{V^{AC}_\ell(p^2)}+g(p^2)\right]^{-1}.
\end{align}
For the particular case of $\ell=0$ it becomes
\begin{align}
\label{200130.15}
T^{AC}_0(p^2)&=\left[\frac{8(mM)^2}{\gamma_E e^2 p^2}
+a(C)-\frac{m\sqrt{-C}}{2\pi}-\frac{im\sqrt{p^2}}{2\pi}\right]^{-1}~.
\end{align}

Let us start the comparison with the exact solution of the AC model by 
considering the spectroscopy. 
The exact pole positions of $\mfs^{AC}_\ell(p)$ are given by the poles of the 
function $\Gamma(z)$
in the numerator of Eq.~\eqref{200130.5}, which occur for $z=-n$,
with $n=0,$~1,~2, $\ldots$ ($n\in \left\{0\right\}\cup \mathbb{N})$.
This gives rise to a third-degree equation for $p$ with the solution, 
\begin{align}
\label{200130.18}
p(\nu)&=(-i)^{1/3}\lambda(\nu)~,
\end{align}
where
\begin{align}
\lambda(\nu)&=\left[\frac{4m M^2}{\alpha}(1+\ell+n)\right]^{1/3}~,
\end{align}
and the pole position depends only on the sum $\nu=n+\ell$.
For every $\nu$ there are three solutions because of the cube root, whose 
explicit expressions are: 
\begin{align}
\label{200130.19}
p_1(\nu)&=i\lambda(\nu)~,\\
p_2(\nu)&=e^{-i\pi/6} \lambda(\nu)~, \nn\\
p_3(\nu)&=-e^{i\pi/6} \lambda(\nu)~,\nn
\end{align}
and the three poles have the same absolute value for their pole positions, 
$|p_i(\nu)|$,
which monotonically increases with $\nu$. 
The first pole at $p_1$ is a bound state in the physical RS, 
while the last two poles  at $p_2$ and $p_3$ are resonances in the second RS 
($\Im p<0$).
Notice that $p_3=-p_2^*$, because in the variable $p^2$ or energy $E$ these two 
pole positions 
are complex conjugate to each other, as required by the Schwarz reflection 
principle applied
to the $S$ matrix of Eq.~\eqref{200130.5}.

We now turn to consider the poles in the unitarized expression of 
$T^{AC}_0(p^2)$, Eq.~\eqref{200130.15} (recall also \eqref{200716.3}).
Let us first study the case in which
\begin{align}
\label{170723.9b}
a(C)=\frac{m\sqrt{-C}}{2\pi}~,
\end{align}
so that, according to Eq.~\eqref{200130.5}, we end up with a third-degree 
secular equation for $p$ without $p^2$ terms, as in the equation for the exact 
solution:
\begin{align}
\label{200131.1}
T^{AC}_0(p^2)&=\left[\frac{2(mM)^2}{\gamma_E \pi\alpha p^2}
-i\frac{m\sqrt{p^2}}{2\pi}\right]^{-1}~.
\end{align}
Thus, no \textit{a priori} undetermined parameter remains. We will show in   
Sec.~\ref{app.200127.1}  that Eq.~\eqref{200131.1} can be directly derived  from 
analyticity and unitarity invoking the Sugawara-Kanazawa theorem 
\cite{Sugawara:1961zz,oller.book}. 

The secular equation for the poles of Eq.~\eqref{200131.1} is  
\begin{align}
\label{200130.20}
&\frac{2(mM)^2}{\gamma_E \pi\alpha p^2}-\frac{imp}{2\pi}=0~,
\end{align}
whose solution is
\begin{align}
\label{200130.21}
p&=\left[-i \frac{4 mM^2}{\alpha\gamma_E}\right]^\frac{1}{3}~.
\end{align}
This expression differs from the exact pole position for $\nu=0$ given in 
Eq.~\eqref{200130.18}
by the factor  $\gamma_E^{-1/3}=1.20$. It is   remarkable that one recovers 
rather accurately the features and parameters of the resonances of the full 
theory with such a simple input (the Born amplitude) in the unitarization 
process.
  Also notice that, in connection with the general treatment of IR divergences 
in Sec.~\ref{sec.200122.1}, one recovers the exact pole positions by  taking $\log 
a=1/2$ in Eq.~\eqref{200810.2}, instead of $\gamma_E/2$, as 
required by the LO expansion of $V^{AC}_0(p^2)$ in Eq.~\eqref{200130.13}. We 
also show below that by including higher-order terms in the expansion of 
$V_0^{AC}(p^2)$ in powers of $\alpha$ one can recover the pole positions in 
Eq.~\eqref{200130.19} for $\nu=0$ with arbitrary precision. From here one also 
learns that by varying $\log a$ with respect to its LO value one may reabsorb 
effectively higher-order corrections and improve the convergence, as well as the 
predictions that result from the unitarization of the Born term. 

Rather than imposing Eq.~\eqref{170723.9b}, which requires the knowledge of
the exact solution $\mfs_\ell^{AC}(p^2)$ , we now determine the poles of the 
unitarized amplitude by using estimates of $a(C)$ and the ultraviolet scale 
$\Lambda^2$ based on naturalness.  This connects to the treatment we performed 
in Sec.~\ref{sec.200531.1} for the graviton scattering and in 
Sec.~\ref{sec.200605.1} for the $\sigma$ meson.
The first type of estimate for the subtraction constant $a(C)$ is based in a 
change of ${\cal O}(1)$ in the subtraction point. Proceeding similarly as in 
Eq.~\eqref{200531.3}, we find for this case that  
\begin{align}
\label{200810.3}
a(C_1)-a(C_2)=\frac{m(\sqrt{-C_1}-\sqrt{-C_2})}{2\pi}~,
\end{align}
where $\sqrt{-C}\sim \Lambda$ and $\Lambda$ is the cutoff of the theory. 
Indeed the value for $a(C)$ used above in Eq.~\eqref{170723.9b} to end with the 
same type of secular equation as in the exact solution is compatible with the 
estimate in Eq.~\eqref{200810.3}.

The other technique used in Sec.~\ref{sec.200531.1} to estimate the natural size 
of $a(C)$ consists of evaluating the unitarity function with a three-momentum 
cutoff  $\Lambda$, obtaining
\begin{align}
\label{200131.2}
g_c(p^2)&=-\frac{m \Lambda}{\pi^2}-i\frac{m\sqrt{p^2}}{2\pi}+{\cal 
O}\left(\frac{p}{\Lambda}\right)~. 
\end{align}
By taking $\sqrt{-C}=2\Lambda /\pi$ we then have that $a(C)=0$. 
Considering also the first iterated contribution involving the unitarity loop 
function $g_c(s)$ by expanding Eq.~\eqref{200130.15b}, we obtain that this 
contribution is suppressed relative to the Born term by $p^2 \Lambda \alpha 
\gamma_E/2\pi m M^2\equiv p^2/\Lambda^2$, where we have identified $\Lambda$ 
with the unitarity cutoff,
\begin{align}
\label{200810.4}
\Lambda&=\left(\frac{2\pi mM^2}{\alpha\gamma_E}\right)^{1/3}~.
\end{align}

The unitarized PWA of Eq.~\eqref{200130.15b} when using $g(s)=g_c(s)$ reads
\begin{align}
\label{200131.3}
T^{AC}_0(p^2)&=\left[\frac{2 (mM)^2}{\pi \gamma_E \alpha p^2}
-\frac{m \mathfrak{b}\Lambda}{\pi^2}-i\frac{m\sqrt{p^2}}{2\pi}\right]^{-1}~,
\end{align}
where $\mathfrak{b}=\Lambda/\left(\frac{2\pi mM^2}{\alpha\gamma_E}\right)^{1/3}$ 
parametrizes the difference between unitarity and fundamental cutoffs, being 
analogous to the parameter $\omega$ of Eq.~\eqref{200206.1} but now for 
three-momentum cutoffs.
The secular equation in $x=p/\Lambda$ for determining the poles of $T^{AC}_0$ 
becomes
\begin{align}
\label{200131.4}
\frac{1}{x^3}-\frac{\mathfrak{b}}{x}-i\frac{\pi}{2}=0~.
\end{align}

\begin{table}
\begin{center}
\begin{tabular}{l|rrrr}
\hline
$\mathfrak{b}$ & $\gamma_E/2$ & 1/2           & 1  & Exact \\
\hline
$p_1$~ $[\Lambda]$  & $i\,0.93$      & $i\, 0.98$     & $i\,1.13$        & 
$i\,0.72$\\ 
$p_2$~ $[\Lambda]$  & $0.74-i\,0.37$ &$0.73-i\,0.33$  & $0.70-i\,0.25$   & 
$0.62-i\,0.36$\\
\hline
$\log a$ & $\gamma_E/2$ & 1/2           & 1  &  \\
\hline
$p_1$~ $[\Lambda]$  & $i\,0.86$      & $i\,0.72$     & $i\,0.57$        & 
$i\,0.72$\\ 
$p_2$~ $[\Lambda]$  & $0.75-i\,0.43$ & $0.62-i\,0.36$ & $0.49-i\,0.28$   & 
$0.62-i\,0.36$\\
\hline
\end{tabular}
\caption{Lightest pole positions in units of $\Lambda=(2\pi 
mM^2/\alpha\gamma_E)^{1/3}$ of the $S$-wave PWA $T_0^{AC}(p^2)$ for the AC 
model. 
The second, third and fourth columns give the values obtained for the unitarized 
 Born-term PWA $V_0^{AC}(p^2)$, and  the exact results are shown in the last 
column. In the rows 2 and 3 a three-momentum cutoff $\mathfrak{b}\Lambda$, with 
$\mathfrak{b}=\gamma_E/2,$ 1/2 and 1, is used in the secular equation 
\eqref{200131.4}. 
For the rows 4 and 5 there is no linear term in the new secular equation 
\eqref{200911.2}  and $\log a$=$\gamma_E/2$, 1/2 and 1. 
For this case the pole positions vary as $(2\log a)^{-1/3}$ with respect to the 
exact values.  
There is a third pole at $p_3=-p_2^*$ which is not shown. 
\label{tab.200131.1}}
\end{center}
\end{table}

In Table~\ref{tab.200131.1} we show the resulting pole positions in units of 
$\Lambda$   for $\mathfrak{b}=\gamma_E/2$, 1/2 and 1, compared to the exact 
results (we do 
not show $p_3$ because it is always $p_2^*$). The pole positions obtained give a 
good approximation for the exact ones with $\nu=0$. 
The poles are rather well reproduced within a deviation typically of a 
$10-30$\%.  One should also emphasize that the sensitivity to the value taken 
for the ultraviolet cutoff  is much weaker for graviton-graviton scattering, 
where the dependence on $\Lambda$ is only logarithmic, cf. Eq.~\eqref{200206.1}, 
 than for the non-relativistic AC model which has a stronger linear sensitivity  
on $\Lambda$. Despite that, we observe a nearly stationary behavior of the pole 
positions for $\mathfrak{b}$ between $[\gamma_E/2,1/2]$.

Let us now consider how the results vary under changes in $\log a$ around 1. For 
concreteness, we take the subtraction constant $a(C)$ as given in 
Eq.~\eqref{170723.9b}, so that we end with a third-degree secular equation without $p^2$ terms.
Now the Born term is given by  Eq.~\eqref{200810.2} and then, from the 
suppression of the first iterated contribution in powers of 
$p^3/{\Lambda'}^{3}$, we have the unitarity scale $\Lambda'$ corresponding to 
\begin{align}
\label{200811.1}
\Lambda'&=\left(\frac{2\pi mM^2}{\alpha\log a}\right)^{1/3},
\end{align}
and the associated dimensionless variable $x'=p/\Lambda'$. 
The new secular equation reads
\begin{align}
\label{200911.2}
\frac{1}{x^{'3}}-i\pi=0~,
\end{align}
and its solution is $x'=(-i/\pi)^{1/3}$. 
The pole positions for $\log a=\gamma_E/2$, 1/2 and 1 are given in units of 
$\Lambda=(2\pi mM^2/\alpha\gamma_E)^{1/3}$ in the rows 5 and 6 and the columns 
2, 3 and 4, respectively, of Table~\ref{tab.200131.1}. They vary as $(\log 
a)^{-1/3}$ and for $\log a=1/2$ they reproduce the exact values, given also in 
the last column. Let us recall that the value $\gamma_E/2$ for $\log a$ is the 
one derived above for the Born-term amplitude at LO, cf. Eq.~\eqref{200128.5}.

 For higher partial waves, $\ell\geq 2$, the partial-wave projected Born 
amplitudes tend to become repulsive and the corresponding poles
move deeper in the complex plane when $\ell$ increases  as $(1+n+\ell)^{1/3}$, 
cf. Eq.~\eqref{200130.18}. This behaviour is similar to what we encountered for 
the graviton-graviton scattering in  Eq.~\eqref{200204.1}.
As already noticed, the expansion of $\mfs_\ell^{AC}(p^2)$
in powers of $\alpha$ generates the same results as our
calculation for the partial-wave projected Born amplitudes 
with the replacement of $\log a$ by $\gamma_E/2$.
For $\ell=2$ this yields $\gamma_E-3/2<0$, for $\ell=4$ one has 
$\gamma_E-25/12<0$, etc.
Nonetheless, the repulsive nature of these interactions is compensated in the AC 
model when considering the full
nonperturbative solutions. In fact,  the pattern of the poles found  does not 
change as $\ell$ increases, for a given value of $\nu$ 
one always has a bound state plus two resonance poles corresponding to the same 
resonance.
Within our approach, based on the unitarization of the partial-wave projected 
Born amplitudes,  for $\nu=0$ we can reproduce this pattern, even 
quantitatively, as shown in Table~\ref{tab.200131.1}.

The quantum-mechanical AC model also allows us to illustrate the procedure for 
including  higher orders in the unitarization process. 
The PWA $T^{AC}_\ell(p^2)$, corresponding to $\mfs^{AC}_\ell(p^2)$, is given in  
Eq.~\eqref{200716.3}. From this equation together with Eq.~\eqref{200130.15b} it 
also follows that $V_\ell^{AC}$ is given by 
\begin{align}
\label{200716.5}
V_\ell^{AC}(p^2)&=\frac{2i\pi}{mp}\frac{\Gamma(1+\ell+i\frac{p^3\alpha}{4m 
M^2})+\Gamma(1+\ell-i\frac{p^3\alpha}{4m 
M^2})}{\Gamma(1+\ell+i\frac{p^3\alpha}{4m 
M^2})-\Gamma(1+\ell-i\frac{p^3\alpha}{4m M^2})}~,
\end{align}
which is a real function because $\Gamma(z^*)=\Gamma(z)^*$. It is also clear 
that $V_\ell^{AC}$ is a function of $p^2$ because it is even under the change 
$p\to -p$. Its expansion in powers of $\delta=p^3\alpha/(4m M^2(1+\nu))$ is 
\begin{align}
\label{200717.1}
V_\ell^{AC}&=-\frac{2\pi}{m p} v(\delta)~,
\end{align}
where
\begin{align}
v(\delta)&=(1+\nu) 
\psi_0(1+\ell)\delta+\frac{(1+\nu)^3}{6}[2\psi_0(1+\ell)^3-\psi_2(1+\ell)]
\delta^3\nn\\
&+\frac{(1+\nu)^5}{120}[
16\psi_0(1+\ell)^5-20\psi_0(1+\ell)^2\psi_2(1+\ell)+\psi_4(1+\ell)]\delta^5+{
\cal O}(\delta^7)~,
\end{align}
and $\psi_n(z)$ is the polygamma function.\footnote{$\psi_n(z)=d^{n+1}\log 
\Gamma(z)/dz^{n+1}$. In particular, $\psi_0(n)=-\gamma_E+\sum_{k=1}^{n-1}1/k$.} 
The interest of having used $\delta$ as an expansion parameter stems from the 
fact that at  the resonance poles characterized by $\lambda(\nu)$ its value is 
$-i$. It is then clear from the expansion in Eq.~\eqref{200717.1} that the 
coefficients grow very fast with $\nu$ (as odd powers of $(1+\nu)$), so that the 
expansion deteriorates quickly as $\nu$ increases. This is why with the 
unitarization of the AC model we can provide an accurate determination  of the 
three lowest-lying $S$-wave poles only ($\nu=0$). Namely, we look for the poles 
of $T_{0;\n}^{AC}$, 
\begin{align}
\label{200717.2}
T_{0;\n}^{AC}=\left[\frac{1}{V_{0;\n}^{AC}}-i\frac{mp}{2\pi}\right]^{-1}~,
\end{align}
that is the PWA  that results by using $V_{0;\n}^{AC}$, which in turn is the  
expansion of $V^{AC}_0$ up to order  $\n$  in odd powers of $\delta$. 
The  secular equation can be written more easily in terms of $v(\delta)$ as  
\begin{align}
\label{200717.2b}
\frac{1}{v(\delta)}+i=0~,
\end{align}
and $v(\delta)$ is expanded up to the required order in $\delta$. 
We show in Table~\ref{tab.200717.1} the pole positions for $p_1$ and $p_2$. It 
is clear that the pole positions get closer to the exact results as the order 
increases, which are again reproduced in the last column for convenience.

\begin{table}
\begin{center}
\begin{tabular}{r|rrrr}
\hline
                  & $\n=1$              & $\n=3$           & $\n=5$          & 
Exact \\ \hline
$p_1$ $[\Lambda]$ & $i\, 0.860$       & $i\,0.729$      & $i\, 0.717$  & 
$i\,0.716$ \\   
$p_2$ $[\Lambda]$ & $0.745-i\, 0.430$ & $0.631-i\,0.364$ & $0.621-i\,0.358$  & 
$0.620-i\, 0.358$\\   
\hline
\end{tabular}
\caption{Lightest pole positions in units of $\Lambda=(2\pi 
mM^2/\alpha\gamma_E)^{1/3}$ of the $S$-wave PWA $T_{0;n}^{AC}(p^2)$ for the AC 
model with $v(\delta)$ calculated at order $\delta^\n$,  
for $\n=1$, 3 and 5. The pole positions are compared with the exact result, 
Eq.~\eqref{200130.18}, given in the last column. 
The convergence in the results as $\n$ increases is clear.
There is a third pole at $p_3=-p_2^*$ which is not shown. 
\label{tab.200717.1}}
\end{center}
\end{table}

A priori, the case for the $\sigma$ resonance is more favourable than the one of 
the AC model. This is because for the latter $|x|\simeq 0.7$ at the pole 
positions while $|x|\simeq 0.2$  for the $\sigma$
(in the chiral limit of QCD), for which the unitarization methods are known to 
be accurate  to obtain $s_\sigma$ \cite{Albaladejo:2012te}. 
As discussed above, this is the same value of $|x|$ for the graviball when the 
fundamental cutoff is taken to be $\pi/G\log a $. Nonetheless, as in the case of 
the $\sigma$ with ChPT, it is necessary to unitarize the higher orders in the 
EFT of gravity to confirm that the results for $s_P$ of the graviball are also 
converging. In this regard, the unitarization of the next-to-leading order 
graviton-graviton scattering amplitude \cite{Dunbar:1994bn} is most significant, 
and deserves a dedicated study.

\subsection{Unitarization of Coulomb scattering}
\label{app.200127.1} 

In the previous sections we have used  unitarization methods to unveil 
non-perturbative 
information about the scattering of three theories with infinite-range 
interactions:
ChPT in the chiral limit, general relativity and the AC model. For completeness, 
and also on its own sake, we now study what these methods imply for the most 
familiar model of this sort:  Coulomb scattering. To connect it with some of the 
techniques above, we can consider it as the  non-relativistic limit of, e.g., 
electron-positron scattering.

The Coulomb potential is 
\begin{align}
\label{170517.3b}
V^C(r)&=\frac{\alpha}{r}~.
\end{align}
To deal firstly with the IR divergences we follow the same logic as for the AC 
model in Eq.~\eqref{200130.7} and consider the screened version for distances 
$r>R$,
\begin{align}
\label{170518.5b}
V^C(r)=\frac{\alpha}{r}\theta(R-r)~.
\end{align}
The Fourier transform, IR-divergent phase and $S$-wave projected potential are 
given by the Eqs.~\eqref{200130.9}-\eqref{200130.12}, by taking the inverse of 
the replacement in Eq.~\eqref{170723.2}, i.e. $\alpha\to \alpha (M/E)^2$. After 
removing the diverging term  we are left with 
the neat $S$-wave projected Born term,
\begin{align}
\label{200128.5}
V^C_0(p^2)&=\frac{2\pi \alpha \gamma_E}{p^2}~.
\end{align}
This result can also be obtained by expanding $\mfs^C_0(p^2)$, 
Eq.~\eqref{200128.1}, to leading order in $\alpha$, analogously to what was done 
above for the AC model, cf. Eq.~\eqref{200130.14}. 
As discussed in the paragraph following Eq.~\eqref{200811.2}, the removal of the 
  divergent Coulomb phase-factor of the seminal works 
\cite{Dalitz:1951,Weinberg:1965nx}
changes the $S$-wave projected Born term   
$F_0^C(p^2)=\frac{4\pi\alpha}{p^2}\log \frac{2p}{\mu}$, obtained  by employing a 
photon mass $\mu\to 0^+$, to 
\begin{align}
\label{200826.1}
V_0^C(p^2)=\frac{4\pi\alpha}{p^2}\log a.
\end{align}
As in the AC model, for $\log a=\gamma_E/2$ we recover Eq.~\eqref{200128.5}.

Let us move on to the  unitarization of $V^C_0(p^2)$ by considering the PWA 
multiplied by $p^2$, namely, $T^C_0(p^2)p^2$. We denote its inverse by
\begin{align}
f(p^2)\equiv (T^C_0(p^2)p^2)^{-1},
\end{align} 
because  in this way
i) the limit for $f(p^2)$ for $p^2\to \infty$ is just given by the inverse of 
the Born term times $p^2$, which is the constant $(2\pi \alpha \gamma_E)^{-1}$, 
cf. Eq.~\eqref{200128.5}. Notice that $\mfs^{C}$, Eq.~\eqref{200128.1}, becomes 
trivial for $p^2\to \infty$; 
ii) the resulting unitary function has a discontinuity along the RHC equal to 
$-im/(\pi p)$, which vanishes for $p\to \infty$.
The Sugawara-Kanazawa theorem \cite{Sugawara:1961zz,oller.book} implies that any 
analytical function $f(p^2)$ in the complex-$p^2$ plane with only a RHC, having 
a finite limit for $p^2\to \infty\pm i\ep$, and such that $f(p^2)$ 
is bounded by a finite power of $p^2$ for  $p^2\to\infty$, can  be represented 
by the following DR:
\begin{align}
\label{200812.3}
f(p^2)&=\bar{f}(\infty)+\frac{1}{\pi}\int_0^\infty dk^2 \frac{ \Delta 
f(k^2)}{k^2-p^2}~.
\end{align}
In this expression, $\Delta f(k^2)$ is the discontinuity of $f(p^2)$ along the 
RHC,  $\Delta f(k^2)=f(k^2+i\ep)-f(k^2-i\ep)$, $k^2\geq 0$, and 
$\bar{f}(\infty)=[f(+\infty+i\ep)+f(+\infty-i \ep)]/2=f(\infty\pm i\ep)$ 
(because there is no LHC). For the case at hand, 
this translates into the DR representation for $(p^2T^C_0(p^2))^{-1}$,
\begin{align}
\label{200812.4}
\frac{1}{p^2 
T^C_0(p^2)}&=\frac{1}{2\pi\alpha\gamma_E}-\frac{m}{2\pi^2}\int_0^\infty 
dk^2\frac{1}{k(k^2-p^2)}~.
\end{align}
The integral is elementary and we end up with
\begin{align}
\label{200128.7}
T^C_0(p^2)&=\left[\frac{p^2}{2\pi\alpha\gamma_E}-i\frac{m p}{2\pi}\right]^{-1}~.
\end{align}
The zeroes of the exact $T^C_0(p^2)$ would imply poles in $f(p^2)$  not included 
in Eq.~\eqref{200128.7}, which is the unitarization of the LO Born-term PWA. 
They can be accounted for by calculating $V^C_0(p^2)$ at higher orders in 
$\Lambda/p^2$, as done below, cf. Eq.~\eqref{200812.2}.

We can also connect with the DR representation of the PWAs for the AC model, cf. 
Eq.~\eqref{200131.1}, by the general rule of substituting $\alpha\to \alpha 
p^4/(2 mM)^2$, as explained when this model was introduced, cf. 
Eq.~\eqref{170723.2}. Thus, by implementing this substitution into 
Eq.~\eqref{200128.7}, we end with the DR representation for $T_0^{AC}(p^2)$ of 
Eq.~\eqref{200131.1}.

From Eq.~\eqref{200128.7}  we can assess the unitarity corrections which enter 
in the combination $(mp/2\pi)V_0^C(p^2)=m\alpha\gamma_E/p$, suggesting the 
three-momentum scale $\Lambda=m\alpha \gamma_E$. It is now clear that the 
unitarization of Coulomb scattering is valid for $|p|\gtrsim \Lambda$, because 
the Born term has a pole at $p=0$. 
The exact function $\Re(T^C_0(p^2)p^2)^{-1}$ wildly oscillates for $|p|\ll 
\Lambda$, while for $|p|\gtrsim \Lambda$ the DR in Eq.~\eqref{200812.4}  
generates a smooth $T^C_0(p^2)$ that matches with the former. 
Contrarily, for  graviton-graviton scattering and the AC model (cf. 
Eqs.~\eqref{200530.1}, \eqref{200530.2} and \eqref{200130.15b}), one has a zero 
at $p=0$ and the unitarization is a resummation valid at low momenta, 
$|p|\lesssim \Lambda$ (here $\Lambda$ is the corresponding cutoff scale in those 
cases).

Let us notice that for pure Coulomb scattering the excited states have less and 
less
binding energy, approaching zero arbitrarily as the degree of excitation 
increases \cite{landau.170517.1}.
Indeed, the Coulomb PWA has an essential singularity at $p= 0$,
as follows from expanding $\mfs^C_\ell(p^2)$ around $p=0$. 
Therefore, the previous equation concerning the unitarized $T^C_0(p^2)$,
and the resulting unique bound state, is not capturing essential features of 
Coulomb scattering in
partial waves in the limit $p\ll \Lambda$ and could be discarded as a sensible 
approximation in such case, but not for $p\gtrsim m\alpha\gamma_E$. Note that 
the  amplitude in Eq.~\eqref{200128.7} has a bound state precisely at 
$p=i\varkappa$ with
$\varkappa=m \alpha \gamma_E$~,
corresponding to a binding energy 
$\ep=\gamma_E\frac{m\alpha^2}{2}$~.  

To assess the precision of the method to generate the exact bound state, one can 
use  the knowledge of $\mfs^{C}(p^2)$, and
  write down the exact $T_\ell^C(p^2)$ and $V_\ell^C(p^2)$. Following closely 
the procedure  explained in Sec.~\ref{sec.200530.1},
\begin{align}
\label{200812.2}
V_\ell^{C}(p^2)&=\frac{2i\pi}{mp}\frac{\Gamma(1+\ell+i\frac{m\alpha}{p}
)+\Gamma(1+\ell-i\frac{m\alpha}{p})}{\Gamma(1+\ell+i\frac{m\alpha}{p}
)-\Gamma(1+\ell-i\frac{m\alpha}{p})}~,
\end{align}
Its expansion in powers of $\delta=m\alpha/p(1+\nu)$ is exactly the same as 
given  in  Eq.~\eqref{200717.1} by introducing the auxiliary function 
$v(\delta)$. 
Hence, the convergence towards the exact pole position $p=i m \alpha$ 
follows the same pattern as explained with respect  to Table~\ref{tab.200717.1} 
for the AC model. 
The difference is that now this discussion is applied to the ground state of the 
system, i.e. the state with the largest binding energy, while for the AC 
scattering it was applied to the poles with the smallest $|p_i|$. For the former 
the unitarization can be considered as a high-energy procedure while for the 
latter it is a low-energy one. 
Since now the absolute units are different to those in Table~\ref{tab.200717.1}, 
we reproduce the corresponding numbers for Coulomb scattering in 
Table~\ref{tab.200812.1} in units of $\Lambda=m\alpha\gamma_E$ for the first 
orders from $\n=1$ to 15. It is clear that the series converge to the exact 
result by increasing the order of the expansion. 

\begin{table}
\begin{center}
\begin{tabular}{r|rrrrrrrrr}
\hline
                  & $\n=1$ & $\n=3$  & $\n=5$ & $\n=7$ & $\n=9$ & $\n=11$ & 
$\n=13$ & $\n=15$ & Exact \\ \hline
$p_1$ $[\Lambda]$ & $i$       & $i\,1.646$      & $i\, 1.729$  & $i\,1.738$ & 
$i\,1.736$ & $i\,1.734$ & $i\,1.733$ & $i\,1.732$ & $i\,1.732$ \\
\hline
\end{tabular}
\caption{Deepest binding momentum in units of $\Lambda=m\alpha\gamma_E$ of the 
$S$-wave PWA $T^{C}(p^2)$ for Coulomb scattering with $v(\delta)$ calculated at 
order $\delta^\n$,  
for $\n=1$, 3, \ldots,15. The pole positions are compared with the exact result 
$i/\gamma_E$ given in the last column. 
The (non-monotonic) convergence in the results as $n$ increases is explicitly 
shown.
\label{tab.200812.1}}
\end{center}
\end{table}

In contrast, for the AC model there is indeed a bound state with finite 
(nonzero) minimal binding energy,
while all the other bound states have increasingly larger binding energies. In 
addition, we have the
resonance poles that also lie further away from zero as $\nu$ increases. 
As a result, the state with the smallest binding energy and the resonance with 
the pole positions closest to the physical
axis are the ones affecting most the physical low-energy region for
scattering. Then, one can interpret the calculated unitarized PWAs as well 
settled from the low-energy point of view. 

\section{$S$-wave 
  scattering for $d>4$ and the graviball pole position}
\label{sec.210115.1}
\def\theequation{\arabic{section}.\arabic{equation}}
\setcounter{equation}{0}
In this section we exploit the fact that for $d\geq 5$ there are no infrared divergences and study the $J=0$ AC-model, Coulomb and graviton-graviton scattering as a function of the number $d$ of space-time dimensions. We show that the poles of the scattering amplitudes obtained at $d=4$ persist for higher dimensions, varying in a smooth continuous way. This property is used below to estimate $\log a$ such that the extrapolation from $d\simeq 5$ to $d=4$ is as smooth as possible. This can also be interpreted as minimizing higher-order uncertainties introduced by the dependence on $\log a$.  This is the essence of the so-called optimized perturbation theory \cite{Stevenson:1981vj,Brodsky:1982gc,Su:2012iy}. This method is tested successfully with the exactly-solvable AC model and then it is used to obtain an ``optimal'' value of $\log a$ for the case of gravity.  

\subsection{Treatment of the infrared divergences from the analytic continuation to $d > 
4$}
\label{sec.200805.2}

For the sake of self-consistency we start by showing how the soft IR divergences in 
the projected Born term can be reabsorbed in the exponencial factor $S_c(s)$ 
within dimensional regularization. An important point to notice 
is that loop contributions with virtual gravitons are IR convergent for 
$d > 4$. This implies that one can eliminate the cutoff $\mfl$ from the ontset by  
regulating IR divergences in dimensional regularization and trade it for other scales required by dimensional analysis.\footnote{Furthermore, 
the one-loop amplitude in pure gravity is free from ultraviolet divergences and 
all its singularities in $d=4$ are of IR 
nature~\cite{Dunbar:1994bn,Dunbar:1995ed,Donoghue:1999qh}.} Indeed, the exponent 
in  Eq.~\eqref{eq:WeinbergGen} is regularized when evaluated in $d=4-2\epsilon$, and one 
can then take already $\mu=0$ and $\mfl=\infty$. 
Summing the contributions involving a pair of either incoming or outgoing particles with initial momenta $p_1$ and 
$p_2$ it gives\footnote{In our notation the Minkowski metric is ${\rm 
diag}(1,-1,-1,-1)$, with a global minus sign relative to 
Ref.~\cite{Weinberg:1965nx}.}
\begin{align}
\label{eq:dimreg0}
\int  \frac{d^dq}{(2\pi)^d} B(q)&=-4\pi G s^2 i\int 
\frac{d^dq}{(2\pi)^d}\frac{1}{(q^2+i\vep)(p_1\cdot q-i\vep)(p_2\cdot 
q+i\vep)}\nonumber\\
&\to 16\pi G s^2 
i\int\frac{d^dq}{(2\pi)^d}\frac{1}{(q^2+i\vep)[(p_1-q)^2+i\vep][(p_2+q)^2+i\vep]
}\nonumber\\
&=-\frac{G 
s}{\pi}\left(-\frac{s}{4\pi}\right)^{-\epsilon}\frac{\Gamma(1+\epsilon)}{
\Gamma(1-2\epsilon)}\Gamma(-\epsilon)^2,
\end{align}
where in the second line we have restored the full $q^2$ dependence of the 
propagators to regulate the ultraviolet divergence in the integral in the first 
line. This leads to the following IR-divergent phase,
\begin{align}
\label{200720.3}
S_c^{\rm dr}(s)=\exp\left[iGs\left(-\frac{1}{\ep}+\log 
\frac{s}{\mu_h^2}+\gamma_E-\log(4\pi)\right)\right]~. 
\end{align}  
In the previous expression we rescaled the dimensions of the Newton's constant 
by inserting a renormalization scale $\mu_h$ and the physical region is 
approached by taking $s+i\vep$ so that $\log(-s)=\log(s)-i\pi$. Note that the 
scale $\mu_h$ introduces an arbitrariness in the finite part related to the fact 
that we are adding hard-graviton modes in the integral~\footnote{This scheme 
dependence in a dimensional treatment of the IR divergence can be made more 
explicit by regulating the hard modes in the first line of 
Eq.~\eqref{eq:dimreg0} also in dimensional regularization~\cite{Naculich:2011ry} 
or with an explicit cutoff~\cite{Ware:2013zja}.\label{foot.200813.1}}.

One then redefines the $S$-matrix by subtracting $S^{\rm dr}_{c}(s)$, which 
provides an IR-divergent contribution to all the PWAs
at $\mathcal O(Gs)$,
\begin{align}
\label{eq:PWA_Sc_dim}
\delta 
F_{\lambda_1'\lambda_2',\lambda_1\lambda_2}^{(J)}=\frac{1}{2^{\lambda/4}}\frac{4 
G s}{\pi}\left(\frac{1}{\ep}-\log \frac{s}{\mu_h^2}-\gamma_E+\log(4\pi)\right),
\end{align}
that should cancel those stemming from the singularity in the forward direction 
of the angular-projection integrals of the Born term.

Extending the projections in partial waves to arbitrary dimensions, as discussed 
in detail in  App.~\ref{app.200805.1},  one obtains
\begin{align}
_S\langle p'J'\lambda_1'\lambda_2'|T| 
pJ\lambda_1\lambda_2\rangle_S=&\frac{1}{2(2\pi)^{d-1}}\left(\frac{s}{4}\right)^{
d/2-2}\times \nonumber\\
&\int 
d\Omega_{d-2}\int^{+1}_{-1}d\cos\theta'(\sin\theta')^{d-4}D_{\lambda\lambda'}^{
(J)}(\mathbf{\hat n}) _S\langle \mathbf{p}\lambda_1'\lambda_2'|T| p \mathbf{\hat 
z}\lambda_1\lambda_2\rangle_S, 
\end{align}
where the normalization of the states has been set as  in $d=4$ (see 
Appendix~\ref{app.200805.1}). For the case $J=0$ ($\lambda=\lambda'=0$), 
$D_{\lambda\lambda'}^{(J)}(\mathbf{\hat n})=1$ and this integral can be easily 
solved in arbitrary dimensions (see e.g.~\cite{Somogyi:2011ir}),
\begin{align}
\label{eq:F0dim}
F_{22,22}^{(0)}(s)&=\frac{2 G 
s}{\pi^2}\left(\frac{s}{16\pi^2}\right)^{d/2-2}\int 
d\Omega_{d-2}\int^{+1}_{-1}d\cos\theta'\frac{(\sin\theta')^{d-4}}{1-\cos\theta'}
\nonumber\\
&=\frac{4 G 
s}{\pi}\left(\frac{s}{4\pi}\right)^{d/2-2}\frac{\Gamma(d/2-2)}{\Gamma(d-3)}.
\end{align}
Expanding in $d=4-2\ep$,
\begin{align}
\label{eq:F0eps}
F_{22,22}^{(0)}(s)&=\frac{4 Gs}{\pi}\left(-\frac{1}{\ep}+\log 
\frac{s}{\mu^2_f}+\gamma_E-\log(4\pi)\right),
\end{align}
where we have regulated the IR divergence stemming from the forward singularity 
at a particular scale $\mu_f$. This is  summed to the divergent contribution 
from $S_c^{\rm dr}(s)^{-1}$ in Eq.~\eqref{eq:PWA_Sc_dim} up to ${\cal O}(Gs)$, 
to give the finite PWA
\begin{align}
\label{eq:V0_dimreg}
V_{22,22}^{(0)}(s)&=\frac{8 Gs}{\pi}\log\left(\frac{\mu_h}{\mu_f}\right).
\end{align}
The  log's  that depend on the energy of the process cancel along with the 
singularities because they must arise from the expansion of the dimensionless 
factor $(s/\mu^2_i)^{-\epsilon}$. Hence, the dependence on energy of the PWA at 
$d=4$ can only be linear in $s$, which is consistent with the argument made  in 
Eq.~\eqref{200121.20} to fix the scale $\mfl$ in the calculation with an 
explicit cutoff. The final coefficient corresponds to $\log a=\log \mu_h/\mu_f$ 
in that language, which also makes obvious the expectation that $\log a\sim {\cal O}(1)$. 

Extending our analysis to integer dimensions $d>4$ opens up a new angle to 
investigate the analytic structure of the graviton-graviton scattering 
amplitude. In particular, the treatment of the divergent phase in $d=4$ should 
not introduce spurious effects and one expects that interesting features in the 
analytic structure of the PWA, such as the presence of poles, remain in $d>4$. 
In this case, the PWA simply corresponds to the one derived from the (IR 
convergent) $\bar S^{(J)}$ and for $J=0$ it is 
equal to Eq.~\eqref{eq:F0dim},
\begin{align}
\label{eq:V0dim_D}
V_{22,22}^{(0),d}(s)&=\frac{\Gamma(d/2-2)}{\Gamma(d-3)}\frac{4G  
s}{\pi}\left(\frac{s}{4\pi \mu_f^2}\right)^{\frac{d-4}{2}}~,
\end{align}
where we have expressed the Newton's constant in $d$-dimensions $G_d$ in terms 
of the  4-dimensional $G$, as $G=G_d\, ({\mu_f^2})^{\frac{d-4}{2}}$.
Note  that the extra functional dependence $s^{d/2-2}$ with respect to $d=4$  
does not introduce a left-hand cut in the PWAs for odd-numbered $d$ by taking 
$p=\sqrt{s}/2$ as the argument.
Besides the additional factor $s^{d/2-2}$, the strength of the interaction gets 
also rapidly diluted with $d>4$ due to the ratio of the $\Gamma$ functions.

\subsection{The graviball in $d>4$}
\label{sec.200815.1}

The graviton-graviton $S$-wave projected Born amplitude without IR singularities for $d>4$ has been
worked out in Eq.~\eqref{eq:V0dim_D}. Analogously as done for $d=4$, we 
can estimate the unitarity cutoff $\Lambda_d$ by taking the first iterated 
expression of the unitarized formula, which now reads
\begin{align}
\label{200815.2}
T^{(0),d}_{22,22}(s)&=\left[V^{(0),d}_{22,22}(s)^{-1}+g(s)\right]^{-1}~.
\end{align}
Therefore, the unitarity cutoff $\Lambda_d^2$ is 
\begin{align}
\label{200814.5}
\Lambda_d^2&=\left((4\pi \mu_f^2)^{\frac{d-4}{2}}2\pi 
G^{-1}\frac{\Gamma(d-3)}{\Gamma(d/2-2)}\right)^{\frac{2}{d-2}}~.
\end{align}
To simplify the dependence on $d$ in this expression we choose units such that 
$2\pi G^{-1}= 1$ and, for illustrative purposes, we take by now 
2$\mu_f^2 G=1$. 
Then,
\begin{align}
\label{200814.6}
\Lambda_d^2&=\left(\frac{\Gamma(d-3)}{\Gamma(d/2-2)}\right)^{\frac{2}{d-2}}~.
\end{align}

\begin{figure}[t]
\begin{center}
\includegraphics[width=130mm]{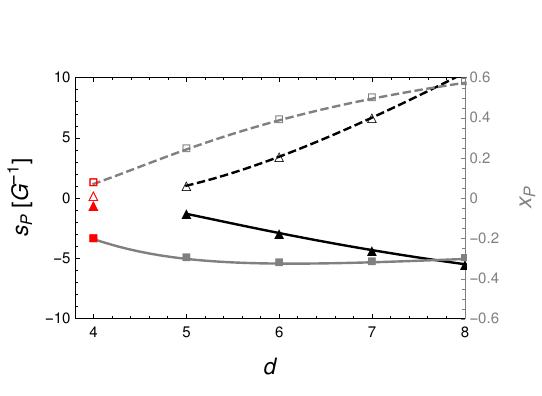}
\caption{{\small 
Real and imaginary parts of $s_P$ (using the left vertical axis) as functions of 
$d$ plotted using solid and dashed (black) lines, respectively. The empty and filled 
triangles show the values for integer $d$ where, for $d=4$, the value $\log a=1$ 
has been chosen. We also plot the real and imaginary parts of $x$ for the graviball ($x_P$, using the 
right vertical axis) with solid and dashed (gray) lines,  respectively. The empty 
and full squares are the corresponding values for integer $d$.}
\label{fig.200814.2}}
\end{center}
\end{figure}

One obtains the secular equation,
\begin{align}
\label{eq:secular_d}
(\omega x_d)^{1-d/2}+\log(-x)-i 2\pi=0,
\end{align}
where $\omega=\Lambda^2/\Lambda_d^2$ corresponds again to the ratio of the 
fundamental and unitarity cutoffs.
For the nominal case of $\omega=1$ this equation is then the same as 
Eq.~\eqref{200814.3} below for searching the pole of the $\sigma$. The resulting poles 
position $s_P=x \Lambda_d^2$ as a function of $d$ are shown in 
Fig.~\ref{fig.200814.2}. The empty and filled triangles represent the real and 
imaginary parts of $s_P$, respectively, and the empty and filled squares 
correspond to the real and imaginary parts of $x$ as a function of $d$. Notice 
that the curves for $s_P$, as a function of $d$, would evolve to zero if 
continued to $d=4$ because $\Lambda_d^2=0$ due to the IR divergence that has  
not been regulated in this expression. We also show the real and imaginary parts 
for $s_P$ at $d=4$ for $\log a=1$.
 We also observe that the real part 
of $s_P$ rapidly increases with $d$ and it becomes larger than the absolute 
value of the imaginary part already for $d\gtrsim 5$, a feature also observed below for the $s_\sigma$  in Fig.~\ref{fig.200814.1}. Therefore, the noteworthy 
fact that the peak in the resonance signal of $1/|D^{0)}(s)|^2$ is much smaller 
than the width, as represented in the right panel of Fig.~\ref{fig.170725.1},  
is very characteristic of $d=4$. Also note that the positions of the poles for 
$d\geq5$ are transplanckian (i.e. heavier than $G ^{-1}$), even though $|x|<1$.   

\subsection{The $\sigma$ meson in $d>4$}
\label{sec.180814.1}


Let us now turn to $\pi\pi$ scattering and the $\sigma$ meson. In this case there are 
no IR divergences so that the extrapolation for integer $d>4$ can be connected 
straightforwardly with the result at $d=4$, a point that will require further 
treatment for the infinite-range interactions, as we treat below.

Making use of Eq.~\eqref{200809.3}, the LO ChPT scalar isoscalar $\pi\pi$ PWA for 
$d$ dimensions in the chiral limit is given by 
\begin{align}
\label{200814.1}
V_0^{{\pi\pi},d}(s)&=
\frac{s}{f_\pi^2}\frac{(p/\mu_f)^{d-4}}{2^{d-2}\pi^{d-3}}\int d\Omega_{d-1}
=
\frac{\Gamma(d/2-1)}{\Gamma(d-2)}\frac{s}{f_\pi^2}\left(\frac{s}{4\pi\mu_f^2}
\right)^{\frac{d}{2}-2}~,
\end{align}
where we have used that the isoscalar $\pi\pi$ scattering amplitude 
$_S\la \vp'|T|\vp\ra_S $ is $s/f_\pi^2$, as derived in Ref.~\cite{Oller:1997ti}, 
and have defined the pion decay constant in $d$-dimensions as 
$f_\pi^d=f_\pi\mu_f^{d/2-2}$. Compared to  Eq.~\eqref{eq:F0dim}, this comes from 
a contact interaction and has no divergences in $d=4$ related to $t$ channel 
exchange. 
The unitarized PWA reads as in Eq.~\eqref{200603.3} but with $f_\pi^2/s$ 
replaced by $1/V_0^{{\pi\pi},d}(s)$. The unitarity cutoff $\Lambda_d$ in this 
case is
\begin{align}
\label{200814.2b}
\Lambda_d^2&=\left((\pi \mu_f^2)^{\frac{d}{2}-2}(4\pi 
f_\pi)^2\frac{\Gamma(d-2)}{\Gamma(\frac{d}{2}-1)} \right)^{\frac{2}{d-2}}~,
\end{align}
and the secular equation, with $x=s/\Lambda_d^2$ reads now
\begin{align}
\label{200814.3}
x^{1-\frac{d}{2}}+\log(-x)-i2\pi=0~.
\end{align}

We can simplify the dependence of $\Lambda_d^2$ (and of $s_\sigma=x\Lambda_d^2$) 
on $d$ by taking energy units such that $4\pi f_\pi=1$.  For illustration, if in 
these units we choose $\mu_f^2=1/\pi$, this scale squared becomes
\begin{align}
\label{200814.2}
\Lambda_d^2&=\left(\frac{\Gamma(d-2)}{\Gamma(\frac{d}{2}-1)} 
\right)^{\frac{2}{d-2}}~.
\end{align}

\begin{figure}[t]
\begin{center}
\includegraphics[width=130mm]{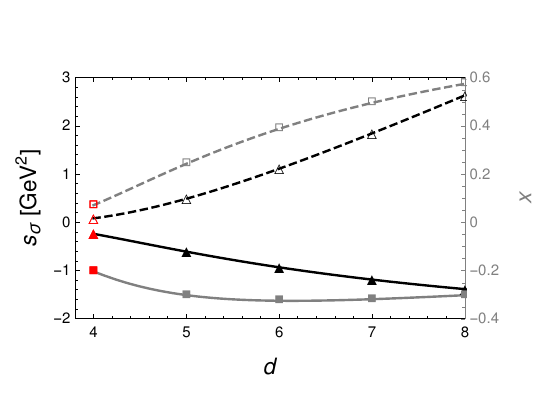}
\caption{{\small Real and imaginary parts of $s_\sigma$ (using the left vertical 
axis) as functions of $d$ plotted using solid and dashed (black) lines, 
respectively. The empty and filled triangles show the values for integer $d$. We 
also plot the real and imaginary parts of $x$  (using the right vertical axis) 
with solid and dashed gray lines respectively. The empty and full squares are the 
corresponding values for integer $d$.}
\label{fig.200814.1}}
\end{center}
\end{figure}
The resulting real (empty triangles) and imaginary part (filled triangles) of 
$s_\sigma$ as a function of $d$ is shown in Fig.~\ref{fig.200814.1} using the 
left vertical axis. We also plot the real and imaginary parts of $x$ using the 
right vertical axis and the full and empty squares, respectively. Notice that in 
all cases $x$ is always less than 1, which means that the pole position is lower 
than the unitarity cutoff of the theory (which is growing with $d$) and for 
$d=5$ $\sqrt{|s_\sigma|}$ is also below the putative cutoff of QCD. The real 
part of $s_\sigma$ grows with $d$ faster than the absolute value of the 
imaginary part, so that already for $d\gtrsim5$ the former becomes larger than 
the latter.

\subsection{The $S$-wave amplitude in $d > 4$ for the AC model}
\label{sec.200812.1}

\begin{figure}
\begin{center}
\includegraphics[width=100mm]{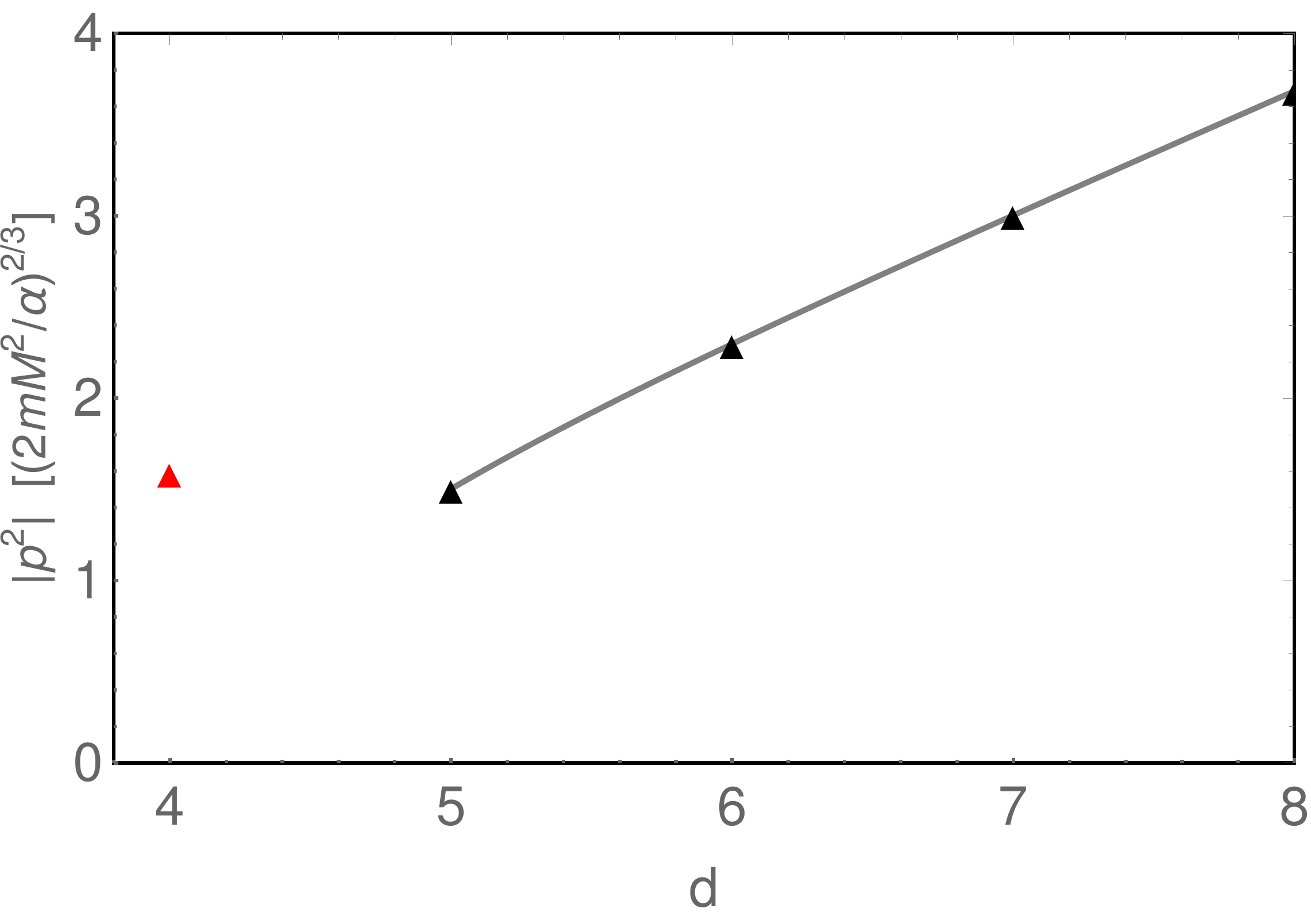}
\caption{{\small The modulus of $p^2$ for the pole positions of the $S$-wave in 
the AC model is shown as a function of $d$ and in units of $(2m 
M^2/\alpha)^{2/3}$.}
\label{fig.200815.1}}
\end{center}
\end{figure}

Let us 
consider the $S$-wave projection of the Born term of the AC model for integer 
dimensions larger than $4$. The corresponding LO tree-level PWA in $S$ wave, 
employing Eq.~\eqref{200809.3}, is
\begin{align}
\label{200813.5}  
F_0^{AC,d}(p^2)&=\frac{\pi\alpha 
p^{d-6}}{(2\pi)^{d-3}\mu_f^{d-4}}\left(\frac{p^2}{2mM}\right)^2\int 
d\Omega_{d-2}\int_{-1}^{+1}d\cos\theta'\frac{(\sin\theta')^{d-4}}{
(1-\cos\theta')}
=\frac{\alpha 
p^{d-6}}{\mu_f^{d-4}\pi^{d/2-3}}\frac{\Gamma(d/2-2)}{\Gamma(d-3)}~.
\end{align}
The unitarity cutoff $\Lambda'_d$ is obtained by multiplying $F_0^{AC,d}(p^2)$ 
by $mp/2\pi^2$, from where we deduce that
\begin{align}
\label{200813.6b}
\Lambda'_d&
=(\pi \mu_f^2)^{\frac{1}{2}}\left(\frac{8\pi m 
M^2/\alpha}{(\pi\mu_f^2)^{\frac{3}{2}}}\frac{\Gamma(d-3)}{\Gamma(d/2-2)}\right)^
{\frac{1}{d-1}}~.
\end{align}
The unitarized expression for the $S$-wave is the same as in 
Eq.~\eqref{200131.1} but substituting $V_0^{AC}(p^2)$ by $F_0^{AC,d}(p^2)$. The 
secular equations that follows is 
\begin{align}
\label{200815.4}
x_d^{d-1}=-\frac{i}{\pi}~,
\end{align}
whose solution is explicitly
\begin{align}
\label{200815.5}
p^2(d)
&=\pi\mu_f^2 
\left(-i\frac{8mM^2/\alpha}{(\pi\mu_f^2)^{\frac{3}{2}}}\frac{\Gamma(d-3)}{
\Gamma(d/2-2)}\right)^{\frac{2}{d-1}}~.
\end{align}
To simplify the dependence on $d$ we choose units such that $2mM^2/\alpha=1$ and 
a value for $\pi\mu_f^2=1$ for illustrative purposes. The running of $p^2(d)$ 
with $d$ then  simplifies to $(-i4\Gamma(d-3)/\Gamma(d/2-2))^{2/(d-1)}$ and 
$|p^2(d)|$ is shown by the triangles of Fig.~\ref{fig.200815.1}. We also give 
the exact result for $d=4$.  We can observe that this curve is qualitatively 
similar to the one for the graviball given above in Fig.~\ref{fig.200814.2}. We 
only show the modulus because all the  poles are obtained by the roots of 
$(-i)^{2/(d-1)}$ times the latter.

\subsection{Unitarization of Coulomb scattering for $d > 4$}
\label{sec.200812.1}

Let us consider electron-positron scattering of initial(final) four-momenta 
$p_i(p'_i)$, $i=1,\,2$, whose non-relativistic limit reduces  to  Coulomb 
scattering. The Mandelstam variable $s$ is given by 
$s=(p_1+p_2)^2=4(\mu^2+p^2)$, where $p$ is the modulus of the CM three-momentum 
and $\mu$ is the mass of an electron. 
According to Ref.~\cite{Weinberg:1965nx} the exponent giving rise to the 
IR-divergent contributions from soft intermediate photons exchanged between 
external lines belonging to the same initial or final state is
\begin{align}
\label{200813.2}
\int\frac{d^dq}{(2\pi)^d}A(q)&=ie^2(s-2\mu^2) \int\frac{d^d q}{(2\pi)^d}
\frac{1}{(q^2+i\vep)(-p_1q+i\vep)(p_2q+i\vep)}\\
&\to 
ie^24(s-2\mu^2)\int\frac{d^dq}{(2\pi)^d}\frac{1}{
(q^2+i\vep)((p_1-q)^2-\mu^2+i\vep)((p_2+q)^2-\mu^2+i\vep)}\nn\\
&=-\frac{2\alpha\Gamma(1+\ep)s^{-1-\ep}}{(4\pi)^{1-\ep}\ep}\int_0^1 dt\, 
(t(t-1)+\mu^2/s)^{-1-\ep}~.\nn
\end{align}
The roots of the denominator in the integrand are
\begin{align}
\label{200813.3}
t_{1,2}&=\frac{1}{2}\pm\frac{1}{2}w,~~~~~ 
w\equiv\frac{p}{\sqrt{\mu^2+p^2}}=\frac{v}{\sqrt{1+v^2}}~.\nn
\end{align}
Notice that $w$ is the relativistic velocity while $v=p/m$ is the 
non-relativistic expression. This decomposition is  appropriate for our 
discussion because at the end we are interested in the non-relativistic limit. 
Making a shift in the integration variable $t\to t-1/2$, we are then left with 
the integral
\begin{align}
\int\frac{d^dq}{(2\pi)^d}A(q)&=
-\frac{2\alpha\Gamma(1+\ep)s^{-1-\ep}}{(4\pi)^{1-\ep}\ep}\left\{
\int_{-1/2}^{1/2}\frac{dt}{t^2-\frac{w^2}{4}} 
-\ep\int_{-1/2}^{1/2}dt\frac{\log(t^2-\frac{w^2}{4})}{t^2-\frac{w^2}{4}}
\right\}+{\cal O}(\ep)~.
\end{align}
Instead of giving the full expressions for the integration, which is somewhat 
lengthy, we give its imaginary part in the leading non-relativistic limit, which 
is the one that actually enters as the exponent of $S_c^C$. It generates the 
global phase
\begin{align}
\label{200813.4}    
S_c^C(p)&=\exp\left[i \gamma\left(-\frac{1}{\ep}
+\log\frac{p^2}{\mu_h^2}+\gamma_E-\log\pi\right)\right]~.
\end{align}
Notice that the terms in parentheses are the same as those in 
Eq.~\eqref{200720.3} for gravity once we replace $s$ by $4p^2$ in 
graviton-graviton scattering. As in this case, we have introduced the 
renormalization scale $\mu_h$ to keep  right the  dimensions when varying $d$. As 
commented above,  regarding Eq.~\eqref{200720.3}, the scale $\mu_h$ introduces 
an arbitrariness in the finite part of the exponent due to having integrated 
over  harder-photon modes, cf. footnote \ref{foot.200813.1}.

Now, we consider the $S$-wave projection of the Coulomb scattering amplitude in 
$d$ dimensions by employing Eq.~\eqref{200809.3}.  We are then left with
\begin{align}
\label{200813.5}  
F_0^{C,d}(p^2)&=\frac{\pi\alpha p^{d-6}}{(2\pi)^{d-3}}\int 
d\Omega_{d-2}\int_{-1}^{+1}d\cos\theta'\frac{(\sin\theta')^{d-4}}{
(1-\cos\theta')}
=\frac{\alpha p^{d-6}}{\pi^{d/2-3}}\frac{\Gamma(d/2-2)}{\Gamma(d-3)}~.
\end{align}
Expanding in powers of $\ep$ we have
\begin{align}
\label{200813.6}  
F_0^{C,d}(p^2)&=\frac{\pi\alpha}{p^2}\left(-\frac{1}{\ep}+\log\frac{p^2}{\mu_f^2
}
+\gamma_E-\log\pi\right)~.
\end{align}
The sum of $F_0^{C,d}(p^2)$ with $-i\pi(S_c^{-1}-1)/mp$  expanded up to ${\cal 
O}(\alpha)$ (let us recall that $m=\mu/2$, the reduced for the electron-position 
system) gives the neat $S$-wave projected Born term for Coulomb scattering
\begin{align}
\label{200813.7}
V^{C}_0(p^2)=\frac{2\pi\alpha}{p^2}\log\frac{\mu_h}{\mu_f}~,
\end{align}
and $\log\mu_h/\mu_f$ corresponds to $\log a^2$, when compared to 
Eq.~\eqref{200826.1}.

We also give from Eq.~\eqref{200813.5} the $S$-wave Born term for integer 
dimensiones $d\geq 5$, which is not affected by the issue of the IR divergences 
in the forward scattering. The dimensions can be kept right e.g. by dividing 
$p^{d-4}$ in this equation by $(m\alpha)^{d-4}$, and we end up with the 
dimensionless Sommerfeld parameter $\gamma$ to the power ${d-4}$.  Then,
\begin{align}
\label{200813.8}  
F_0^{C,d}(p^2)&=\frac{\alpha 
\gamma^{4-d}}{\pi^{d/2-3}p^2}\frac{\Gamma(d/2-2)}{\Gamma(d-3)}~.
\end{align}
Of course, any other choice would differ by some coupling that could vary with 
$d$ (but not with $p$).

We do not study for this case the evolution of the poles in $S$ wave with $d$ 
because for $d=5$ the secular equation for $x$, 
\begin{align}
\label{200815.6}
x^{d-5}=-\frac{i}{\pi}~,
\end{align}
has obviously no solution. The reason is the change in the character of the 
theory, because as $d$ increases the theory can be unitarized as a low-energy 
EFT, while at $d=4$ this procedure is valid for $p\gtrsim \Lambda$.

\subsection{Maximal-stability estimate of $\log a$}
\label{sec.210115.2}

The  value of $\log a$ 
comes from the separation of hard and soft modes in  the IR divergent 
part of loop diagrams. As discussed in Sec.~\ref{sec.200122.1} this   points towards using $\log a \gtrsim 1$, though 
its value can not be fixed by perturbation theory. We saw in 
Sec.~\ref{sec.200530.1} that for the AC model, the results from the Born 
approximation suggest a slightly smaller value: the LO one, $\gamma_E/2$, gets 
``dressed'' by higher-order corrections, so that at the end $\log a=1/2$ is the 
one reproducing the exact positions of the lightest poles.  
To get more insight on which values of $\log a$ may be {\it a priori} adequate 
without knowing higher-order contributions (like in our present study for the 
graviball), we  investigate these theories in $d>4$, where the IR divergences 
eventually disappear, and study their evolution to $d=4$. 
We first discuss the method applied to the AC-scattering model and show that, 
indeed, it gives a value $\log a\simeq 1/2$. Then, we apply the analogous method 
to the $J=0$ graviton-graviton scattering and obtain the larger result $\log 
a\simeq 1$. 

For $d\to 4$ the effect of the IR divergences affecting the partial-wave 
projection of the  Born terms in the AC model requires including 
${S_c^{AC}}(p^2)$. This can be obtained in $d>4$ by using the one for QED, 
$S^C_c$,  calculated in App.~\ref{sec.200812.1}, and applying the substitution 
$\alpha\to \alpha p^4/(2mM)^2$ in Eq.~\eqref{200813.4}.
After replacing $p^2$ by its value at the pole positions in $d$ dimensions 
$p^2(d)$ given in Eq.~\eqref{200815.5} we are left with the combination 
\begin{align}
\label{200816.1}
\frac{2}{d-4}+\gamma_E-4\log 
a+\frac{2}{d-1}\log\left(-i\frac{8mM^2/\alpha}{(\pi 
\mu_f^2)^\frac{3}{2}}\frac{\Gamma(d-3)}{\Gamma(d/2-2)}\right)~,
\end{align}
in the exponent of $S_c^{AC}$, where we have used that $\log a=\frac{1}{2}\log 
\mu_h/\mu_f$ for Coulomb scattering. 
In the following we take units such that $2mM^2/\alpha=1$ and introduce the 
variable $y=\pi\mu_f^2$.

We now study when the pole term in the previous equation is comparable in size 
with the other terms. This gives a critical  dimension, $d_c$, below which the 
pole term (associated to the IR divergences) is prominent and one cannot 
extrapolate the theory by directly evaluating the $S$-wave projection of the 
Born term in $d$ dimensions. This is certainly expected to happen at some  
$d_c\lesssim 5$, since for integer dimensions one could in principle apply an 
essentially  analogous unitarization procedure as for $d=4$, cf. 
Sec.~\ref{app.170723.2}. Specifically,
the criterion that we will use for the onset of large IR corrections, giving us 
$d_c$, is 
\begin{align}
\label{200816.3}
\frac{2}{d-4}&=\left|4\log 
a-\gamma_E-\frac{2}{d-1}\log\left(\frac{-i4\Gamma(d-3)}{y^{3/2}\Gamma(d/2-2)}
\right)\right|~.
\end{align}

For each value of $d_c(\log a,y)$ we then quantify the difference 
with respect to $d=4$ by calculating the relative difference between the 
unitarity-cutoff scales at $d_c$ and $d=4$. Namely, 
\begin{align}
\label{200829.3}
r(\log a,y)&=\frac{|{\Lambda'}_d^2-{\Lambda'}^{2}|}{{\Lambda'}^{2}}~,
\end{align}
with $\Lambda'_d$ and $\Lambda'$ given in Eqs.~\eqref{200813.6b} and 
\eqref{200811.1}, respectively.
This definition is based on the fact that the unitarity-cutoff scale 
$\Lambda'_d$
controls the strength of the interaction as $d$ varies.  
The relative difference is mostly determined by the rapid variation of 
${\Lambda'}_d^2$ for $d$ close to 4, which is much affected by the IR 
divergences. The latter, in turn,  manifest through  the pole term 
$\Gamma(d/2-2)$ in the denominator of Eq.~\eqref{200813.6b}. 

\begin{figure}
\begin{center}
\includegraphics[width=0.5\textwidth]{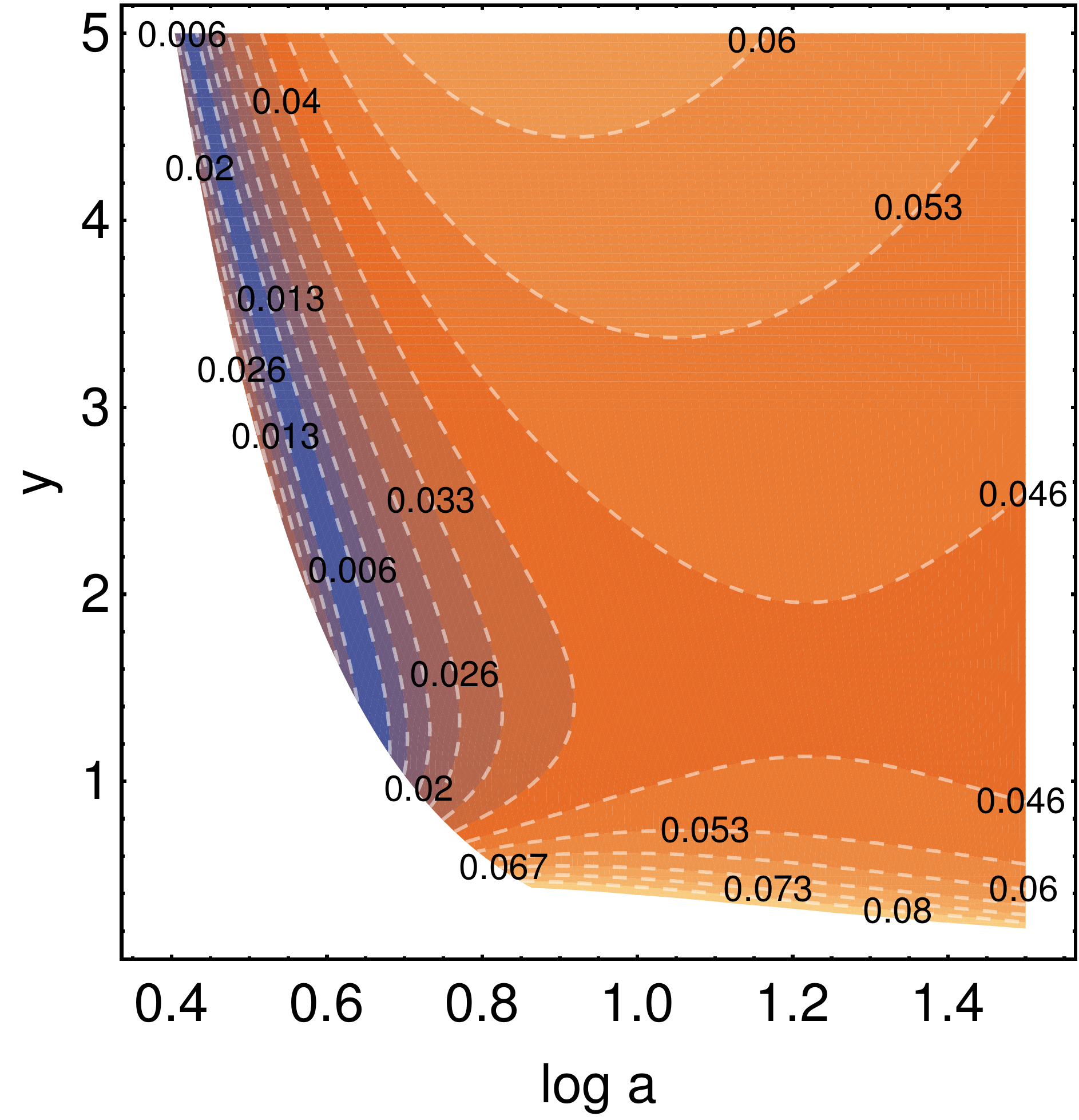}
\caption{{\small AC model. We show the relative difference 
$r=|{\Lambda'}_{d_c}^2-{\Lambda'}^2|/{\Lambda'}^2$ as a function of $\log a$ 
($x$ axis) and $y$ ($y$ axis). Notice the minima for $\log a\simeq 0.4-0.6$.}
\label{fig.200823.2}}
\end{center}
\end{figure}

We then propose to look for the combination of $\log a$ and $y$ that minimizes 
the difference $r$ with the constraint $d_c<5.$
This minimization requirement of Eq.~\eqref{200829.3} is equivalent to a 
criterion of \textit{maximal smoothness} in the theory when extrapolating from 
$d_c$ to $d=4$. We  plot in Fig.~\ref{fig.200823.2} the relative difference $r$ 
in the $(\log a,y)$ plane.
We indeed observe a continuum of local minima for $\log a\sim 0.4-0.6$, which is 
the value we have already determined by reproducing the lightest pole positions 
with $\nu=0$ in the AC-scattering model~\footnote{One actually finds a minimum 
at $\log a\simeq 1/2$ if one  relaxes the criterion in Eq.~\eqref{200816.3} by 
changing slightly the balance between the $2/(d-4)$ piece and the rest of the 
contributions in the definition of $d_c$.}.Also note that the range of $y$ 
considered corresponds to $\mu_f
^2\sim 0.1 - 0.4$ in units of $\Lambda'^2$, which are reasonable values for an 
IR regularization scale. 

To summarize, the method is based on the study of the scattering for $d>4$ and 
requiring maximum smoothness when passing from $d_c$ to $d=4$ to determine $\log 
a$. This procedure is reminiscent of optimized perturbation theory 
\cite{Stevenson:1981vj,Brodsky:1982gc,Su:2012iy} which implements the principle 
of minimal sensitivity to fix scale ambiguities in perturbation theory to 
improve its convergence properties.  One can also see similarities between the 
principle of maximum smoothness and some techniques used in statistical 
mechanics. In our case, by minimizing $r(\log a,y)$ one is reducing the needed 
number of terms in an expansion of the $S$-wave projected Born term (which fixes 
$\Lambda'_d$), in powers of $d-5$, since $d_c$ is typically close to 5, from 
$d=5$ to 4. Although we know $V_d^{(0)}(s)$, we pretend to be influenced by just 
a few terms in its expansion, since higher-order terms are increasingly more 
sensitive to the singularity at $d=4$ of $\Gamma(d/2-2)$ in 
Eq.~\eqref{eq:V0dim_D}. This is then effectively similar to standard application 
of  dimensional continuation  in  statistical mechanics \cite{itzi,son,ton}, 
where typically only a few terms in the density expansion are available and from 
which the optimal solution is sought.

Encouraged by the successful application of this method to the AC toy model we 
apply it now to graviton-graviton scattering. 
Using the same unitarization procedure for the Born term as in $d=4$, but with 
the corresponding $V_{22,22}^{(0),d}(s)$ and $\Lambda_d^2$ in 
Eqs.~\eqref{eq:V0dim_D} and~\eqref{200814.5}, respectively, one obtains the 
secular equation in Eq.~\eqref{eq:secular_d}. 
Therefore, a resonance also arises from the gravitational interactions in  
$d>4$, whose position is given as $s_P(d)=x_d\Lambda_d^2$. In the following we 
take $\omega=1$, which is our benchmark scenario, and we refer to the 
discussions at the beginning of this section for other values of $\omega$.

\begin{figure}[H]
\begin{center}
\includegraphics[width=0.5\textwidth]{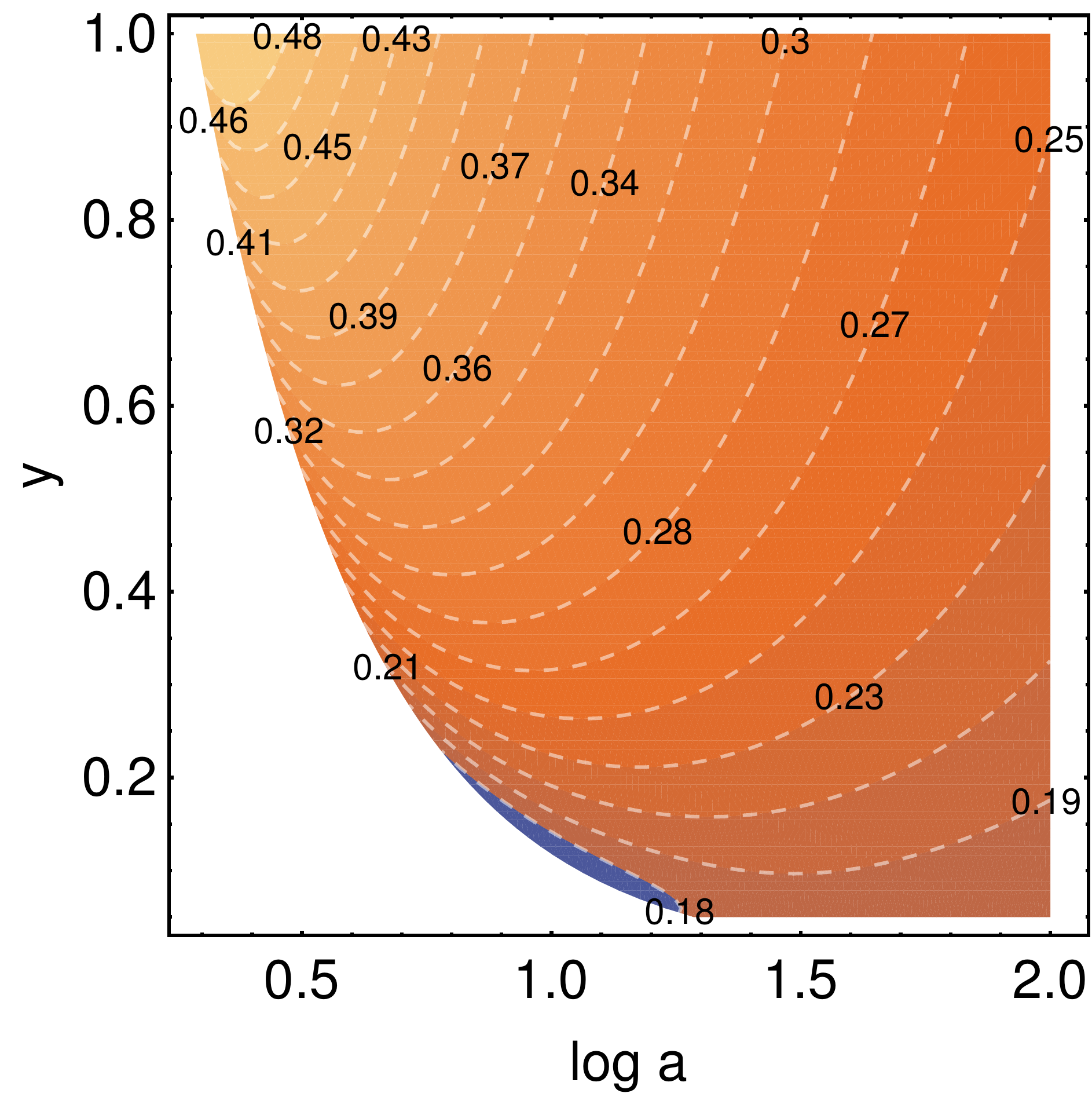}
\caption{{\small Graviton-graviton scattering: The relative difference 
$r=|\Lambda_{d_c}^2-{\Lambda}^2|/{\Lambda}^2$ is plotted as a function of $\log 
a$ ($x$ axis) and $y$ ($y$ axis). We restrict our analysis to the values 
$0.05\leq y\leq1$ and $\log a\leq2$.
\label{fig.200823.1}}}
\end{center}
\end{figure}

We first  determine the values of the critical dimension, $d_c$, for which the 
pole term in the exponent of $S_c(s)$, calculated in Eq.~\eqref{eq:dimreg0}, is 
comparable to the other contributions. In this expression we replace $s$ by 
$s_P$  in the $s$-dependent $\log$, so that this exponent adopts the form
\begin{align}
\label{200814.7}
i G s\left(\frac{2}{d-4}+\gamma_E-2\log 
a+\frac{2}{d-2}\log\frac{\Gamma(d-3)}{2y\Gamma(d/2-2)}+\log x_d
\right)~,
\end{align}
where $y=\mu_f^2 G$ and we have also taken into account here that 
$\log\mu_h/\mu_f=\log a$, as deduced after Eq.~\eqref{eq:V0_dimreg}. Notice that 
the dependence on $x_d$  in the previous equation is mild because it is only 
inside the $\log(x_d)$.
Our criterion for the definition of $d_c$ is the one obtained from
\begin{align}
\label{200814.8}
\frac{2}{d-4}&=\left|2\log 
a-\gamma_E-\frac{2}{d-2}\log\frac{\Gamma(d-3)}{2y\Gamma(d/2-2)}-\log 
x_d\right|~,
\end{align}
which is analogous to Eq.~\eqref{200816.3} for the AC model. Note the change 
of numerical factor multiplying $\log a$ between the two equations. 

Now, as for the AC model, for every $d_c$ we define the ``distance'' $r$ in 
terms of the relative difference between the unitarity cutoffs at $d_c$ and 
$d=4$,
\begin{align}
\label{200829.2}
r(\log a,y)=\frac{|{\Lambda}^2_{d_c}-{\Lambda}^2|}{{\Lambda}^2}~,
\end{align} 
where $\Lambda_d$ can be found in  Eq.~\eqref{200814.5}, which is 
dominated by a zero as $d\to 4$ due to the IR divergences.
We consider $r$ as a good indicator of the difference between
$d_c$ and $d=4$ because the unitarity cutoff controls the strength of the 
interactions.
We also notice that the difference between these scales at different $d$ has the 
advantage of being independent of $\omega$, 
while a similar definition in terms of the difference between pole positions at 
different dimensions would
depend on this parameter.

We then minimize $r$ by adjusting $\log a$ and $y$ with the constraint $d_c<5$. 
This is shown in Fig.~\ref{fig.200823.1} where $r$ is plotted in the $(\log 
a,y)$ plane. A minimum is clearly visible for $\log a \simeq 1$. According to 
Eq.~\eqref{eq:grav.pole.app} this leads to a  pole at $d=4$ whose position is
\begin{align}
\label{eq:graviball.pole}
s_P\simeq (0.22-i\,0.63)~G^{-1},     
\end{align}
providing further support for the existence of the subplanckian graviball in 
pure gravity discussed in Sec.~\ref{sec.200131.1}. It is also important to 
emphasize that these conclusions are robust with respect to moderate variations 
of the criteria used to obtain the optimal and minimal $\log a$; namely by e.g. 
studying alternative ``distances'' such as  $|s_P(d_c)-s_P|/|s_P|$ (for 
$\omega$=1) or by changing the balance between pole and finite contributions to 
define $d_c$ in Eq.~\eqref{200814.8}. 
We also stress that the same method
that led us to the right value of $\log a\simeq 1/2$ for the AC model gives a 
larger value $\log a\simeq 1$ for graviton-graviton scattering. 

Finally, it is important to point out that for $\log a\simeq 1$ the unitarity 
cutoff $\Lambda\simeq \pi G^{-1}$, and a dimensional estimate of the 
higher-order corrections to $s_P$ gives a typical size $|x|\simeq 2/3\pi\approx 20\%$.
Let us stress that this is the same estimate as for the $\sigma$ meson in 
QCD, whose pole position $s_\sigma$  shows a clear convergent pattern after 
unitarizing the subleading $\pi\pi$ PWAs 
\cite{Oller:1997ti,Albaladejo:2012te,Pelaez:2015qba}. It is, then, of great 
interest to investigate the unitarization of the NLO 
results~\cite{Dunbar:1994bn,Dunbar:1995ed} for graviton-graviton scattering in 
order to test the robustness of the LO prediction and to calculate the NLO 
correction to $s_P$.

%
\section{Conclusions and outlook}
\label{sec:conclu}
\def\theequation{\arabic{section}.\arabic{equation}}
\setcounter{equation}{0}

We have developed a unitarization formalism for calculating partial-wave 
amplitudes
in the presence of infinite-range forces. 
This formalism has been applied to pure general relativity to study 
graviton-graviton scattering in partial waves by unitarizing the Born terms. We 
have also applied it to other exactly solvable potentials, where we were able to 
retrieve non-trivial non-perturbative information about the scattering process.

In order to deal with the IR divergences associated with infinite-range 
interactions, 
we have defined a new $S$ matrix by removing a global phase factor whose 
presence is
well known since the seminal paper by S.~Weinberg on IR divergences for QED and 
in quantum
gravity~\cite{Weinberg:1965nx}.
This factor also emerges by studying the asymptotic dynamics at times $t\to \pm 
\infty$ and should be
removed to construct a well-behaved $S$ matrix, as established by the formalism 
of P.~Kulish and L.~Faddeev~\cite{Kulish.200121}.

Within our prescription, we arrive at the conclusion that the graviton-graviton 
$J=0$ partial-wave amplitude has a pole
at $s_P=(\varkappa-i2/(3\pi))\Lambda^2$, with $\varkappa$ quite smaller than the 
imaginary part $2/(3\pi)$ and where $\Lambda^2\propto G^{-1}$ is the ultraviolet 
cutoff scale.
This corresponds to a graviton-graviton scalar resonance with vacuum quantum 
numbers
$J^{PC}=0^{++}$, that we dubbed the {\it graviball}. This resonance  peaks at a 
value of $s$  significantly below $\Lambda^2$, because  $\varkappa\ll 1$.  
 This can be understood in simple terms if we think of the Lorentzian function 
stemming from the pole, cf. eq.~\eqref{200602.2}.
For the actual modulus squared of the partial-wave amplitude there is some extra 
$s$ dependence along the real $s$ axis,
but the feature that the peak lies at values of $s$ considerable smaller than 
$|s_P|$  still holds.
The presence of such a peaked resonant structure  would generate large 
corrections for graviton-graviton  scattering in $S$-wave within the
effective field theory of gravity. 
Indeed, given the  smallness of $\varkappa$,  the consequences of the graviball 
would be completely analogous to the well-known phenomenon appearing in the 
scalar isoscalar meson-meson sector
due to the lightest resonance of QCD, the $\sigma$.

The presence of the graviball is robust against the introduction of new `light' 
degrees of freedom (masses below $G^{-1/2}$). In this case, given $N$ new 
fields, the position of the pole $s_P$ decreases as $\sim1/N$. The graviball 
then becomes correspondingly lighter and narrower, with its resonance effects 
taking place at lower energies. If $\Lambda^2$  also scales like $\sim1/N$, as 
expected from general arguments  
\cite{gomez,Dvali:2007wp,Dvali:2014ila,han.200204.1,Dvali:2001gx}, then the 
relative position between the resonance and the cutoff would not change. On the 
other hand, in a self-healing scenario in which $\Lambda^2$ remained 
$\mathcal{O}(G^{-1})$ as $N$ increases the position of the resonance would drift 
towards energies much lower than the cutoff.

An intrinsic limitation of our scheme is related to the presence of an 
undetermined quantity $\log a$, coming from an artificial separation of hard and 
soft scales in loop diagrams. The non-perturbative result should substitute this 
uncertainty by a number which we expect to be $O(1)$. We have explored this 
ansatz, together with the other possible drawbacks of our method, by applying 
our unitarization prescription to restore two-body unitarity in a quantum 
mechanical toy-model with a structure resembling general relativity (the AC 
scattering model). We found that our methods do, indeed, reproduce the first 
bound and resonance states of the spectrum of this model for natural values of 
$\log a$. Similarly, we studied the consequences for the unitarization of 
Coulomb potential. For this case it is interesting to point out that  the 
unitarization method yields an approximation valid at higher energies  than the 
typical scale $\Lambda$ for the problem. In this regime, we also managed to 
reproduce the existence and energy of the deepest bound state. 

Finally, we have presented a first attempt to  constrain the values of the $\log 
a$ from studying the presence of the resonances at  $d>4$, and imposing the 
value of $\log a$ that minimizes the effect coming from the IR divergences. 
Quite remarkably, these studies reproduce the resonance of the toy model, and 
generate a value $\log a\approx 1$ for graviton-graviton scattering. If this is 
confirmed by other calculations, it would lead to the conclusion that a 
graviball exists at energies below the nominal cutoff of the theory, $G^{-1}$.

There are several directions worth exploring in the future. On one hand, it 
would be important to understand if the resonance survives after introducing 
higher order corrections to the graviton-graviton scattering. These processes at 
one and two-loop level are known in the literature with different matter 
content, e.g. \cite{Dunbar:1994bn,Abreu:2020lyk,Goroff:1985th,Bern:2020gjj}. 
Similarly, string theory amplitudes may be relevant  for this purpose, see 
e.g.~\cite{Alonso:2019ptb,Schwarz:1982jn}, as may be  
other inputs from other proposals to UV complete theories of gravity, 
e.g.\cite{weinberg:safe,Niedermaier:2006wt,Codello:2008vh,Falls:2014tra,
Blas:2009qj,Steinwachs:2020jkj}.
We  stress that a  very well defined program is to extend our  unitarization 
formalism to  coupled channels involving other (massive) particle species. 
Similarly, even if our methods are not able to cure  all the problems with 
gravitational scattering  amplitudes that hampered 
the application of the $S$-matrix theory 
to gravitational 
amplitudes~\cite{Giddings:2009gj,Bellazzini:2019xts,Tokuda:2020mlf,
Alberte:2020jsk}, they 
may eventually suggest possible new approaches.  Finally, it seems pertinent to 
explore the connection to other non-perturbative approaches to graviton-graviton 
scattering. In particular, the question whether the bound states investigated in 
\cite{Dvali:2014ila,Dvali:2008fd,Guiot:2020pku}  may  be accessible by our 
methods, or if any hint of the graviball may appear in AdS/CFT analysis (see e.g. the recent work in ref.~\cite{Guerrieri:2021ivu}).

We would like to mention also that having extended the nonperturbative unitarization methods familiar in hadron physics to infinite-range interactions allows one to apply them to account for the Coulomb contributions in two-body  scattering of charged particles, like $\pi^+\pi^-$, $\pi^\pm p$ or $K^\pm p$, at the same time as the strong interactions are unitarized. This should translate in a more accurate decription of the data affected by Coulomb scattering near threshold, as well as in a better determination of the free parameters fitted in those processes. E.g. in $K^\pm p$ scattering there are many data that exhibit large Coulomb corrections just above theshold that could be treated in an improved manner employing the present approach \cite{Guo:2012vv}. 

Before closing, let us remark that our analysis has dealt with the {\it 
theoretical} aspects of the graviball. However, there may also be {\it 
observational} consequences worth exploring after assuming that the resonance 
survives in UV-complete models. This may happen in the processes with the 
highest possible energies in the gravitational degrees of freedom. The natural 
candidate would be the impact in the studies of primordial
gravitational waves and its experimental searches. For instance, the presence of 
a scalar resonance that
strongly couples to graviton-graviton could produce a reduction of tensor perturbations
in the background of gravitational waves expected from inflationary models, with 
the subsequent
reduction of expected $B$-mode signals in the polarization of the CMB. This may 
be difficult in standard models of inflation, where the viable tensor-to-scalar
ratio imply a 
 hierarchy between the scale of inflation and $G^{-1}$, e.g. 
\cite{Antoniadis:2014xva}. This result may be altered in theories with more 
degrees of freedom, where the resonance may happen at parametrically smaller 
values, while the scale of inflation may remain the same  \cite{Kleban:2015daa}. 
However, since the   cut-off of the theory may also scale with $\sqrt{N}$, this 
possibility requires more investigation.
Also, in theories with large enough number of light degrees of freedom  
\cite{Dvali:2007wp,Arkani-Hamed:2016rle,Dvali:2007hz,Dvali:2007wp} (or with 
large extra dimensions)
  the resonance may also modify the gravitational phenomena at scales as low as 
$\mu m$ (corresponding to a UV cutoff around the TeV scale) 
\cite{ArkaniHamed:1998rs,Antoniadis:1998ig,Dvali:2001gx,Csaki:2004ay,
Randall:1999ee,Randall:1999vf,Giudice:2016yja}.
 In these cases, the fundamental cutoff of the theory evolves with $N$ as 
$\Lambda^2=G^{-1}/N$. 
Within our calculation, this implies that $x$ and $\omega$ stay put but then 
$|s_P|=|x|\Lambda^2$ in absolute terms decreases as $1/N$. This opens 
interesting phenomenological possibilities, for tests of gravity at the $\mu$m 
scales and even at the level of collider physics, that are left for future 
studies.

\section*{Acknowledgements}

The authors are grateful to R.~Alonso, A.~Dobado, J. Donoghue, D. G. Figueroa, 
M.~Herrero-Valea, J.~Penedones, S. Patil and G. Dvali for valuable discussions. 
This work has been supported in part by the  FEDER (EU) and MEC (Spain) Grants 
FPA2016-77313-P, PID2019-106080GB-C22 and PGC2018-102016-A-I00, and the 
``Ram\'on y Cajal'' program RYC-2016-20672.


\appendix


\section{Graviton states and partial-wave amplitudes}
\label{app_states}
\def\theequation{\Alph{section}.\arabic{equation}}
\setcounter{equation}{0} 

To fix notation we first define in detail the one- and two-graviton states that 
we use. The general procedure that we follow to define our states is similar to 
that in Ref.~\cite{martin.200705.1}.
 The one-graviton states $|\vp,\lambda\ra$ result from the action of a standard 
rotation $R(\hvp)$ on a state with the same
helicity and three-momentum along the $\vz$ axis:
\begin{align}
\label{200116.2}
|\vp,\lambda\rangle &= R(\hvp)|p\vz,\lambda\rangle~.
\end{align}
The standard rotation transforms $\vz$ to $\vp$, namely, $R(\hvp)\vz = 
\hat{\vp}$.
In our convention we define it explicitly as
\begin{align}
\label{200116.3}
R(\hvp)=&R_z(\phi)R_y(\theta)~,\\
\vp=&p(\sin\theta \cos\phi,\sin\theta \sin\phi,\cos \theta)~,\nn
\end{align}
where $\theta$ and $\phi$ are the polar and azimuthal angles for $\vp$.

The two-graviton states in the center-of-mass frame (CM), $|\vp,\lm_1\lm_2\ra$, 
with momentum $\vp$ and helicities $\lambda_1$ and
$\lambda_2$, are defined as
\begin{align}
\label{200116.4}
|\vp,\lm_1\lm_2\ra&=|\vp,\lm_1\ra|-\vp,\lm_2\ra~.
\end{align}
It is straightforward to show that they can also be expressed in terms of the 
standard rotation $R(\hvp)$ as
\begin{align}
\label{200116.5}
|\vp,\lm_1\lm_2\ra&=R(\hvp)|p\vz,\lm_1\ra |-p\vz,\lm_2\ra~.
\end{align}
The expansion of these states into states with well defined total angular 
momentum $J$ and third component $J_z=M$, proceeds by noticing that the states 
$|p\vz,\lm_1\lm_2\ra$ have $M=\lambda_1-\lambda_2$. Therefore,
\begin{align}
\label{200116.6}
|p\vz,\lm_1\lm_2\ra&=2\pi\sum_J \sqrt{2J+1}|pJ\lm,\lm_1\lm_2\ra,
\end{align}
where $\lm=\lm_1-\lm_2$
and $2\pi\sqrt{2J+1}$ is set as normalization factor, cf. Eq.~\eqref{200116.8} 
below.
After acting on both sides of Eq.~\eqref{200116.6} with the standard
rotation $R(\hvp)$, we arrive at the expression of the plane-wave states in 
terms of those
in the spherical basis,
\begin{align}
\label{200528.1}
|\vp,\lm_1\lm_2\ra&=2\pi\sum_J\sum_{M=-J}^J(2J+1)D^{(J)}_{M\lm}(\phi,\theta,
0)|pJM,\lm_1\lm_2\ra~.
\end{align}
Here, $D^{(J)}_{M\lambda}(\phi,\theta,0)$ is the rotation matrix specified by 
the Euler angles corresponding to the standard rotation $R(\hvp)$ 
for the irreducible representation of the rotation group with angular momentum 
$J$~\cite{rose.170517.1}. By inverting this expression, one finds 
$|pJ\lm,\lm_1\lm_2\ra$ in terms of the $|\vp,\lm_1\lm_2\ra$ states as, 
\begin{align}
\label{200116.7}
|pJM,\lm_1\lm_2\ra&=\frac{\sqrt{2J+1}}{8\pi^2}\int_0^{2\pi}d\phi\int_{-1}^{+1}
d\!\cos\theta \, 
  D^{(J)}_{M\lm}(\phi,\theta,0)^* |\vp,\lm_1\lm_2\ra~.
\end{align}
The normalization is chosen such that
\begin{align}
\label{200116.8}
\la 
p'J'\lm',\lm_1'\lm_2'|pJ\lm,\lm_1\lm_2\ra&=\frac{2}{\pi}\delta_{\lm'_1\lm_1}
\delta_{\lm'_2\lm_2}~, 
\end{align}
where a global $(2\pi)^4\delta^{(4)}(P_f-P_i)$  has been factorized out because 
of the total four-momentum
conservation.
In obvious notation, $P_f$ and $P_i$ are the total four-momenta of the (left) 
final and
(right) initial states, respectively.

Now we take explicitly into account the Bose-Einstein symmetric character of the 
two-graviton states which is indicated
in the following by a subscript $\S$ in the ket,
\begin{align}
\label{200116.9}
|\vp,\lm_1\lm_2\ra_\S&=\frac{1}{\sqrt{2}} 
\big[|\vp,\lm_1\lm_2\ra+|-\vp,\lm_2\lm_1\ra\big]~,
\end{align}
which can also be expressed in terms of the standard rotation $R(\hvp)$ as
\begin{align}
\label{200116.10}
|\vp,\lm_1\lm_2\ra_\S&=R(\hat{\vp})|p\vz,\lm_1\lm_2\ra_\S~.
\end{align}
Its expansion in the  basis with well-defined angular-momentum can be obtained
by proceeding analogously to the derivation of Eq.~\eqref{200528.1} . It reads, 
 
\begin{align}
\label{200116.11}
|\vp,\lm_1\lm_2\ra_\S&=\sum_{J,M}2\pi\sqrt{2J+1}
D^{(J)}_{M\lambda}(\phi,\theta,0)\frac{1}{\sqrt{2}}
\big[|pJM,\lm_1\lm_2\ra+(-1)^J|pJM,\lm_2\lm_1\ra\big]\, .
\end{align}
In the following we denote by $|pJM,\lm_1\lm_2\ra_\S$ the states
\begin{align}
\label{170517.2}
|pJM,\lm_1\lm_2\ra_\S=\frac{1}{\sqrt{2}}\left( 
|pJM,\lm_1\lm_2\ra+(-1)^J|pJM,\lm_2\lm_1\ra \right)~,
\end{align}
which is (anti)symmetric under the exchange of the helicities of the two 
gravitons if $J$ is (odd) even.

It is straightforward to invert Eq.~\eqref{200116.11} to find
\begin{align}
\label{200117.1}
|pJM,\lm_1\lm_2\ra_\S
&=\frac{\sqrt{2J+1}}{8\pi^2}\int_0^{2\pi}d\phi\int_{-1}^{+1}d\!\cos\theta\,D^{
(J)}_{M\lm}(\phi,\theta,0)^* 
|\vp,\lm_1\lm_2\ra_\S~.
\end{align}

Let us discuss the formal definition of the partial waves amplitudes  (PWAs)
for graviton-graviton scattering. 
The PWA expansion of a two by two scattering process follows
from Eq.~\eqref{200116.11}, which is more conveniently  applied when the final 
three-momentum $\vp'$
lies in the $xz$ plane with $\phi=0$, namely 
$\vp'=\vp'_{xz}=(\sin\theta,0,\cos\theta)$. Thus,
\begin{align}
\label{200528.2}
{_\S\la}\vp',\lm'_1\lm'_2|T|p\hvz,\lm_1\lm_2{\ra_\S}&=4\pi^2\sum_J(2J+1)\sum_{
M=-J}^Jd^{(J)}_{M\lm'}(\theta)
\underbrace{d^{(J)}_{M\lm}(0)}_{=\delta_{M\lm}}{_\S\la} p J,\lm'_1\lm'_2|T|p 
J,\lm_1\lm_2{\ra_\S}\\
&=4\pi^2\sum_J(2J+1)d^{(J)}_{\lm\lm'}(\theta) {_\S\la} p J,\lm'_1\lm'_2|T|p 
J,\lm_1\lm_2{\ra_\S}~,\nn
\end{align}
and only the Wigner (small) d-matrix $d^{(J)}_{\lm\lm'}(\theta)$ enters in the 
angular projection.\footnote{We have dropped the label $M$ of the third 
component of total angular momentum 
because the PWAs do not depend on it due to the Wigner-Eckart theorem applied to 
rotational symmetry.}  
We can invert Eq.~\eqref{200528.1} by taking into account the orthogonality 
property of the Wigner d-matrix
functions 
\cite{rose.170517.1},
\begin{align}
\label{200528.3}
\int_{-1}^1 {d\cos\theta}\, 
d_{\lm\lm'}^{(j_1)}(\theta)d_{\lm\lm'}^{(j_2)}(\theta)&=\frac{2}{2j_1+1}\delta_{
j_1j_2}~.
\end{align}
Then, the partial-wave amplitude 
$\bar{T}^{(J)}_{\lm'_1\lm'_2,\lm_1\lm_2}(s)\equiv {_\S\la} 
pJ,\lm_1'\lm_2'|T|pJ,\lm_1\lm_2\ra_\S$  is given by 
\begin{align}
\label{200119.4b}
\bar{T}^{(J)}_{\lm'_1\lm'_2,\lm_1\lm_2}(s)&=
\frac{1}{8\pi^2} \int_{-1}^{+1} d\!\cos\theta' \, d^J_{\lm\lm'}(\theta')\,
_\S\la \vp'_{xz},\lm'_1\lm'_2|T|p\vz,\lm_1\lm_2\ra_\S~. 
\end{align}

The symmetric states $|pJM,\lm_1\lm_2\ra_S$ for $\lm_1=\lm_2$ can only have 
$J=$even and then $|pJM,\lm_1\lm_1\ra_S=\sqrt{2}|pJM,\lm_1\lm_1\ra$. Therefore, 
these  states are normalized to $4/\pi$. However, the symmetric states with 
$\lm_1\neq \lm_2$ can sustain both $J$ even and odd and they are normalized to 
$2/\pi$. Because of this different normalization by a factor of $2$, the 
relation between the $S$- and $T$-matrix in PWAs is different,  and this is the 
reason behind the factor $2^{|\lm|/4}$ in Eq.~\eqref{200121.11}.

\section{Partial-wave amplitudes in arbitrary dimensions}
\label{app.200805.1}
\def\theequation{\Alph{section}.\arabic{equation}}
\setcounter{equation}{0}   

In this section we derive the PWAs as a function of the space-time dimensions 
$d$ for graviton-graviton and spinless-particle scattering. 
The needed  angular integrations in $d=4-2\ep$ dimensions are taken from 
Ref.~\cite{Somogyi:2011ir}. 
One of them is the area  of the unit sphere in $d-1$ spatial dimensions 
$\Omega_{d-1}$, which reads
\begin{align}
\label{200720.4}
\Omega_{d-1}=(4\pi)^{d/2-1}\frac{\Gamma(d/2-1)}{\Gamma(d-2)}~.
\end{align}

We start the discussion by considering the expansion of the plane-wave 
two-graviton states
in the basis of states with definite $J$,
\begin{align}
\label{170513.11b}
|\vp,\lm_1\lm_2\ra&=\sum_{J,M}C_J D^{(J)}_{M\lm}(\hvp,0)|pJ M,\lm_1\lm_2\ra~,
\end{align}
which is analogous to Eq.~\eqref{200528.1} but now the coefficient $C_J$ depends 
on  $d$. The second argument is set equal to zero because the $\hvz$ axis is 
always rotated to $\hvp$ by the standard rotation, which is completely specified 
by $\hvp$ once we set, by convention, a first rotation around the $\hvz$ axis 
equal to the identity (analogously to $d=4$). For brevity in the notation we 
remove this zero in the following  and keep only $\hvp$ in the argument of  the 
standard rotation.

To write the states with definite $J$ in terms of the plane-wave ones from 
Eq.~\eqref{170513.11b}, we use the orthogonality property of the rotation 
matrices in $d$ dimensions, which is a consequence of the unitary character of 
the rotations. 
It can be written in $d$ dimensions as, 
\begin{align}
\label{200721.3}
\int d\Omega_{d-1}
D^{(J')}_{M'\lm}(\hvp)^* D^{(J)}_{M\lm}(\hvp)=
\frac{\Omega_{d-1}}{2J(d)+1}\delta_{JJ'}\delta_{MM'}~. 
\end{align}
It is clear that this orthogonality relation requires $\Omega_{d-1}$ because for 
$J=0$ the rotation matrix is the identity matrix, and then we simply have the 
volume of the unit sphere in $d-1$ dimensions. The factor $2J+1$ in $d=4$ would 
become dependent on $d$ for general $J$, and this is why we have written it as $ 
2J(d)+1 $, in order to keep track of this fact for $J\neq 0$. We do not really 
need to be more specific about it because it will indeed disappear from the 
final expression for the calculation of the PWAs in terms of he scattering 
amplitudes. Also,   we are only interested in 
$J=0$, and for this case the coefficient in front of the Kronecker delta 
functions in the right-hand side  of Eq.~\eqref{200721.3} is $\Omega_{d-1}$. 
Multiplying Eq.~\eqref{170513.11b} by $D^{(J)}_{M\lambda}(\hvp)^*$ and applying 
the orthogonality relation of Eq.~\eqref{200721.3} one can express 
$|pJM,\lm_1\lm_2\ra$ as  
\begin{align}
\label{200721.4}
|pJM,\lm_1\lm_2\ra&=\frac{2J(d)+1}{\Omega_{d-1} C_J}\int d\Omega_{d-1}
D^{(J)}_{M\lm}(\hvp)^* |\vp,\lm_1\lm_2\ra~.
\end{align}
Regarding the normalization of the states with definite $J$ we keep the same one 
as in $d=4$, given in Eq.~\eqref{200116.8}, by fixing properly $C_J$. 
This choice guarantees that we do not have to change the unitarity loop function 
$g(s)$ because its discontinuity does not change and then we have the same DR in 
the variable $s$.

The one-graviton states are normalized such that $\la 
\vp'\lm'|\vp\lm\ra=2p(2\pi)^{d-1}\delta(\vp'-\vp)\delta_{\lm'\lm}$, 
The two graviton states, after removing the global factor 
$(2\pi)^d\delta^{(d)}(P'-P)$ associated with the free-motion of the CM with the 
total momentum $P$, are correspondingly normalized to 
$2^{d+1}\pi^{d-2}/p^{d-4}$. As a result,
\begin{align}
\label{200721.6}
\la pJ'M',\lm_1'\lm_2'|pJM,\lm_1\lm_2\ra&
=\frac{(2J(d)+1)2^{d+1}\pi^{d-2}}{\Omega_{d-1}C_J^2p^{d-4}}
\delta_{\lm'_1\lm_1}\delta_{\lm'_2\lm_2}\delta_{J'J}=\frac{2}{\pi}\delta_{
\lm'_1\lm_1}\delta_{\lm'_2\lm_2}\delta_{J'J}~,
\end{align}
The coefficient $C_J$ is then 
\begin{align}
\label{200721.7}
C_J&=\sqrt{\frac{(2J(d)+1)2(2\pi)^{d-1}}{\Omega_{d-1}p^{d-4}}}~.
\end{align}

Let us now consider the calculation of the PWAs.  By
directly implementing Eq.~\eqref{170513.11b} with the coefficient $C_J$ just 
calculated, one has
\begin{align}
\label{200721.8}
_S\la\vp',\lm_1'\lm_2'|T|p\vz,\lm_1\lm_2\ra_S&=
\sum_{J',M'}\sum_{J,M}C_J C_{J'}
D^{(J')}_{M'\lambda'}(\hvp)^*
\underbrace{_S\la pJ'M',\lm_1'\lm_2'|T|pJ \lm,\lm_1\lm_2\ra_S}_{\propto 
\delta_{M'\lm}\delta_{J'J}}\\
&=\sum_J\frac{(2J(d)+1)2(2\pi)^{d-1}}{\Omega_{d-1}p^{d-4}}
D_{\lm\lm'}^{(J)}(\hvp)^*\,_S\la p J,\lm'_1\lm'_2|T|p J,\lm_1\lm_2\ra_S~.\nn
\end{align}
The next step is to isolate the PWA by using the orthogonality properties of the 
rotation matrices in $d$ dimensions, cf. Eq.~\eqref{200721.3}. Then,
\begin{align}
\label{200721.9}
_S\la pJ,\lm_1'\lm_2'|T|pJ,\lm_1\lm_2\ra_S&=
\frac{p^{d-4}}{2(2\pi)^{d-1}}\int d\Omega_{d-1}
D^{(J)}_{\lambda\lambda'}(\hvp)\,
_S\la\vp',\lm'_1\lm'_2|T|p\vz,\lm_1\lm_2\ra_S~.
\end{align}
Let us also mention that if the Born term term can be expressed in terms of 
products of momenta, as in gravity \cite{Dunbar:1994bn}, we can then use the 
same expression as in $d=4$ in Eq.~\eqref{200721.9}, except for an overall 
change of dimensions in the coupling.

We now derive the expression for the PWAs with varying dimensions for the 
scattering of two {\it massive spinless} particles. The normalization of the 
one-particle states is the Lorentz invariant one 
$2E(p)(2\pi)^{d-1}\delta(\hvp'-\hvp)$, where $E(p)=\sqrt{\mu^2+\vp^2}$ and $\mu$ 
is the mass of either particle. For simplicity, in the discussion we 
particularize  to equal mass scattering, the one we need here, though its 
generalization to particles with different masses is straightforward. 
We also introduce the standard rotation $R(\hvp)$ that takes $\hvz$ to $\hvp$. 
With the one-particle states defined such that $|\vp\ra=R(\vp)|\hvz\ra$, then we 
also have a relation analogous to Eq.~\eqref{200116.10} for a state of two 
spinless particles, $|\vp\ra\otimes|-\vp\ra$,  
\begin{align}
\label{200808.1}
|\vp\ra\otimes |-\vp\ra=R(\hvp)|p\hvz\ra\otimes |-p\hvz\ra~.
\end{align}
In order to simplify the notation in the following we simply 
denote a two particle state also by $|\vp\ra$, and only use this symbol 
only referred to this state. Then, a $|p\hvz\ra$ state is an eigenvector of the 
third component of the angular-momentum operator with null eigenvalue. 
Therefore, we can write it in terms of partial-wave states with definite angular 
momentum $J$ and third component $M$, $|pJM\ra$, as
\begin{align}
\label{200808.2}
|p\hvz\ra&=\sum_{J}C_J|pJ0\ra~,
\end{align}
From here on the procedure is completely analogous to the one developed for the 
graviton-graviton states. Instead of Eq.~\eqref{170513.11b} we have now
\begin{align}
\label{200808.3}
|\vp\ra&=\sum_{J}\sum_{M=-J}^JC_J D^{(J)}_{M0}(\hvp)|pJ M\ra~.
\end{align}
After using the orthogonality relation between the rotation matrices,  
Eq.~\eqref{200721.3}, we can invert the expansion in 
Eq.~\eqref{200808.3} with the result,
\begin{align}
\label{200808.4}
|pJ M\ra&=\frac{2J(d)+1}{\Omega_{d-1}C_J}\int 
d\Omega_{d-1}D^{(J)}_{M0}(\hvp)^*|\vp\ra~.
\end{align}
The coefficient $C_J$ is fixed as usual by imposing the normalization of the 
partial-wave states to be the same as in $d=4$. Namely, 
\begin{align}
\label{200808.5}
\la p J'M'|pJM\ra&=\frac{4\pi\sqrt{s}}{p}\delta_{J'J}\delta_{M'M}~.
\end{align}
For the kinematics of two {\it relativistic} particles of equal mass $\mu$ the 
normalization of the two-body state $\la \vp'|\vp\ra$ is 
$2^{d+1}\pi^{d-2}E(p)/p^{d-3}$. This allows us to fix $C_J$, analogously as it 
was done above in Eq.~\ref{200721.6}, 
with the result
\begin{align}
\label{200809.1}
C_J=\sqrt{\frac{(2J(d)+1)2^{d-2}\pi^{d-3}}{\Omega_{d-1}p^{d-4}}}~.
\end{align}
We are then ready to obtain the expressions for the decomposition in PWAs and 
the calculation of the latter ones, cf. Eqs.~\eqref{200721.8} and 
\eqref{200721.9} above, which now read, respectively,
\begin{align}
\label{200809.2}
_S\la\vp',\lm_1'\lm_2'|T|p\vz,\lm_1\lm_2\ra_S&=
\sum_J\frac{(2J(d)+1)2^{d-2}\pi^{d-3}}{\Omega_{d-1}p^{d-4}}
D_{00}^{(J)}(\hvp)^*\,_S\la p J0|T|p J0 \ra_S~, \\
\label{200809.3}
_S\la pJ0|T|pJ0\ra_S&=
\frac{p^{d-4}}{2^{d-2}\pi^{d-3}}\int d\Omega_{d-1}
D^{(J)}_{00}(\hvp)\,
_S\la\vp'|T|\vp\ra_S~.
\end{align}
In the case of two {\it non-relativistic} spinless particles the total energy of 
the system as a function of $p$ now reads $\vp^2/2m$ with $m=\mu/2$, the reduced 
mass of the two particles. The normalization of the different states also 
changes: The one-particle states have the normalization 
$(2\pi)^3\delta(\vp'-\vp)$, and the normalization for the two-particle states in 
the plane-wave basis is $(2\pi)^{d-2}\delta(\hvp'-\hvp)/mp^{d-3}$. Regarding the 
states in the plane-wave basis the normalization is fixed to 
\begin{align}
\label{200809.4}
\la p J'M'|pJM\ra&=\frac{\pi}{mp}\delta_{J'J}\delta_{M'M}~.
\end{align}
One has the same equations \eqref{200808.1}--\eqref{200808.4} as above, and from 
there one deduces that the  coefficient $C_J$ is the same as in 
Eq.~\eqref{200809.1} for this case too. Therefore, the decomposition in PWAs and 
their calculations is again given by Eqs.~\eqref{200809.2} and \eqref{200809.3}, 
respectively.
\bibliography{gravball.bib}
\bibliographystyle{apsrev4-1}

\end{document}